%% file: main.tex
\documentclass{jpp}
\usepackage{graphicx}
\usepackage{color,soul}
\usepackage{hyperref}

\usepackage[utf8]{inputenc}
\usepackage[T1]{fontenc}
\usepackage{amsmath}
\usepackage{longtable}
\usepackage[table,xcdraw]{xcolor}
\usepackage{comment}

\providecommand{\DIFdel}[1]{} % Don't show deleted text

\usepackage{titletoc}

\usepackage[sectionbib]{bibunits}
\defaultbibliographystyle{jpp}
\defaultbibliography{main}

\shorttitle{Learning collision operators using differentiable simulators}
\shorttitle{Learning collision operators from plasma phase space data}

\shortauthor{
    D. D. Carvalho, 
    P. J. Bilbao, 
    W. B. Mori, 
    L. O. Silva
    and E. P. Alves}

\title{Learning collision operators from plasma phase space data using differentiable simulators}
\author{
    Diogo~D.~Carvalho\aff{1,2}
  \corresp{\email{diogo.d.carvalho@tecnico.ulisboa.pt}},
    Pablo~J.~Bilbao\aff{1,3},
    Warren~B.~Mori\aff{4},
    Luis~O.~Silva\aff{1}
    \and E.~Paulo~Alves \aff{2,4}
}

\affiliation{
\aff{1} GoLP/Instituto de Plasmas e Fus\~ao Nuclear, Instituto Superior T\'ecnico,
Universidade de Lisboa, 1049-001 Lisbon, Portugal
\aff{2}Mani L. Bhaumik Institute for Theoretical Physics, University of California, Los Angeles, California, USA
\aff{3}The Rudolf Peierls Centre for Theoretical Physics, University of Oxford, Oxford OX1 3NP, UK
\aff{4}Department of Physics and Astronomy 
University of California, Los Angeles, California, USA

}

\begin{document}

\maketitle

% max 250 words

\begin{abstract}
We propose a methodology to infer collision operators from phase space data of plasma dynamics. Our approach combines a differentiable kinetic simulator, whose core component in this work is a differentiable Fokker-Planck solver, with a gradient-based optimisation method to learn the collisional operators that best describe the phase space dynamics. We test our method using data from two-dimensional Particle-in-Cell simulations of spatially uniform thermal plasmas, and learn the collision operator that captures the self-consistent electromagnetic interaction between finite-size charged particles over a wide variety of simulation parameters. We demonstrate that the learned operators are more accurate than alternative estimates based on particle tracks, while making no prior assumptions about the relevant time scales of the processes and significantly reducing memory requirements. We find that the retrieved operators, obtained in the non-relativistic regime, are in excellent agreement with theoretical predictions derived for electrostatic scenarios. Our results show that differentiable simulators offer a powerful and computational efficient approach to infer novel operators for a wide rage of problems, such as electromagnetically dominated collisional dynamics and stochastic wave-particle interactions.
\end{abstract}

\begin{bibunit}

\section{Introduction}

Plasma physics has seen significant advances in the ability to simulate nonlinear plasma dynamics from first principles. However, capturing the complex multi-scale dynamics of plasmas, especially in non-equilibrium and strongly coupled regimes, remains challenging~\citep{kushner2020plasma}. One such challenge is the development of collisional operators that capture the macroscopic dynamics in regimes where analytical theory does not exist or is expected to fail, e.g., regimes where collisional dynamics are mediated by electromagnetic interactions (relativistic temperatures)~\citep{braams1987differential, pike2016transport} or when assumptions about the predominance of small-angle scattering events break down (strongly coupled regimes)~\citep{baalrud2012transport, baalrud2014extending}. These regimes occur for example in inertial confinement fusion (ICF) experiments~\citep{atzeni2004physics}, dense astrophysical objects (e.g. white dwarfs, neutron stars, core of gas giants)~\citep{chabrier2006dense}, and in dusty~\citep{merlino2004dusty} and ultracold~\citep{killian2007ultracold} astrophysical and laboratory plasmas. It is therefore important to design first-principles numerical simulations to probe deviations from the theory and to extract improved models which can serve as the building blocks for future theoretical developments.

For weakly coupled plasmas, where small-angle Coulomb collisions dominate the dynamics, the Fokker-Planck (FP) collision operator is expected to provide a good description of the collisional dynamics~\citep{landau1936kinetische,rosenbluth1957fokker}. The operator can be written as~\citep{rosenbluth1957fokker}:
\begin{equation}
    \left(\frac{\partial f}{\partial t}\right)_{FP} = - \nabla_{\boldsymbol{v}}\boldsymbol{\cdot}(\mathsfbi{A}f) + \frac{1}{2}\nabla_{\boldsymbol{v}}\boldsymbol{\cdot}\left[\nabla_{\boldsymbol{v}}\boldsymbol{\cdot}(\mathsfbi{D}f)\right]
\label{eq:fp_equation}
\end{equation}
where $\boldsymbol{v} \in \mathbb{R}^N$ is the velocity vector in the $N$-dimensional phase space, $f(\boldsymbol{v}, t): \mathbb{R}^N \times \mathbb{R} \rightarrow \mathbb{R}$ is the underlying spatially distribution function (for simplicity here we assume that $f$ does not depend on the position $\boldsymbol{r}$), $\mathsfbi{A}(\boldsymbol{v}, t): \mathbb{R}^N \times \mathbb{R} \rightarrow \mathbb{R}^N$ is the advection (drift) vector, and $\mathsfbi{D}(\boldsymbol{v}, t): \mathbb{R}^N \times \mathbb{R} \rightarrow \mathbb{R}^{N\times N}$ is the diffusion matrix. 

The FP operator is often coupled with the Vlasov and Maxwell systems of equations to self-consistently model collisional transport and the kinetic evolution of particle distribution functions in phase space~\citep{bell2006fast, tzoufras2011vlasov, thomas2012review, tzoufras2013multi, bell2024fastvfp}. Furthermore,
it has been shown that other stochastic acceleration processes mediated by turbulence (and not collisional dynamics) can also be studied using this description~\citep{wong2020first, wong2025energy, camporeale2022data}.

Significant theoretical work has been made to determine the mathematical form of the advection and diffusion coefficients that capture the dynamics of interest~\citep{rosenbluth1957fokker, braams1987differential, reynolds1997velocity}. Usually, these efforts result in collisional integrals, which can be costly to compute and to treat analytically. When theory is not available, both numerical and experimental tools have been used to estimate such coefficients, e.g.,  particle track information from numerical simulations is commonly used to infer the coefficients for a setup of interest~\citep{hockney1971measurements, matsuda1975collisions, wong2020first, wong2025energy}. While particle tracks provide all the required information to estimate the coefficients, it is often unfeasible to store and post-process all the necessary data for large-scale simulations and broad parameter scans. Furthermore, to accurately estimate the coefficients (at run-time), and, therefore, avoid storing the full tracks, one needs to a priori have an estimate of all the acceleration (forces) time scales involved (not only collisions). In particular, one must guarantee that the time-step over which velocity change statistics are computed fulfils the Markovian assumption underpinning the FP equation.

In this work, we propose a general-purpose framework to learn collision operators from phase space data (in this paper, obtained from kinetic numerical simulations) by rewriting the problem as an optimization task, which we tackle using a differentiable kinetic simulator and gradient-based minimization methods. The simulator evolves phase space distributions using a differentiable FP solver, where the advection and diffusion tensors are learned from data. 
By making use solely of phase space data, we greatly reduce the memory requirements associated with storing particle tracks. Furthermore, the differentiable simulator allows us to extract operators that accurately capture the long-term dynamics without prior knowledge of the involved time scales. The technique is general and can be used with other sets of phase space information, e.g., from experimental~\citep{bergeson2025experimental} or observational~\citep{camporeale2022data} data, and the simulator can be extended to include different collision operator forms, external fields, or other options deemed relevant for the problem under study.

As a first test case, we use these tools to extract the collision operator for an electromagnetic Particle-in-Cell (PIC) code~\citep{birdsall2018plasma}, which evolves the self-consistent fields of finite-size particles, and compare the results against theory originally proposed by~\cite{langdon1970nonphysical} and \cite{langdon1970theory}, and against coefficients derived from particle tracks. This is a good test bench since theoretical predictions assume that a FP-type collision operator correctly captures the self-consistent collisional dynamics of the PIC algorithm. We note that the collision operator we aim to extract is not explicitly solved by the PIC algorithm, e.g., via a Monte-Carlo module \citep{takizuka1977binary} or a grid-based Langevin operator~\citep{manheimer1997langevin}. The learned operator is instead a consequence of the self-consistent interactions between the finite-size particles and the generated fields in a regime where small-angle scattering events are expected to dominate.

This paper is structured as follows. In Sections~\ref{sec:pic_collision_operator} and \ref{sec:related_machine_learning_work} we review the underpinning theoretical and numerical results. In Section~\ref{sec:methods} we describe the methods deployed in this paper to learn collision operators. In Section~\ref{sec:results} we present the main results for the collision operator learned for plasmas of finite-size particles. Finally, in Section~\ref{sec:conclusions} we state the conclusions and describe directions for future work.

\subsection{PIC Collision Operator}
\label{sec:pic_collision_operator}

The PIC algorithm~\citep{birdsall2018plasma} solves the Klimontovich equation~\citep{klimontovich1967statistical} for finite-size particles,\footnote{Also referred to as macroparticles or superparticles in the literature.} coupled with the Maxwell system of equations (or a subset) to self-consistently model the fields generated by the movement of the particles and advance their position over time. Since particles have a finite size, the electrostatic force between two particles is reduced at short distances (comparable to the particle radius), which significantly reduces the short-range collisional interactions compared to point-like particles; we note that the long-range interactions are correctly captured. Hence, the effective Coulomb collision operator for PIC-simulated plasmas is modified relative to that of real plasmas.

Theoretical work on the collision operator for plasmas composed of finite-size particles was originally made by \citet{okuda1970collisions}, \citet{langdon1970nonphysical}, and \citet{langdon1970theory}. \citet{langdon1970nonphysical} derived the first description of a FP collision operator for electrostatic PIC simulations, which includes the effect of aliasing due to the finite grid size, discrete time-step, shape function, among others. More recently,~\citet{touati2022kinetic} thoroughly re-derived these results and extended the description to include the effect of varying macroparticle weights.

Several numerical experiments were conducted to complement and verify some predictions using electrostatic PIC codes. \citet{hockney1971measurements} empirically studied collisional relaxation rates and inferred a scaling law for 2-D thermalization times for the Nearest Grid Point (NGP, zeroth-order interpolation) and Cloud-in-Cell (CIC, first-order interpolation) shape functions. \citet{matsuda1975collisions} performed direct comparisons between the predicted advection for Gaussian-shaped particles moving at $v=v_{th} = \sqrt{k_B T/m}$ and diffusion for particles at rest $v=0$ for both 2-D and 3-D simulations. They found that for ions the theoretical predictions were accurate within a $20\%$ error, while for electrons the predictions could deviate from the result up to a factor of $200\%$. They also demonstrated that, when imposing a strong external magnetic field, the collisionality of the simulated plasma decreases significantly if the gyroradius is smaller than the particle size. More recently, \citet{gatsonis2009three} and \citet{averkin2018parallel} studied deflection and slow-down times in 3-D PIC codes with unstructured Delaunay–Voronoi grids, and \citet{jubin2024numerical} studied numerical relaxation rates for 2-D low-temperature plasma setups using CIC shape functions while proposing mitigation strategies to reduce undesired thermalization effects due to finite-size particle collisions. Finally,~\citet{park2025numerical} complemented the work of~\citet{jubin2024numerical} by thoroughly exploring 1-D, 2-D, and 3-D scenarios.

In this work, we extend these previous results by providing a thorough comparison between predicted and measured advection and diffusion coefficients for a broad range of particle velocities ($v_{x,y}/v_{th} \in [0, 5])$ and simulation parameters (shape functions, grid resolution, particles per cell, thermal velocity). Furthermore, we conduct these tests for the first time using an electromagnetic PIC code~\citep{fonseca2002osiris}. We note that the theory of collisions in plasmas is based purely on electrostatic interactions, so this work provides an important first step on possible modifications of the collisional operators when including self-consistent electromagnetic interactions.  For all scenarios, we consider the 2-D version of OSIRIS~\citep{fonseca2002osiris} that utilizes a finite-difference time-domain (FDTD) Yee solver~\citep{yee1966numerical}, and a plasma consisting of a single mobile electron species moving over a uniform neutralizing ion background. 

Under these conditions, and when neglecting the effect of temporal and spatial aliasing, the theoretical advection $\mathsfbi{A}(\boldsymbol{v})$ and diffusion $\mathsfbi{D}(\boldsymbol{v})$ operators for the finite-size e-e collisions are given (in S.I. units) by (cf. derivation in Supplemental Material~\ref{app:derivation_pic_collision_operator}):
\begin{equation}
\mathsfbi{A}(\boldsymbol{v}) = \mathsfbi{A}^{(1)}(\boldsymbol{v}) + \mathsfbi{A}^{(2)}(\boldsymbol{v})
\label{eq:a_theory}
\end{equation}
\begin{equation}
\mathsfbi{A}^{(1)}(\boldsymbol{v}) = -\frac{\omega_{p} v_{th}}{N_{Dmac} 2^{5/2} \pi^{3/2}}
\int_{\mathbb{R}^2} \mathbf{d\boldsymbol{\tilde{k}}}
\frac{S_\rho^m(\boldsymbol{\tilde{k}})^4}{\left| \epsilon(\boldsymbol{\tilde{k}} \boldsymbol{\cdot} \boldsymbol{\tilde{v}}, \boldsymbol{\tilde{k}}) \right|^2 \tilde{k}_s^4}
\frac{\boldsymbol{\tilde{k}}_s(\boldsymbol{\tilde{k}}_s \boldsymbol{\cdot} \boldsymbol{\tilde{k}})(\boldsymbol{\tilde{k}} \boldsymbol{\cdot} \boldsymbol{\tilde{v}})}{\tilde{k}^3}
e^{-(\boldsymbol{\tilde{k}} \boldsymbol{\cdot} \boldsymbol{\tilde{v}})^2 / 2\tilde{k}^2}
\label{eq:a1_theory}
\end{equation}
\begin{equation}
\mathsfbi{A}^{(2)}(\boldsymbol{v}) = \frac{1}{2v_{th}}\nabla_{\boldsymbol{\tilde{v}}} \boldsymbol{\cdot} \mathsfbi{D}(\boldsymbol{v})
\label{eq:a_2_theory}
\end{equation}
\begin{equation}
\mathsfbi{D}(\boldsymbol{v}) =  \frac{\omega_{p} v_{th}^2}{N_{Dmac} 2^{3/2} \pi^{3/2}}
\int_{\mathbb{R}^2} \mathbf{d\boldsymbol{\tilde{k}}}
\frac{S_\rho^m(\boldsymbol{\tilde{k}})^4}{\left| \epsilon(\boldsymbol{\tilde{k}} \boldsymbol{\cdot} \boldsymbol{\tilde{v}}, \boldsymbol{\tilde{k}})\right|^2 \tilde{k}_s^4} 
\frac{(\boldsymbol{\tilde{k}}_s \boldsymbol{\otimes} \boldsymbol{\tilde{k}}_s)}{\tilde{k}}
e^{-(\boldsymbol{\tilde{k}} \boldsymbol{\cdot} \boldsymbol{\tilde{v}})^2 / 2\tilde{k}^2}
\label{eq:d_theory}
\end{equation}
where the pre-factor constants correspond to the electron plasma frequency $\omega_p = \sqrt{n_e e^2 /\varepsilon_0 m_e}$ (with $n_e$ being the electron number density, $e$ the electron charge, $\varepsilon_0$ the vacuum permittivity, and $m_e$ electron mass), and the number of macroparticles per Debye square $N_{Dmac} = n_{mac}\lambda_D^2$ where $\lambda_D = v_{th}/\omega_p$ is the electron Debye length and $n_{mac}=N_{ppc}/(\Delta_x\Delta_y)$ is the macroparticle number density (with $N_{ppc}$ being the number of particles per cell and $\Delta_{x,y}$ the grid resolution). The integral is defined over $\mathbb{R}^2$ and all vector quantities have two directions (e.g. $\boldsymbol{\tilde{v}} = (\tilde{v}_x, \tilde{v}_y)$, $\mathbf{d\boldsymbol{\tilde{k}}} = \mathrm{d}\tilde{k}_x\mathrm{d}\tilde{k}_y$,); $\boldsymbol{\tilde{v}} = \boldsymbol{v}/v_{th}$ is the particle velocity normalized to the thermal velocity, $\boldsymbol{\tilde{k}} = \boldsymbol{k}\lambda_D$ is the wave-vector normalized by the inverse of the electron Debye length, $\boldsymbol{\tilde{k}}_s = \boldsymbol{k_s} \lambda_D$ is a dimensionless vector associated with the FDTD field solver used where $\boldsymbol{k}_s =  (k_x \mathrm{sinc}(k_x\Delta_x/2), k_y\mathrm{sinc}(k_y\Delta_y/2))$ with $\mathrm{sinc}(x) = \sin(x)/x$. Finally, $S_\rho^m(\boldsymbol{\tilde{k}}) = \mathrm{sinc}(\tilde{k}_x \Delta_x/2\lambda_D)^{m+1}\mathrm{sinc}(\tilde{k}_y \Delta_y/2\lambda_D)^{m+1}$ is the charge deposition shape function of order $m$,
and $\epsilon$ the dielectric function defined as (cf. derivation in Supplemental Material~\ref{app:derivation_pic_collision_operator}):
\begin{equation}
\epsilon(\boldsymbol{\tilde{k}} \boldsymbol{\cdot} \boldsymbol{\tilde{v}}, \boldsymbol{\tilde{k}}) = 1 -
\frac{S_\rho^m(\boldsymbol{\tilde{k}})^2 (\boldsymbol{\tilde{k}}_s \boldsymbol{\cdot} \boldsymbol{\tilde{k}})}{2 \tilde{k}_s^2 \tilde{k}^2}
Z'\left( \frac{\boldsymbol{\tilde{k}} \boldsymbol{\cdot} \boldsymbol{\tilde{v}}}{\sqrt{2}\tilde{k}} \right)
\label{eq:dieletric_function}
\end{equation}
where $Z'(\xi) = \pi^{-1/2} \int_\mathbb{R} dt \,e^{-t^2}/(\xi - t)^2$ is the first derivative of the plasma dispersion function~\citep{fried2015plasma}. A more in-depth discussion regarding the derivation of the advection and diffusion coefficients (based on the more general form of the PIC collision operator derived by~\cite{touati2022kinetic}) is provided in Supplementary Material~\ref{app:derivation_pic_collision_operator}.

\subsection{Related Machine Learning Work}
\label{sec:related_machine_learning_work}

The machine learning community has proposed several approaches to learning operators that recover observed dynamical data (a type of \textit{inverse problem}). Common neural network (NN) based approaches include Neural Operators such as Deep Operator Networks~\citep{lu2019deeponet}, Fourier Neural Operators~\citep{li2020fourier}, among many others~\citep{kovachki2023neural} designed to learn mappings between function spaces. While these approaches are extremely flexible, since the operator form is not hard-constrained, they require substantial amounts of training data to obtain an accurate general operator. Furthermore, the learned operator is, by construction, not interpretable. On the other hand, Physics-Informed Neural Networks~(PINNs)~\citep{raissi2019physics} or discrete grid-based alternatives~\citep{karnakov2022optimizing}, can also be used to extract operators at a significantly reduced data cost. However, they require the underlying form of the operator to be defined \textit{a priori}, and the full solution of the differential equation must also be learned during the fitting process. A middle ground between traditional numerical solvers and data-driven methods is found in Universal  Differentiable Equations (UDEs)~\citep{rackauckas2020universal}, where learnable models (e.g. NNs) are integrated within a prescribed differentiable simulator. Therefore, UDEs combine the benefits of previous extensive research on traditional numerical solvers with the flexibility of data-driven approaches. This is an important combination for problems where we want to strongly enforce known physical biases and numerical constraints, and are not interested in creating a surrogate model for the differential equation solution (since the cost of integrating the dynamics with the simulator is not significant). Our work belongs to this family of algorithms: we combine a differentiable Fokker-Planck solver with learnable advection/diffusion operators.  Interpretable symbolic regression methods can also be used for this task~\citep{brunton2016discovering, rudy2017data, gurevich2024learning}. However, to correctly recover the operator, its expression must be a linear combination of the library of pre-defined terms. This is a problem in situations where we do not have an intuition on what the underlying form of the operator is, or how its coefficients shold be parameterised, since there exist infinite possible combinations of input variables that can be added to the library. 

The majority of recent work at the intersection of machine learning and collision operators for plasma physics simulations has focused on reducing the associated computational cost of evaluating the collision operator using surrogate models. For example, surrogate models were trained for the Landau-Fokker-Planck~\citep{miller2021encoder, lee2023oppinn, noh2025fpl}, Rosenbluth-Fokker-Planck~\citep{chung2023data}, and Boltzmann~\citep{xiao2021using, xiao2023relaxnet, holloway2021acceleration, miller2022neural, lee2024structure, lee2025fourierspecnet} collision operators.
Unlike our work, the underlying simulation data used for training all these models was explicitly solving the prescribed collision operator. The goal of these approaches was to reduce the computational cost of evaluating the operator, not to inform the discovery of a new model (the goal of the present work).

PINNs have also been extensively used to model equations that contain Fokker-Planck-type operators. This includes works on the Vlasov-Fokker-Planck~\citep{hwang2020trend}, the Vlasov-Poisson-Fokker-Planck~\citep{lee2021model}, the Landau-Fokker-Planck~\citep{chung2023data}, and the Fokker–Planck~\citep{xu2020solving, chen2021solving, zhai2022deep, zhang2022physically, wang2025tensor} equations. Similarly, PINNs have been used to solve the Boltzmann equation for weakly ionized plasmas~\citep{zhong2022low, kawaguchi2022physics}, the simplified Boltzmann-BGK equation~\citep{lou2021physics, li2024solving, oh2025separable}, and relativistic Fokker–Planck equations describing runaway electron dynamics in magnetic confinement fusion devices~\citep{mcdevitt2023physics, mcdevitt2025efficient, mcdevitt2025physics}.

We emphasize that, with the exception of~\citet{chen2021solving}, all previous examples focus on solving the equation of interest (a \textit{forward problem}) using a prescribed collision operator (with fixed, non-learnable coefficients) rather than on extracting a new operator from observed dynamics (an inverse problem), as we aim in this paper. While~\citet{chen2021solving} proposed using PINNs to simultaneously address the forward and inverse problems, the simulation data in that work was generated by explicitly solving the equation with the learned collision operator, whereas in our work the data is obtained from first-principles simulations, that capture the electromagnetic interactions between particles self-consistently.

An exception to this line of research is the work of~\citet{camporeale2022data}, who used PINNs to learn FP operators that capture observed electron dynamics in the Earth's radiation belts. The underlying motivation and method are the same as ours, with the exception that we use a differentiable simulator to integrate the dynamics. We believe our approach to be more advantageous for our case study since we can generate as much kinetic simulation data as required, and the differentiable simulator allows us to quickly extract the collision operator without having to train a PINN as in~\citet{camporeale2022data}. Additionally, in this work, we highlight the reason why the PINNs in~\citet{camporeale2022data} do not find a unique solution to the advection/diffusion operators and propose an approach to mitigate this problem.

Another very promising related work is that of~\citet{zhao2025data, zhao2025fast} which recently proposed an approach to learn collision operators from molecular dynamics (MD) data. Similarly to our work, they infer a collision operator from kinetic data, with the operator written in its integral formulation. They demonstrate the capability of this formulation to accurately capture the results of MD simulations for both strongly coupled and weakly correlated plasmas initially outside thermal equilibrium. Unlike our approach here, they still use particle tracks during training while we learn directly from phase space data. Our approach, as previously explained, is more efficient in terms of storage requirements and can be more easily generalizable to other datasets. Furthermore, while their approach minimizes the prediction error over one time-step, ours optimizes for long-term stability since we use a differentiable simulator to integrate the dynamics over multiple time-steps. Lastly, while the formulation of the operator proposed in~\citet{zhao2025data} is tuned specifically for collisional dynamics, the FP description proposed in this work is applicable to other processes.

Differentiable simulators have been previously used in plasma physics to tackle distinct problems. Examples include the discovery of long-lived plasma wave packets~\citep{joglekar2022unsupervised}, learning a fluid closure for non-linear Landau damping~\citep{joglekar2023machine}, mitigating laser-plasma instabilities~\citep{joglekar2024generative}, and simulating the kinetic dynamics of a one-dimensional plasma with arbitrary degrees of collisionality~\citep{carvalho2024learning}. While it can be argued that a collisional operator was being learned in~\citet{carvalho2024learning}, the main motivation of the work was reducing the run-time of the simulator. Furthermore, the Graph Neural Network (GNN) model used only captures collisions between a reduced number of neighbouring electrons and does not provide an interpretable description as presented in this work.

\section{Methods}
\label{sec:methods}

Our aim is to extract advection and diffusion operators that are capable of reproducing the collisional dynamics of a plasma composed of finite-size particles. Based on the generated simulated data, we will use two different methods to estimate the coefficients. The first method makes use of particle tracks and is the standard approach from previous literature~\citep{hockney1971measurements, matsuda1975collisions}. The second is the novel contribution of the work, and consists of using only the distribution function (phase space) evolution of subpopulations to estimate the operator using a differentiable simulator framework. We will examine the benefits/tradeoffs between the two approaches later in this work.

\subsection{Particle-in-Cell Simulations}

Throughout this work, we perform 2-D simulations using the electromagnetic PIC code OSIRIS~\citep{fonseca2002osiris}. We consider a single mobile electron species moving over a fixed uniform ion background inside a periodic box. In these simulations, only electron-electron collisions are captured. The initial electron velocities are sampled from a thermal distribution with zero net velocity and standard deviation $v_{th}$ in all directions, and the simulation window is set to $10^3\lambda_D\times10^3\lambda_D$. For all scenarios, we use a charge-conserving current deposition scheme~\citep{esirkepov2001exact}, a second-order FDTD electromagnetic field solver~\citep{yee1966numerical}, and the standard Boris pusher~\citep{boris1972proceedings} to advance particle momenta and position. 

To generate a dataset of simulations with different advection and diffusion operators, we vary four parameters: the number of particles per cell~($N_{ppc}$), the current deposition shape function~($S_J^m$, where $m$ is the interpolation order), the ratio of the grid-resolution over the electron Debye length~($\Delta_x/\lambda_D$ = $\Delta_y/\lambda_D$), and the electron thermal velocity~($v_{th}/c$). In total, we performed approximately 100 simulations corresponding to different parameter combinations and preserved a subset of approximately 75 simulations for which no significant PIC heating was observed.\footnote{We preserved all simulations for which the total energy variation ratio threshold $\Delta E/E < 10^{-2}$ was not exceeded. We discarded simulations above this threshold because in this work we are only focusing on learning FP operators that do not change over time.} The detailed list of simulation parameters used and the heating tests performed are provided in Supplementary Material~\ref{app:simulation_parameters}.

For each simulation, we stored the raw data of all macroparticles (approximately $10^6-10^7$ macroparticles depending on the numerical parameters, we store both their position and momenta) at equally spaced time intervals to collect a minimum of 100 snapshots over the full simulation. This information is used to construct the phase space information of different subpopulations in post-processing, as described in Supplementary Material~\ref{app:subpopulation_sampling}. We note that we only store the data of all macroparticles to be able to compare the operators recovered from the two methods: from macroparticle particle tracks (i.e. macroparticle position and momenta history), and from the phase space evolution of subpopulations. In future works, this memory-intensive step will not be required.

\subsection{Advection-Diffusion Coefficients from Particle Tracks}
\label{sec:ad_tracks}

The advection and diffusion coefficients for particles moving at a specific velocity $\boldsymbol{v}$ can be estimated by tracking the rate of the average velocity drift and spread over time:
\begin{equation}
    \mathsfi{A}_i(\boldsymbol{v}) = \frac{<\Delta v_i>_{\boldsymbol{v}}}{\Delta t}
\label{eq:A_i}
\end{equation}
\begin{equation}
    \mathsfi{D}_{ij}(\boldsymbol{v}) = \frac{<\Delta v_i\Delta v_j>_{\boldsymbol{v}}}{\Delta t}
\label{eq:D_ij}
\end{equation}
where $\Delta v_i = v_i{(t_0 + \Delta t)} - v_i{(t_0)}$ corresponds to the change in the velocity along the $i$-axis of a macroparticle with $\boldsymbol{v}{(t_0)} = \boldsymbol{v}$ over a time-interval $\Delta t$, and $<\cdot>_{\boldsymbol{v}}$ corresponds to an average over all macroparticles with the selected initial velocity. 

\begin{figure}
    \centering
    \includegraphics[width=0.80\linewidth]{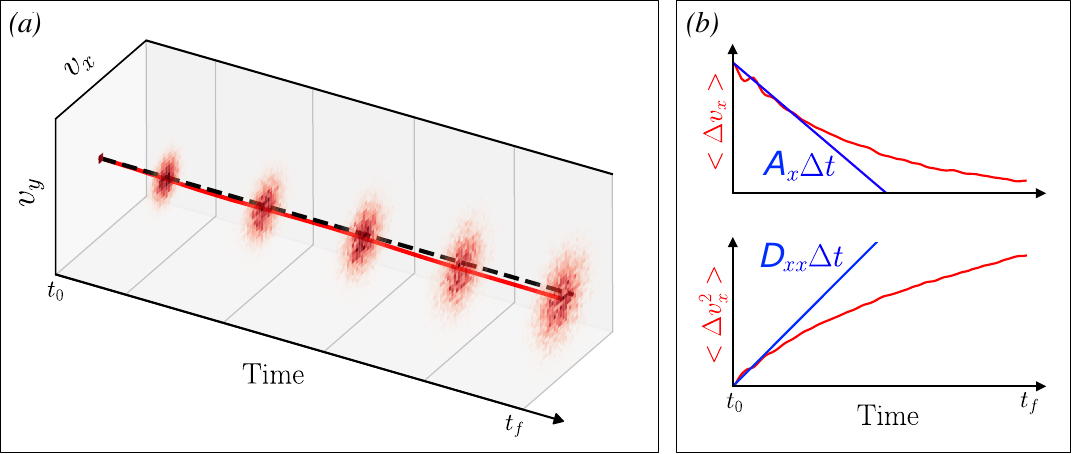}
    \caption{Illustration of how advection and diffusion coefficients can be inferred from 2-D particle tracks. (a) By following a group of particles with similar initial velocities over time, we observe that advection leads to an average velocity drift ($<\Delta v_i>$), while diffusion leads to an increased spread of the distribution ($<\Delta v_i\Delta v_j>$). Drift is visible by noting that the average particle velocity (red line) is changing when compared to the initial velocity (dashed black line). Diffusion is clearly illustrated by the increased distribution spread (highlighted via the phase space histograms). (b) In practice, the advection-diffusion values are estimated by measuring the rate of change of $<\Delta v_i>$ and $<\Delta v_i\Delta v_j>$ with respect to a period of time where the evolution is linear. To obtain an accurate estimate of the coefficients it is crucial that statistics are computed during the linear evolution phase.}
    \label{fig:tracks}
\end{figure}

In practice, we relax the above condition to consider a finite region of phase space over which the average is computed (Figure~\ref{fig:tracks}).
In this work, we divide the $\mathcal{V}: [-5,5] v_{th} \times [-5,5]v_{th}$ phase space into equal-sized bins resulting in a $N_v\times N_v$ grid with $N_v=51$. To increase the statistics, we divide the simulation into contiguous, equally sized time intervals so that the average is computed across particles trajectories starting at the desired region of phase space at the beginning of any of the time intervals ($t_0 = \{k \Delta t:$ $k\in \{0, 1, ..., t_{max}/\Delta t\}\}$). The equations~\eqref{eq:A_i} and~\eqref{eq:D_ij} are then rewritten as:
\begin{equation}
    \mathsfi{A}_i(\boldsymbol{v}_{bin}) = \frac{<\Delta v_i>_{\boldsymbol{v}(t_0) \in \boldsymbol{v}_{bin}}}{\Delta t}
\label{eq:a_tracks}
\end{equation}
\begin{equation}
    \mathsfi{D}_{ij}(\boldsymbol{v}_{bin}) = \frac{<\Delta v_i\Delta v_j>_{\boldsymbol{v}(t_0) \in \boldsymbol{v}_{bin}}}{\Delta t}
\label{eq:d_tracks}
\end{equation}
where $\boldsymbol{v}_{bin}$ represents all velocities within the phase space bin and $<\cdot>_{\boldsymbol{v}(t_0) \in \boldsymbol{v}_{bin}}$ corresponds to an average over all macroparticle trajectories starting inside the bin at the beginning of any time-interval.

The time-interval over which statistics are computed affects the retrieved values of the advection-diffusion coefficients. If $\Delta t$ is too small, the particle trajectories are correlated, and the Markovian assumption underlying the FP derivation breaks down. If $\Delta t$ is too large, the evolution of $<\Delta v_i>$ and $<\Delta v_i \Delta v_j>$ stops being linear (see Figure~\ref{fig:tracks}). Note that these time scales are  simulation dependent since they are defined by the collisional timescales (which we want to infer) and collective plasma processes. To tackle this problem, we then devised and automatised process where, for each simulation: 1)~we compute the coefficients using different values of $\Delta t$; 2) for each value of $\Delta t$, we evaluate the corresponding advection-diffusion coefficients accuracy in reproducing the phase space dynamics of different subpopulations of particles; 3) we pick the best performing coefficients to compare against the machine learning based approaches. A broader discussion on the consequences of estimating the advection diffusion coefficients with this technique, is provided in Supplementary Material~\ref{app:ad_from_tracks}.

\subsection{Advection-Diffusion Coefficients from Phase Space Evolution}
\label{sec:ad_phase_space}

\begin{figure}
    \centering
    \includegraphics[width=\linewidth]{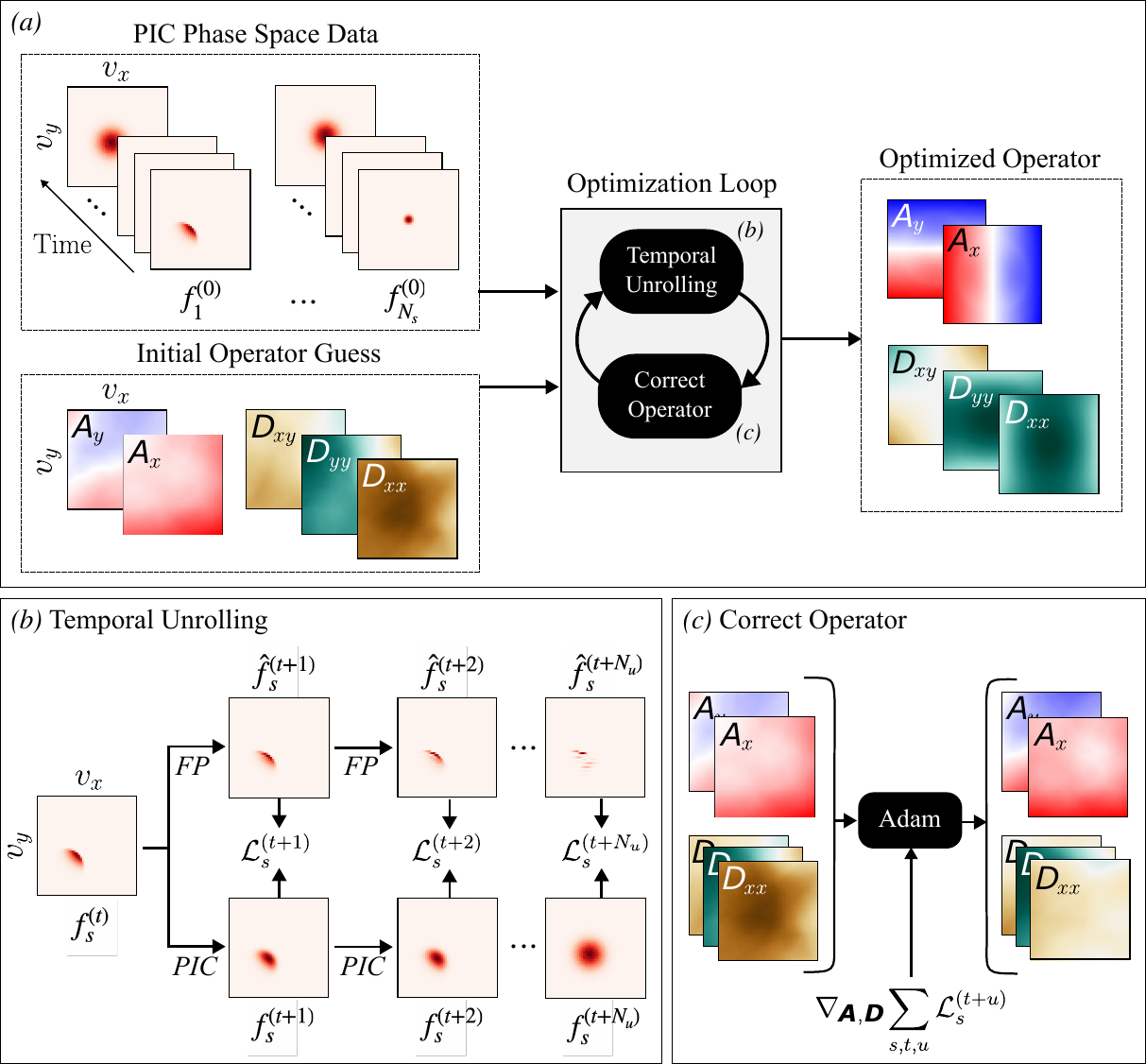}
    \caption{(a) Illustration of how advection and diffusion coefficients are inferred from the 2-D phase space evolution of $N_s$ subpopulations using a differentiable Fokker-Planck (FP) solver. The evolution of the phase space and the changes in the operator are exaggerated for visualization purposes. (b) Using the FP solver and the current operator state, we advance the phase space over $N_u$ time steps and compare against the PIC results via a scalar error metric $\mathcal{L}$. This operation is performed over all training subpopulations $s\in[1, N_s]$ and initial time-steps $t\in[0,N_t - N_u]$. (c) We then update the operator using the gradient-based optimizer Adam~\citep{kingma2014adam}, to minimize the unrolled error across all subpopulations and time-steps. The two operations are performed sequentially in a loop until the results have converged. }
    \label{fig:fp_solver}
\end{figure}

Determining the advection and diffusion operators that describe the phase space dynamics can be written as an optimization task:
\begin{equation}
    \min_{\mathsfbi{A},\mathsfbi{D}} \sum_u \mathcal{L}\left(\hat{f}^{(t+u)}\left(\mathsfbi{A}, \mathsfbi{D}, f^{(t)}\right), f^{(t+u)}\right)
\end{equation}
where $f^{(t+u)}\in \mathbb{R}^{N_v} \times \mathbb{R}^{N_v}$ corresponds to the true distribution function at time-step $t+u$, $\hat{f}^{(t+u)}\in \mathbb{R}^{N_v} \times \mathbb{R}^{N_v}$ the predicted distribution function when evolving the phase space dynamics of $f^{(t)}$ over $u$ time-steps using the estimated $\mathsfbi{A}$, $\mathsfbi{D}$ operators, and $\mathcal{L}$ a error (loss) function computed over the full phase space (e.g. mean absolute error).

For a given single ``trajectory'' of $f$, this is an ill-posed inverse problem, since there is a family of solutions that results in equivalent dynamics for a given initial distribution function. This is caused by the presence of velocity space gradients in~\eqref{eq:fp_equation} which allows perturbations to the advection/diffusion coefficients to cancel each other (cf. discussion in Supplementary Material~\ref{ap:non_uniqueness_ad}). This non-uniqueness issue had been observed, e.g., in previous work by~\citet{camporeale2022data}, where PINNs initialized with different random seeds recover significantly different advection-diffusion operators that describe the phase space dynamics with similar levels of accuracy. Furthermore, to determine the coefficients in a particular region of phase space, it is necessary to have access to data in that region, i.e., $\partial f/\partial t \neq 0$ at some time-step.\footnote{For  a thermal distribution, $\partial f/\partial t \approx 0$ everywhere, so no dynamics are observed in phase space.}

To address these two challenges, we propose to employ the phase space evolution of different subpopulations. Ideally, these subpopulations are sampled at initialization and cover different regions of the phase space. Furthermore, since we want operators that are stable in the long-term prediction of the dynamics, we optimize for the evolution of the distribution function over multiple time steps; this is a common strategy known as temporal unrolling (see e.g.~\citet{brandstetter2022message, list2025differentiability}). Both these strategies serve the same underlying purpose: to generate data that can constrain the space of possible solutions. The optimization task is then rewritten as:
\begin{equation}
    \min_{\mathsfbi{A},\mathsfbi{D}} \sum_s^{N_s} \sum_t^{N_t - N_u}\sum_u^{N_u} \mathcal{L}\left(\hat{f}_s^{(t+u)}\left(\mathsfbi{A}, \mathsfbi{D}, f_s^{(t)}\right) - f_s^{(t+u)}\right)
\end{equation}
where $f_s$ corresponds to the distribution function of subpopulation $s$, and we make explicit the sum over $N_s$ subpopulations, $N_t$ simulation time-steps, and $N_u$ time-steps into the future for which we generate a prediction $\hat{f}^{(t+u)}$; we are softly enforcing the physics via the loss function, similarly to PINNs or equivalent algorithms, with the difference that these methods minimize the theoretical PDE residual evaluated on the predicted solution at predefined spatio-temporal positions while here we are mininizing the error between PDE trajectories.

We solve the optimization task using a differentiable simulator coupled with a gradient-based optimization method (Figure~\ref{fig:fp_solver}). We implement the full differentiable simulator and its core 2-D Fokker-Planck solver in PyTorch~\citep{ansel2024pytorch}, which we use to compute the forward dynamics of different subpopulations, and we use the Adam optimizer~\citep{kingma2014adam} to update the advection and diffusion operators such that the difference between the predicted and real distribution function evolution is minimized. To facilitate experiment tracking and performance comparisons over a large range of training and test setups we make use of MLflow~\citep{zaharia2018mlflow}.

Using this methodology, the advection and diffusion operators can be parameterized using a discrete or continuous function. The discrete case is equivalent to computing coefficients from particle tracks (Section~\ref{sec:ad_tracks}), i.e., we discretize the operators over a fixed grid $\mathsfbi{A}_i \in \mathbb{R}^{N_v \times N_v}$ and $\mathsfbi{D}_{ij} \in \mathbb{R}^{N_v \times N_v}$. A continuous operator can be obtained by instead using a smooth function approximation. In our case, we use Neural Network (NN) models, i.e., $\mathsfi{A}_i(\boldsymbol{v}) = \mathrm{NN}_{\mathsfi{A}_i}(\boldsymbol{v})$ and $\mathsfi{D}_{ij}(\boldsymbol{v}) = \mathrm{NN}_{\mathsfi{D}_{ij}}(\boldsymbol{v})$. Furthermore, a continuous description allows us to extract more general operators that might depend on other system parameters, since these can be provided as inputs to the NN. In our case study, these will correspond to numerical simulation parameters, e.g., $\mathsfi{A}_i(\boldsymbol{v}) = \mathrm{NN}_{\mathsfi{A}_i}(\boldsymbol{v}, N_{ppc}, \Delta_x/\lambda_D,m, v_{th}/c)$. For other cases, it might be more relevant to include meaningful physical parameters or even the simulation time itself if the coefficients are expected to change over time. 

It is possible to guarantee known constraints in the advection and diffusion operators, for both the discrete and continuous cases, by enforcing known symmetries. We will elaborate further on this topic when comparing the performance of the different approaches in later sections of the manuscript. More details about the symmetries and the overall FP solver implementation are also provided in Supplementary Material~\ref{app:fp_solver_implementation}.

\section{Results}
\label{sec:results}

We now evaluate the performance of the advection and diffusion operators extracted using different methods. They are named accordingly: \textit{Tracks} - computed using particle tracks; \textit{PS-Tensor} - computed using the phase space evolution and a discrete operator description; \textit{PS-NN} - computed using phase space evolution and a continuous operator parameterized by a NN; \textit{PS-NN-Multi} - similar to PS-NN but in this case we trained a single NN model on multiple simulations at once (i.e., predicted coefficients are conditioned on simulation parameters, $\mathsfi{A}_i(\boldsymbol{v}) = \mathrm{NN}_{\mathsfi{A}_i}(\boldsymbol{v}, N_{ppc}, \Delta_x/\lambda_D,m, v_{th}/c)$ and $\mathsfi{D}_{ij}(\boldsymbol{v}) = \mathrm{NN}_{\mathsfi{D}_{ij}}(\boldsymbol{v}, N_{ppc}, \Delta_x/\lambda_D,m, v_{th}/c)$ with simulation parameters pre-processed as detailed in Supplementary Material~\ref{app:fp_solver_implementation}).

The neural network models used to parameterize the advection and diffusion operators ($\mathrm{NN}_{\mathsfi{A}_i}(.), \mathrm{NN}_{\mathsfi{D}_{ij}}(.)$) are all Multi-Layer Perceptrons (MLPs) with Leaky ReLU activation functions after each hidden layer. For the PS-NN method, we use 2 hidden layers, while for PS-NN-Multi we use 3 hidden layers. All hidden layers have 128 neurons. 
Further details on the optimization loop (e.g., curriculum, learning rates, duration) can be found in Supplementary Material~\ref{app:training_procedure}. For reference, extracting the operator from one simulation takes 5-10 minutes on a single Nvidia Titan X GPU using the default optimization strategy without resorting to PyTorch Just-in-Time compilation capabilities.

To assess the generalization potential of operators recovered from phase space dynamics, we will make use of two distinct sets of populations: $Train$ - 9 subpopulations (sampled from a normal distribution at different regions in phase space); $Test$ - 19 subpopulations (distinct normal distributions, rings, quadrants, etc). Illustrations of the different subpopulations are provided in Figures~\ref{fig:pdf_examples} to \ref{fig:pdf_examples_index_51} in Supplementary Material~\ref{app:subpopulation_sampling}.

\subsection{Example for a Single Simulation}
\label{sec:example_single_simulation}

To illustrate the overall aspect of the PIC collision operators recovered by the different methods, we start by analyzing the results for a single simulation, simulation $index=0$ ($N_{ppc}=4$, $m=1$, $\Delta_x /\lambda_D = 1$, $v_{th}=0.01c$), which provides a good representation of the standard behaviour in our simulation dataset.

\begin{figure}
    \centering
    \includegraphics[width=\linewidth]{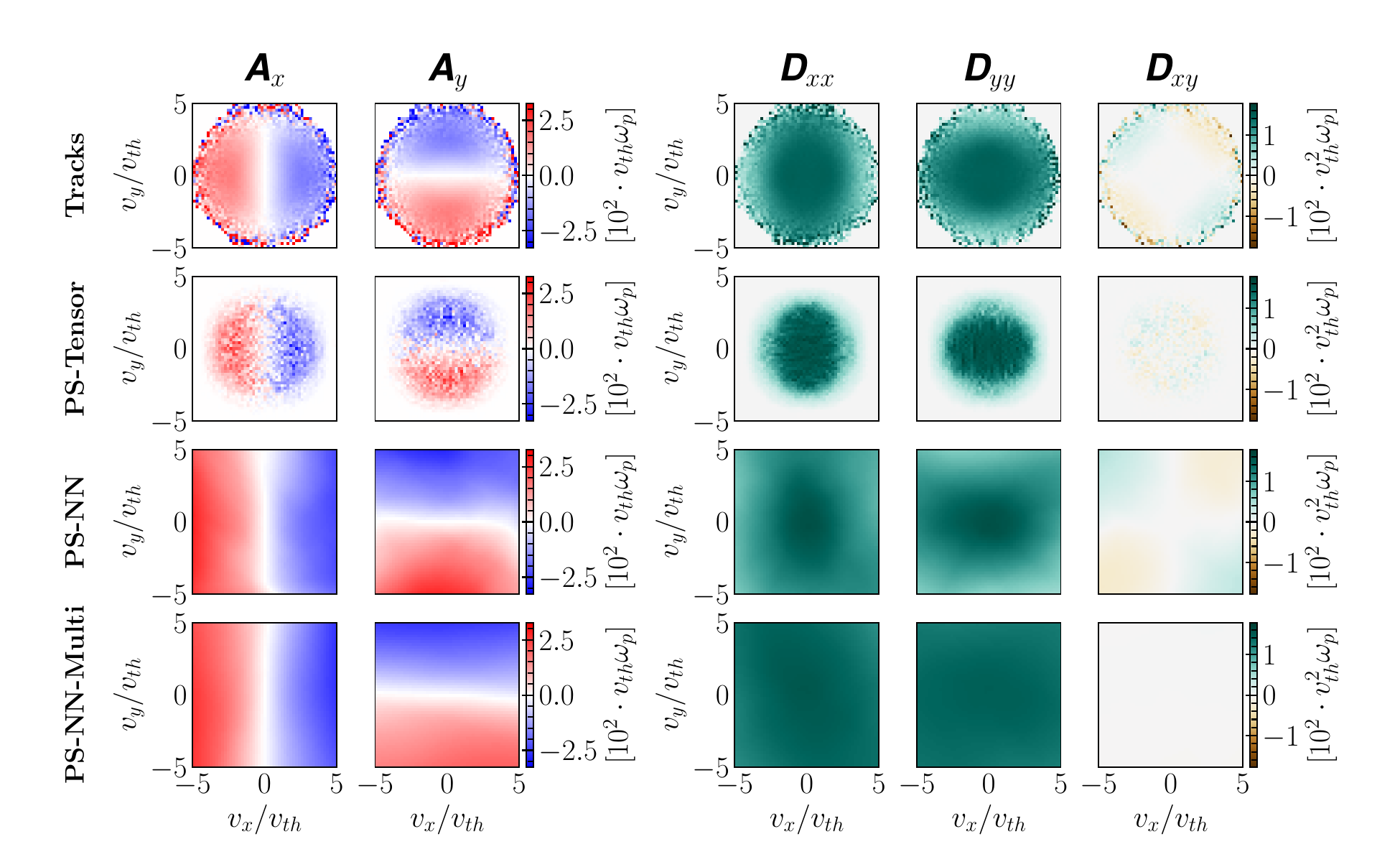}
    \caption{Advection and diffusion coefficients retrieved with different approaches for simulation $index=0$ ($N_{ppc}=4$, $m=1$, $\Delta_x/\lambda_D=1, v_{th}=0.01c)$. \textit{Track} operator is computed from statistics over all macroparticles, \textit{PS} operators are computed from the phase space evolution of 9 subpopulations using a differentiable FP solver and a discrete (PS-Tensor) or continuous (PS-NN, PS-NN-Multi) approximator. Note that all these operators perform similarly when predicting the phase space evolution of most tested subpopulations. More examples for other simulation parameters are provided in Supplementary Material~\ref{app:example_single_simulation}.
    }
    \label{fig:AD_comparison_index_0} 
\end{figure}

In Figure~\ref{fig:AD_comparison_index_0}, we show the operators retrieved using the different methods. 
Overall, we observe the expected behaviour. Advection $\mathsfi{A}_i$ is positive for $v_i < 0$ and negative for $v_i > 0$, revealing a drag that forces particles to $v_i=0$. Diagonal diffusion terms $\mathsfi{D}_{ii}$ are greater or equal to zero everywhere (otherwise the system would be unstable), and slower particles diffuse faster. Finally, we observe a correlation between diffusion along the $x$ and $y$ axis ($\mathsfi{D}_{xy} \neq 0$) for particles moving at an angle. The higher the velocity and the closer the angle is to $45^\circ$, the higher the cross diffusion term.

The operator obtained from particle tracks (Tracks) is smooth except for regions where statistics are poor (high $v$). This leads, in fact, to numerical instabilities if the operator is used as is during the time integration. To tackle this issue, we always use a smaller time step when later integrating the dynamics with these operators (we use $\Delta t = \Delta t_{dump}/10$ where $\Delta t_{dump}$ is the phase space diagnostic period, with a broader discussion on other considered alternatives provided in Supplementary Material~\ref{app:fp_solver_implementation}).

When we consider a discrete operator learned from phase space dynamics (PS-Tensor), values are only defined within the phase space region where data was available during training. The retrieved advection-diffusion values are similar to the ones estimated from tracks, however the overall behaviour is noisier. Unlike for operators learned from tracks, the noise level does not cause the integrator to become unstable. This is one benefit of learning the operator with the integrator in the loop.

Regarding the NN-based models (PS-NN, PS-NN-Multi), since they are continuous operators, they can extrapolate outside the training data region and bias the solutions to be smooth. However, for high $v$ values, we can still clearly see the presence of artifacts (e.g., $\mathsfi{A}_y$ is not zero at $v_y=0$ and $v_x \simeq 5v_{th}$). Furthermore, while PS-NN seems to be able to capture more accurately the overall form of the operator (if we consider the Tracks operator as the correct reference), the operator trained on multiple simulations (PS-NN-Multi) seems to lack expressiveness on its diffusion term ($\mathsfi{D}_{ii}$ values do not significantly decrease at high $v$ and $\mathsfi{D}_{ij}$ terms are not visible). We conjecture this happens due to the limited size of the latter NN model, whose expressivity could be increased by training on larger amounts of simulation data while increasing its size. Another option, which will be explored in later sub-sections, is enforcing further symmetries to facilitate the learning procedure.

\begin{figure}
    \centering
    \includegraphics[width=0.7\linewidth]{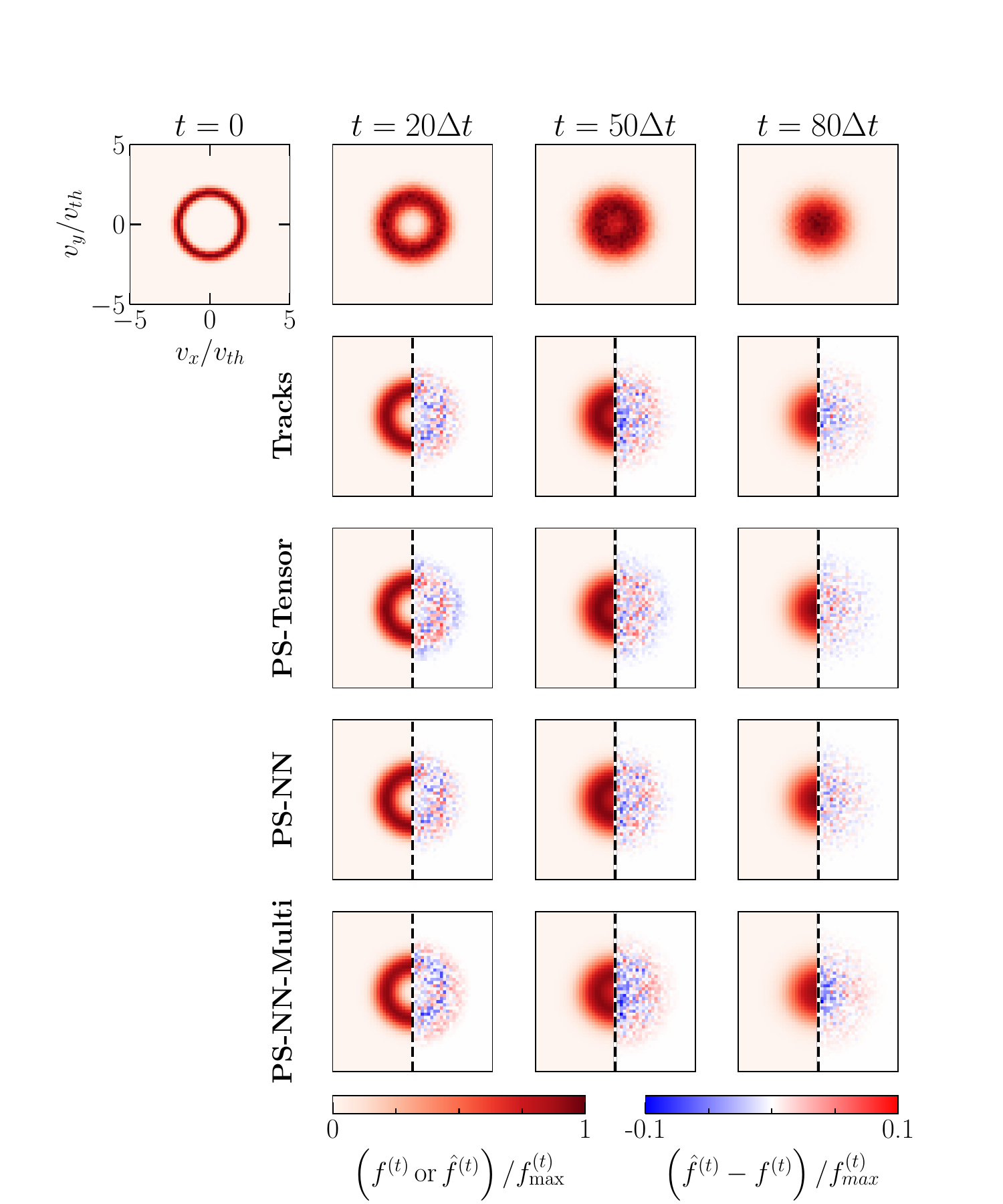}
    \caption{Phase space evolution for a ring subpopulation using operators recovered in Figure~\ref{fig:AD_comparison_index_0}. This subpopulation was not used during the training of  PS-Tensor, PS-NN, and PS-NN-Multi models. The top row corresponds to the observed dynamics in the PIC simulation ($f^{(t)}$). The remaining rows represent the predicted phase space evolution on the left ($\hat{f}^{(t)}$ for $v_x/v_{th}\in [-5,0]$) and the difference to the PIC data on the right ($\hat{f}^{(t)} - f^{(t)}$ for $v_x/v_{th}\in [0,5]$). Values are normalized to the peak of the PIC distribution function at time $t$ ($f^{(t)}_{max}$).  All operators approximate the dynamics relatively well, and overall, the random distribution of errors can be attributed to the granularity of the original distribution function. However, both Tracks and PS-NN-Multi seem to systematically slightly underestimate advection (central blue error region and outer red halo). Examples for other subpopulations are provided in Supplementary Material~\ref{app:example_single_simulation}.}
    \label{fig:ex_rolllout_dif_0_ring_normal_2_0.2}
\end{figure}

To illustrate the capabilities of the previous operators to predict long-term phase space dynamics, we show in Figure~\ref{fig:ex_rolllout_dif_0_ring_normal_2_0.2} the relaxation of a ring subpopulation (part of the test set).
The top panels showcase the dynamics retrieved from the PIC simulation. The bottom panels show the integration of the dynamics using the operators from Figure~\ref{fig:AD_comparison_index_0} and the respective prediction error. The errors, per phase space bin, are below $5\%$ of the maximum distribution function value for the corresponding timestep. Random distribution of errors at later times for PS-Tensor/PS-NN can be attributed to the granularity of the original distribution function due to the finite particle number. Systematic errors at later times for Tracks/PS-NN-Multi seem to indicate a slight underestimation of advection. Nonetheless, all models behave fairly well despite their different forms. In the following sub-section, we will show that this conclusion holds across many simulations and subpopulations using more systematic tests. Finally, additional examples of collision operators retrieved for other simulation parameters and the predicted evolution of different subpopulations are provided in Supplementary Material~\ref{app:example_single_simulation}.

\subsection{Long-term Accuracy of Different Methods}
\label{sec:accuracy_different_methods}

To systematically measure the long-term accuracy of the operators retrieved using different methods, and for each subpopulation, we take the phase space at the initial time-step and use the FP solver to evolve it until the end of the simulation ($N_t\approx 100)$ by applying the extracted operator. We refer to this process of simulating the phase space dynamics of an initial subpopulation from $t=0$ to $t=t_{max} = N_t\Delta t$ as performing a \textit{simulation rollout}. For the operators retrieved using the differentiable simulator we use the same time-step as the phase space diagnostic ($\Delta t=\Delta t_{dump}$) when integrating the dynamics over time. For operators obtained from particle tracks we reduce the integrator time-step ($\Delta t=\Delta t_{dump}/10$) to avoid numerical instabilities caused by the noisy estimates of the coefficients at high $v$. 

As an accuracy metric, we define the phase space mean absolute error (MAE) averaged over a simulation rollout ($\mathrm{MAE{\!-\!}Rollout}$) as:
\begin{equation}
    \mathrm{MAE{\!-\!}Rollout} = \frac{1}{N_t}  \sum_{t}^{N_t} \left|\hat{f}^{(t)}\left(\mathsfbi{A}, \mathsfbi{D}, f^{(0)}\right) - f^{(t)}\right|
\end{equation}
where the sum over phase space dimensions is implicit. Since by definition $\left\lVert f^{(t)} \right\rVert_1 = 1$ the $\mathrm{MAE{\!-\!}Rollout}$ value can be interpreted as the average relative error over all time-steps (e.g., $\mathrm{MAE{\!-\!}Rollout}=0.03$ implies that on average, we have a $3\%$ error across all time steps).

\begin{figure}
    \centering
    \includegraphics[width=0.6\linewidth]{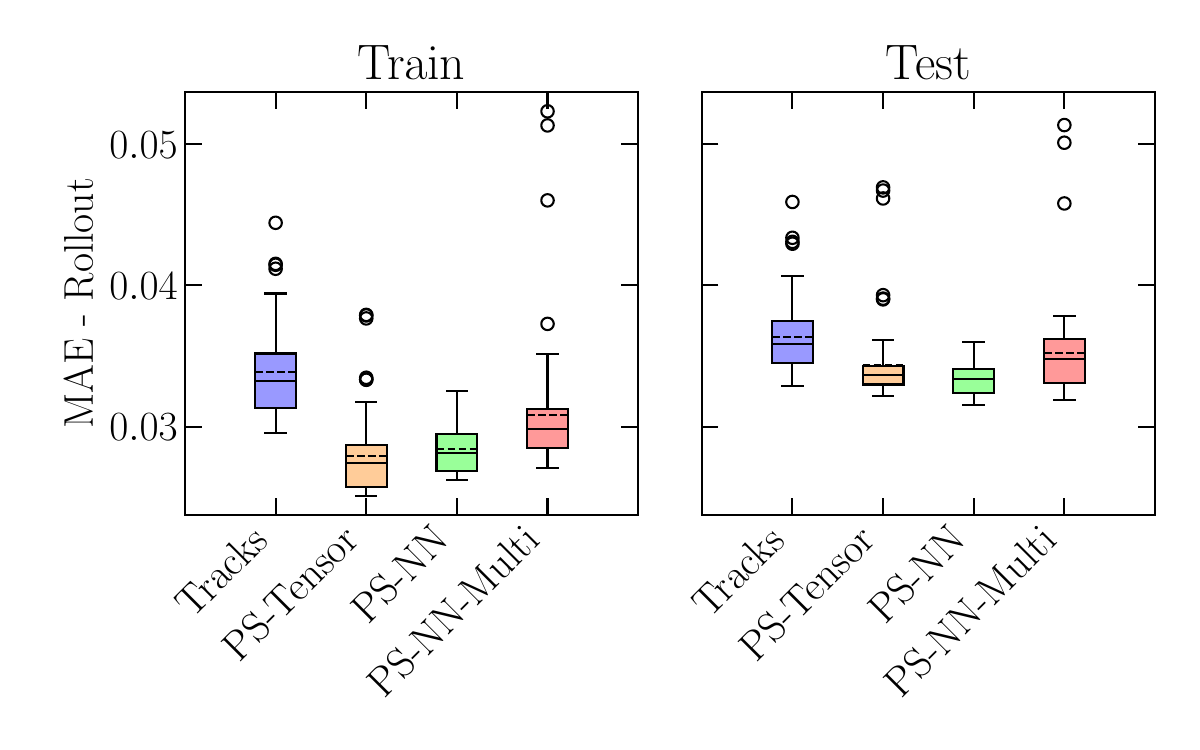}
    \caption{Distribution of rollout errors for advection and diffusion models obtained from particle tracks (Tracks) or phase space evolution of subpopulations (PS-Tensor, PS-NN, PS-NN-Multi). Boxplots represent statistics over the full dataset of PIC simulations (see table~\ref{tab:simulation_parameters} for more details) averaged over initial subpopulations (Train - 9 subpopulations; Test - 19 subpopulations, more information in Supplementary Material~\ref{app:subpopulation_sampling}). The mean error is shown with a dashed line, the median with a full line. The filled area corresponds to values between the first and third quartiles ($Q_1$ and $Q_3$). Whiskers represent the lowest values up to $Q_1 - 1.5(Q_3 - Q_1)$ and $Q_3 + 1.5(Q_3 - Q_1)$. Dots represent values outside this range.  The performance of the different methods is, on average, equivalent on the test data. These results demonstrate that estimating the operators from phase space information using a differentiable simulator is a viable alternative approach to particle tracks.}
    \label{fig:l1_model_comparison_boxplot}
\end{figure}

The statistics obtained over the full dataset of simulations for training and test subpopulations are shown in Figure~\ref{fig:l1_model_comparison_boxplot}.
We observe that methods based on the differentiable simulator perform on average better than the values estimated from the statistics on both the train and test subpopulations. Furthermore, PS-NN models trained on a single simulation seem to be more resilient to outliers than the alternative approaches. We attribute this to the enforcement of smooth coefficient values in regions where there is little statistics available, i.e., higher $v$. In fact, subpopulations that contain a larger number of particles at high $v$, which are mostly present in the test set and not the train set, are the reason why the error is slightly higher for test scenarios (even for operators extracted from particle tracks). We elaborate further on this topic in Supplementary Material~\ref{app:rollout_errors_per_subpopulation}.

Regarding the outliers, they are consistently associated with the same simulations, where a smaller number of macroparticles was used. This makes it harder both for the statistical method and the subpopulation approach to correctly capture dynamics in the full phase space since there is a lack of statistics at high $v$. A more detailed analysis of these results is provided in Supplementary Material~\ref{app:rollout_errors_per_simulation}.

\subsection{Impact of Temporal Unrolling during Training}
\label{sec:impact_temporal_unrolling}

\begin{figure}
    \centering
    \includegraphics[width=0.6\linewidth]{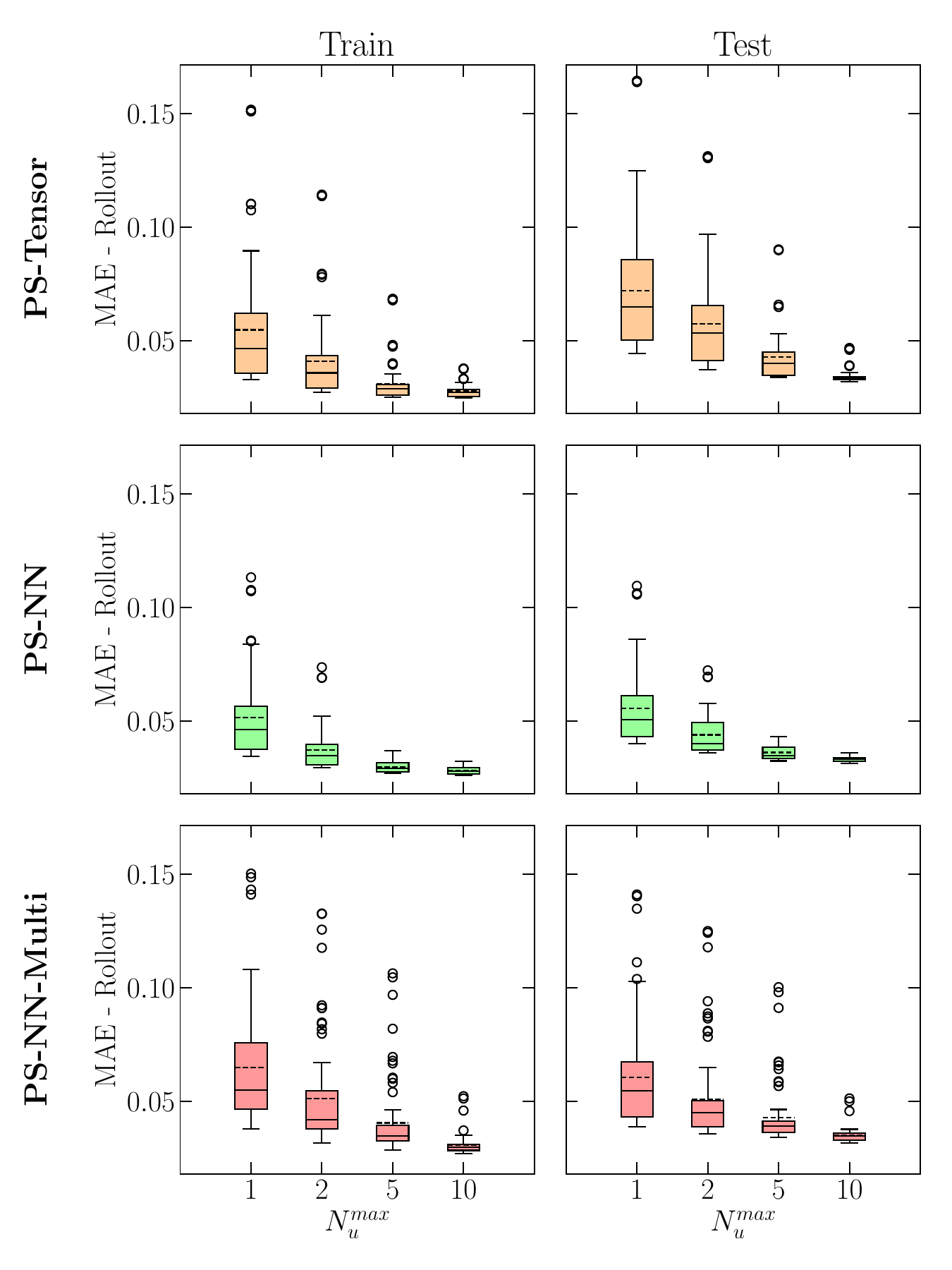}
    \caption{Impact of maximum temporal unroll length during training ($N_u^{max}$) on the long-term rollout error of different models. Rollout length is $N_t \approx 100$ across the full dataset of subpopulations and simulations. Larger temporal unrolling at train time consistently leads to improved rollout performance across all models, showcasing the importance of optimizing the operators for long-term prediction.}
    \label{fig:l1_model_comparison_boxplot_unroll}
\end{figure}

To justify a differentiable simulator approach to optimize the coefficients over longer prediction times, we compare, in Figure~\ref{fig:l1_model_comparison_boxplot_unroll}, the rollout error of operators extracted using different maximum training temporal unroll lengths $N_u^{max}$. It is clear that increasing the rollout duration during training significantly improves long-term performance across all models.
We expect the optimal value of this rollout length to be problem-dependent, for instance, with respect to the noise/smoothness of the dynamics, relative time scales between collisional dynamics and other processes, as well as the phase space diagnostic frequency. Future works should tune the rollout length to maximize long-term accuracy, bearing in mind that increasing this length leads to significant increases in training memory requirements and duration.  In our case, we did not find meaningful improvements when using $N_{u}^{max} > 10$ and this is why we set this chosen maximum value.

\begin{figure}
    \centering
    \includegraphics[width=\linewidth]{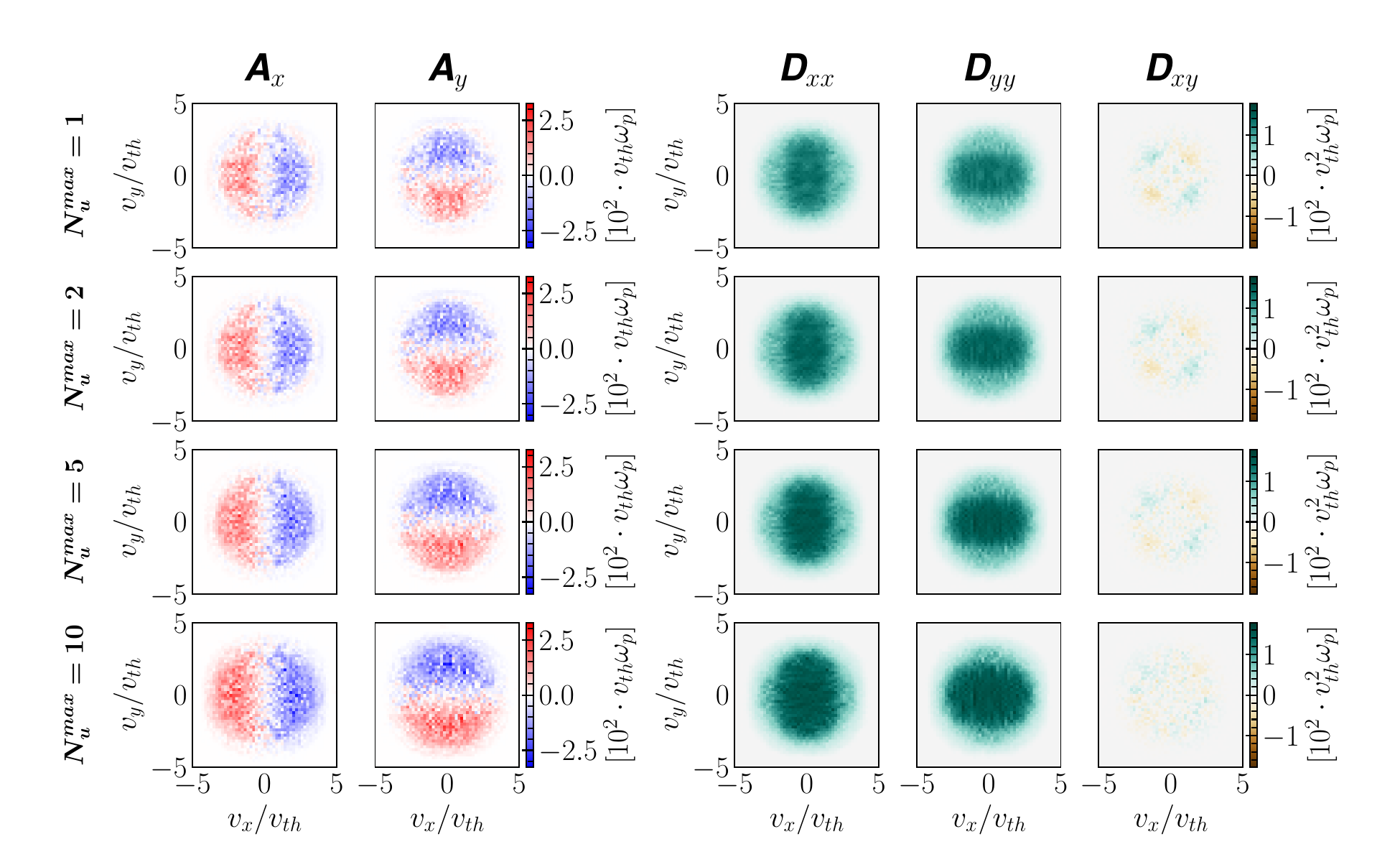}
    \caption{Illustration of the impact of maximum training temporal unroll length on PS-Tensor models. The example shown corresponds to simulation $index=0$ ($N_{ppc}=4$, $m=1$, $\Delta_x/\lambda_D=1, v_{th}=0.01c)$. It is clear that increasing $N_u^{max}$ leads to a smoother operator defined over a larger region of the phase space while also significantly changing the average advection and diffusion values in regions of $v\approx 0$.}
    \label{fig:AD_comparison_unroll_tensor_index_0}
\end{figure}

To visualize the impact of the training temporal unrolling in the final operator extracted, we showcase an example for a PS-Tensor model in Figure~\ref{fig:AD_comparison_unroll_tensor_index_0}. Increasing the temporal interval not only promotes smoother solutions but also allows to capture the correct values for a larger range of the phase space. The impact is not as visually clear in NN-based models, for which we provide some examples in Supplementary Material~\ref{app:impact_temporal_unrolling}, but the results in Figure~\ref{fig:l1_model_comparison_boxplot_unroll} demonstrate that indeed the operators are being meaningfully modified.

\subsection{Enforcing Known Symmetries}
\label{sec:enforcing_known_symmetries}

Given known physical priors, such as symmetries, it is possible to enforce these into the learned advection-diffusion models. This is a common strategy in the physics-inspired machine learning literature which is particularly useful when handling inverse problems in the presence of sparse noisy data, such as the case under study in this work.

To test the impact of imposing such symmetries, we define a hierarchy of increasingly more restrictive models: 
$AD$ - No symmetries imposed (same as before); 
$AD_T$ - Enforces $x/y$ coefficients to be equivalent (i.e. $\mathsfi{A}_y(v_x, v_y) = \mathsfi{A}_x(v_y,v_x)$ and $\mathsfi{D}_{yy}(v_x, v_y) = \mathsfi{D}_{xx}(v_y,v_x)$);
$AD_{Sym}$ - Extends $\mathrm{AD}_T$ to include (anti-)symmetry along $v_x=0$ (i.e. $\mathsfi{A}_x(v_x,v_y) = - \mathsfi{A}_x(-v_x, v_y)$, $\mathsfi{D}_{xx}(v_x,v_y) = \mathsfi{D}_{xx}(-v_x,v_y)$, and $\mathsfi{D}_{xy}(v_x,v_y) = -\mathsfi{D}_{xy}(-v_x,v_y)$;
$AD_{\parallel, \perp}$ - Considers only parallel and perpendicular coefficients (i.e. $\mathsfi{A}_\parallel(v)$, $\mathsfi{D}_\parallel(v)$, and $\mathsfi{D}_\perp(v)$ which are projected to obtain $\mathsfi{A}_i$ and $\mathsfi{D}_{ij}$). Further details on the implementation of these models are provided in Supplementary Material~\ref{app:fp_solver_implementation}.

\begin{figure}
    \centering
    \includegraphics[width=0.6\linewidth]{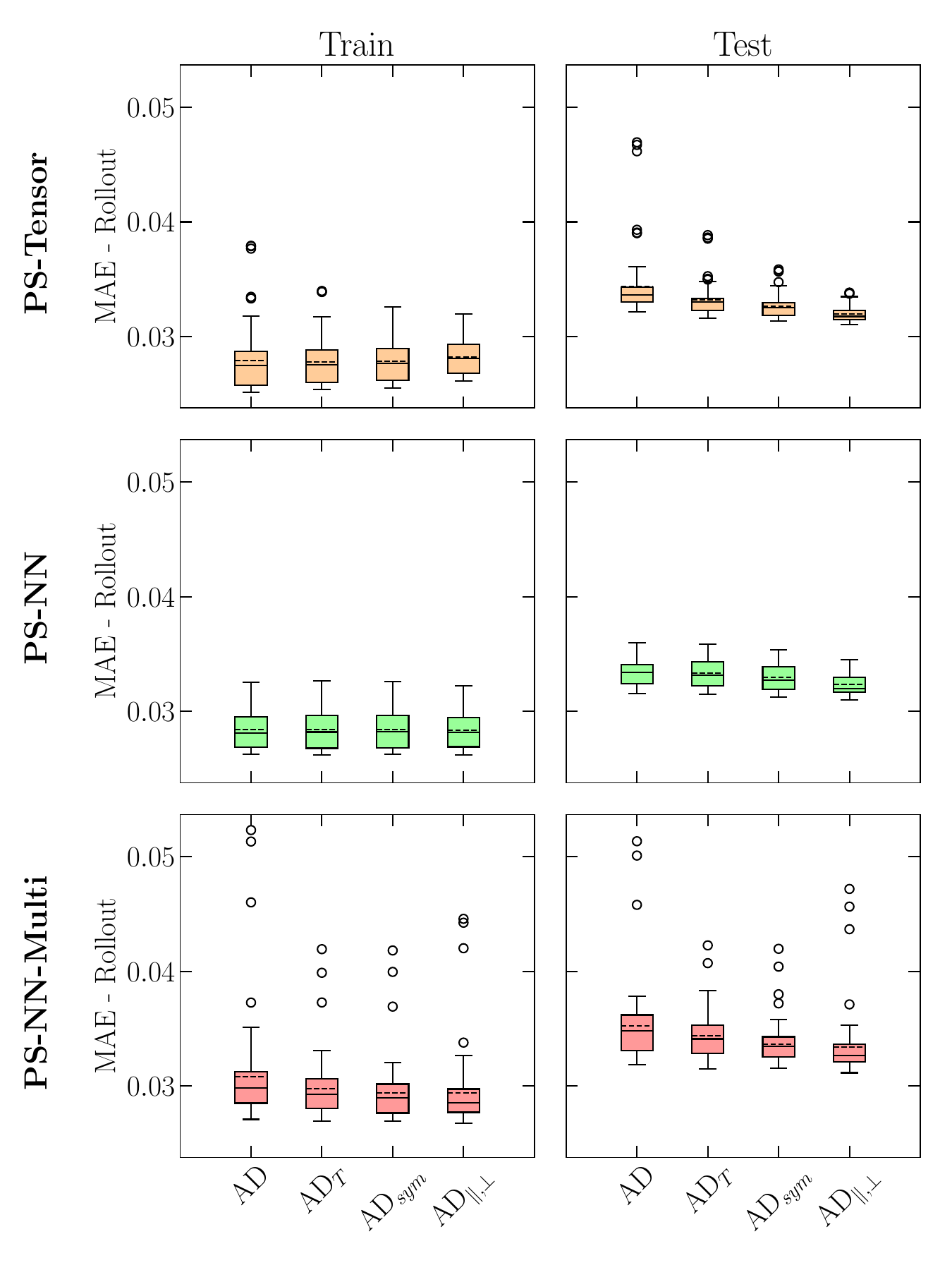}
    \caption{Impact of enforcing increasingly stricter symmetries on the rollout error. Introducing symmetries can lead to a slight increase in training error (removing some possible overfit), but consistently reduces the average test error and mitigates the appearance of outliers.}
    \label{fig:l1_model_comparison_boxplot_sym}
\end{figure}

The rollout performance of the different symmetries is shown in Figure~\ref{fig:l1_model_comparison_boxplot_sym}. 
The impact on the training subpopulations is not significant. In fact, a slight increase in average error is observed for PS-Tensor and PS-NN models, while for PS-NN-Multi, we observe a slight decrease. On the other hand, it is clear that imposing symmetries leads to an improvement in performance for test subpopulations. We attribute this improvement to a more accurate estimate of the coefficients at larger values of $v$ since the symmetries remove significant degrees of freedom, which artificially increases the statistics for estimating the coefficients in a given region of phase space. This leads to overall smoother operators. Notably, PS-Tensor (discrete) models end up performing slightly better than NN (continuous) models, which strengthens the argument that the smoothing bias imposed by the NN is no longer as important.

\begin{figure}
    \centering
    \includegraphics[width=\linewidth]{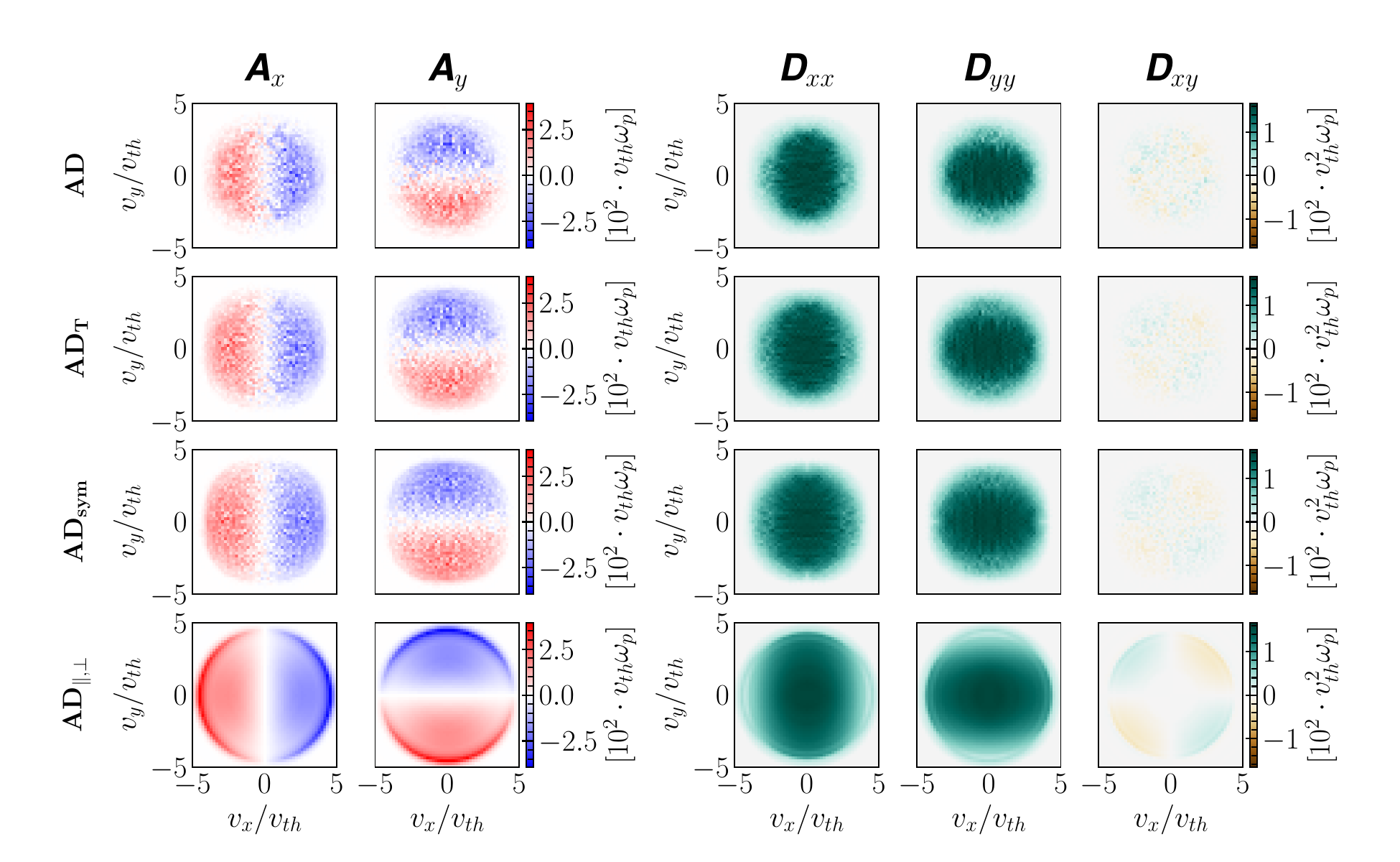}
    \caption{Impact of enforcing symmetries into Tensor models. Example shown corresponds to simulation $index=0$ ($N_{ppc}=4$, $m=1$, $\Delta_x/\lambda_D=1, v_{th}=0.01c)$. Enforcing symmetries allows us to recover smoother and more accurate coefficients for a larger phase space region. Artifacts are nonetheless present at high $v$ regions since very limited statistics are available at train time.}
    \label{fig:AD_comparison_sym_tensor_index_0}
\end{figure}

To illustrate these points, we provide in Figure~\ref{fig:AD_comparison_sym_tensor_index_0} the impact of symmetries in a PS-Tensor operator.
It is clear that, as we enforce stronger symmetries, a larger region of phase space is covered and smoother descriptions are recovered. However, due to the very limited train statistics at high $v$, artifacts (even if smooth) are still present (the thicker high advection regions and a small zero diffusion halo at high $v$). For NN-based approaches, we do not observe differences as striking as these when imposing more symmetries. However, it is still clear that symmetries improve results at high $v$, and PS-NN-Multi models can better capture the overall form of the operator. Additional examples are provided in Supplementary Material~\ref{app:enforcing_known_symmetries}.

The results shown in this subsection clearly highlight the benefits of enforcing known symmetries (which are problem dependent) for the task of retrieving collisional operators from sparse noisy phase space data. Enforcing stronger symmetries allows for the recovery of operators with less numerical artifacts and improves the operator generalization to unseen dynamics.

\subsection{Comparisons against theory}
\label{sec:comparisons_against_theory}

Finally, we compare the retrieved operators against theoretical predictions. To compute the theoretical curves, we numerically evaluate $\mathsfbi{A}, \mathsfbi{D}$ directly from equations~\eqref{eq:a_theory} and \eqref{eq:d_theory}. The derivative of the plasma dispersion function in~\eqref{eq:dieletric_function} is evaluated using the implementation available in PlasmaPy~\citep{plasmapy2024}. For all scenarios we consider that the test macroparticle moves with velocity $\boldsymbol{v} = v\hat{\boldsymbol{e}}_x$ (we do not observe any meaningful difference when setting the velocity at a different angle) such that $\mathsfi{A}_{x}(\boldsymbol{v}) = \mathsfi{A}_\parallel(v)$, $\mathsfi{A}_{y}(\boldsymbol{v}) = \mathsfi{A}_\perp(v) = 0$, $\mathsfi{D}_{xx}(\boldsymbol{v}) = \mathsfi{D}_\parallel(v)$, $\mathsfi{D}_{yy}(\boldsymbol{v}) = \mathsfi{D}_\perp(v)$, $\mathsfi{D}_{xy}(\boldsymbol{v}) = 0$.

\begin{figure}
    \centering
    \includegraphics[width=0.9\linewidth]{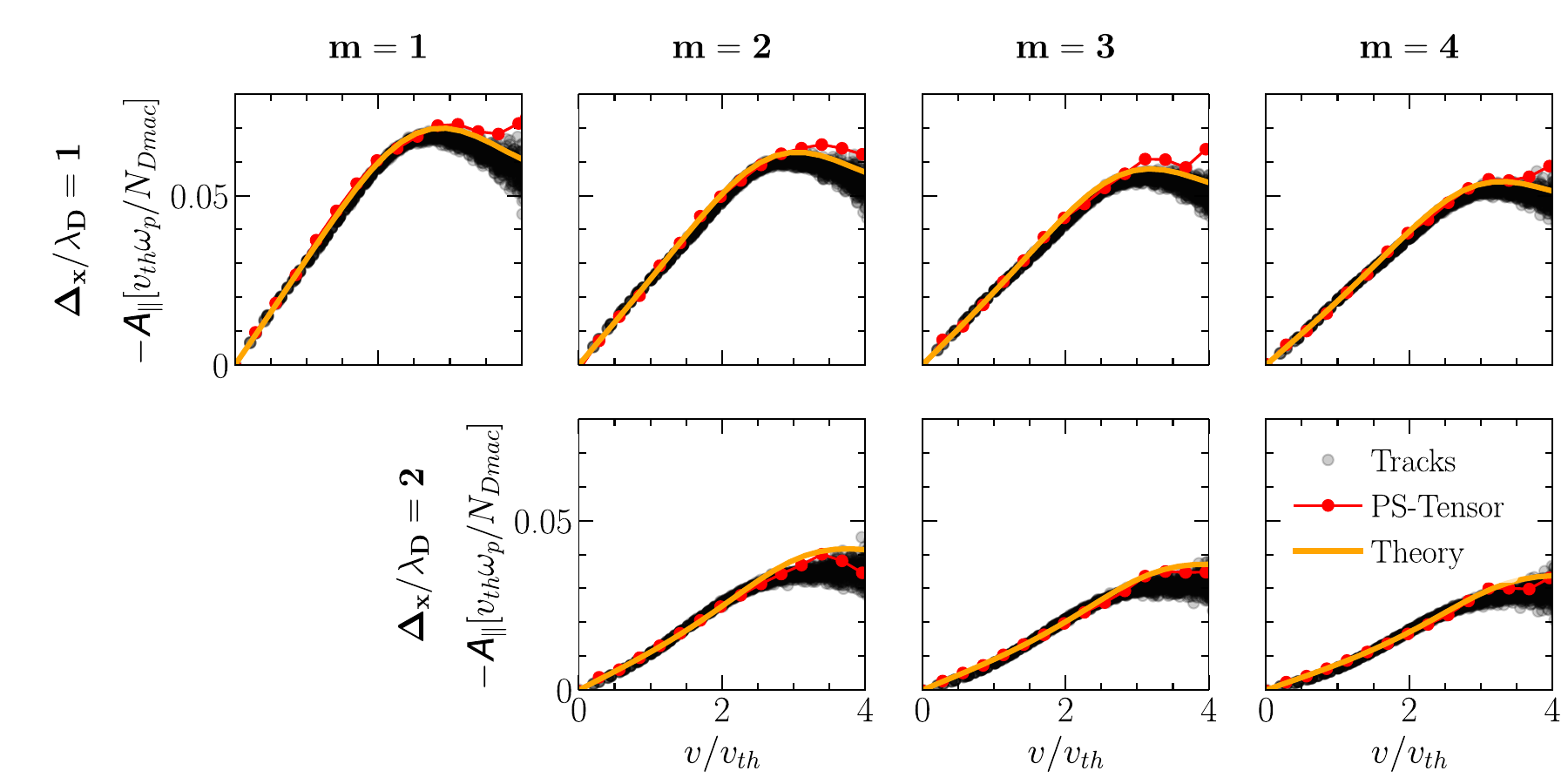}
    \includegraphics[width=0.9\linewidth]{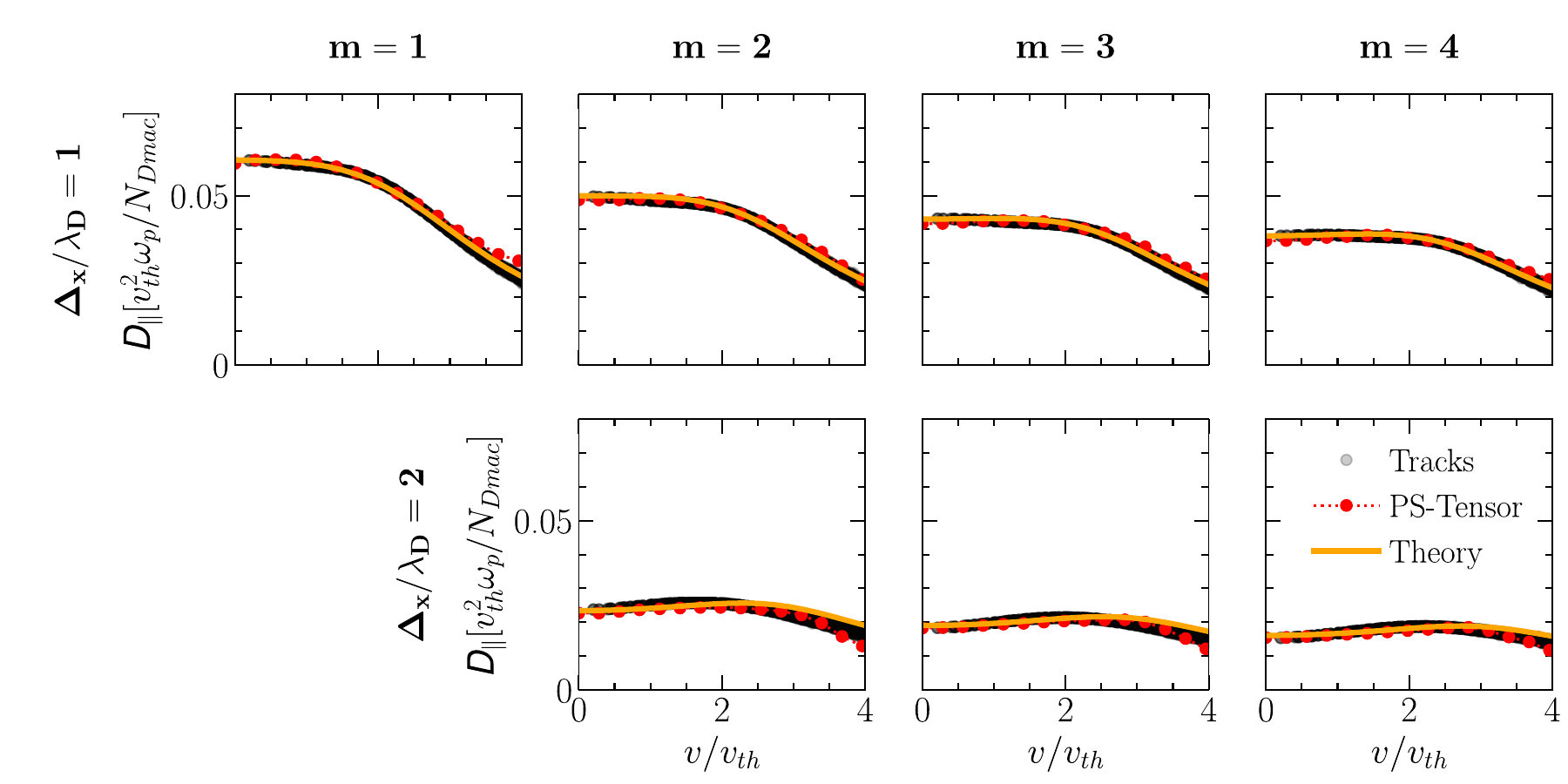}
    \includegraphics[width=0.9\linewidth]{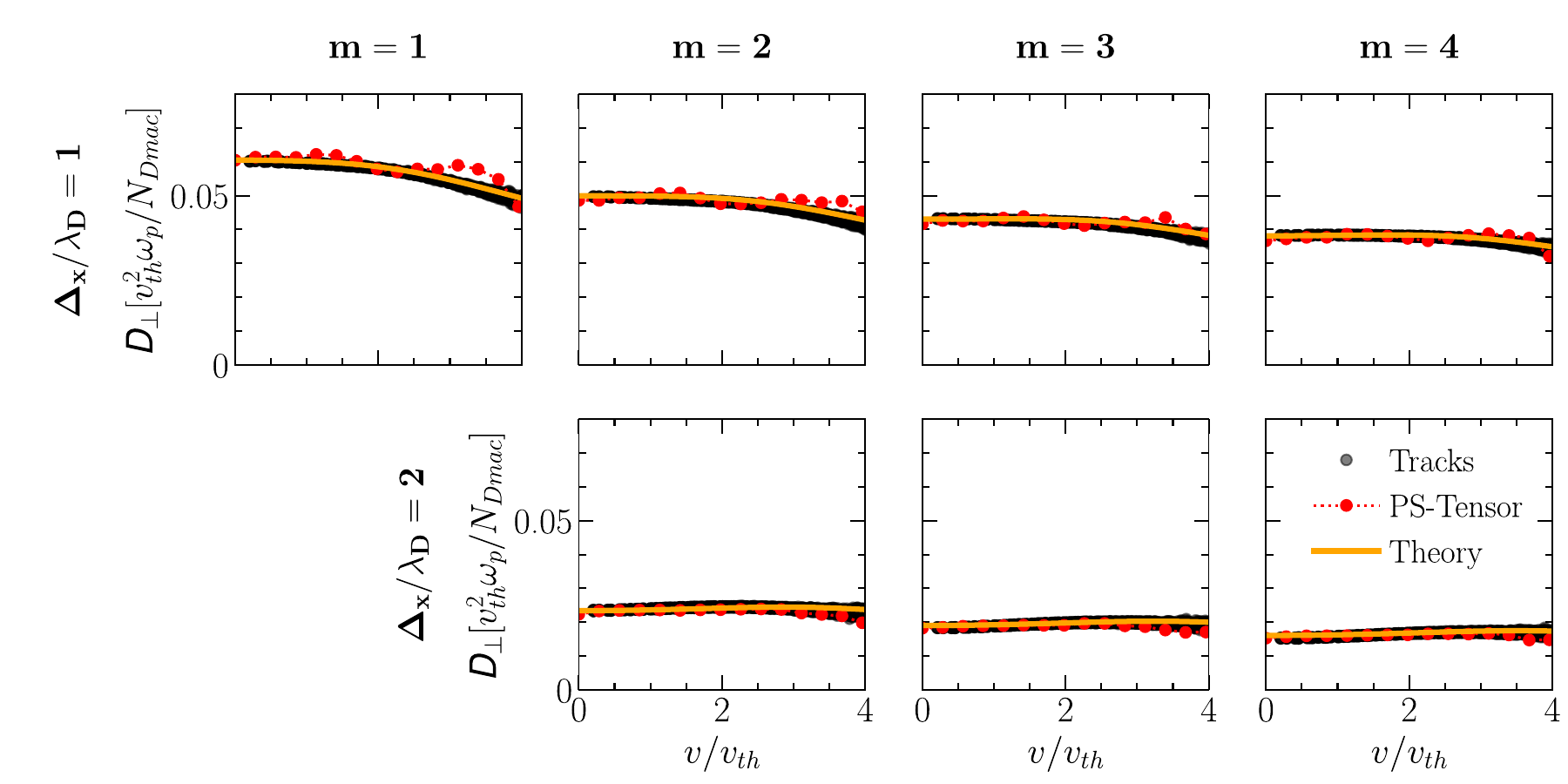}
    \caption{Comparison between theoretical values for advection / diffusion coefficients and those obtained from particle tracks and the PS-Tensor method for different shape functions and grid resolutions. All simulations shown use $N_{ppc}=25$ and $v_{th} = 0.01c$. There is overall an excellent agreement with theory, particularly up to $v/v_{th} = 3$. PS-Tensor is expected to not correctly capture well values above this threshold since training data did not contain significant statistics. Particle tracks measurements are also noisier after this value similarly due to lack of statistics.}
    \label{fig:theory_comparison}
\end{figure}

A comparison of the theoretical curves against the results obtained from particle tracks and PS-Tensor for simulations with $N_{ppc}=25$ and $v_{th} = 0.01c$ are provided in Figure~\ref{fig:theory_comparison}. 
We show only the region up to $v/v_{th} = 4$ since estimates beyond this value are not meaningful due to lack of statistics. Additionally, the PS-NN results are not included solely for readability purposes since they are equivalent to the PS-Tensor case.

Overall, we observe an excellent agreement between the theoretical predictions and the estimated values across different grid resolutions and shape functions. Differences are only visible for $\mathsfi{A}_\parallel$ when $v/v_{th} > 3$, a region where tracks measurements are also noisier and PS-Tensor shows some numerical artifacts at high $v$ (see e.g. Figure~\ref{fig:AD_comparison_sym_tensor_index_0}). Regarding both $\mathsfi{D}_\parallel$ and $\mathsfi{D}_\perp$ much better agreements are observed across all values of $v$. 

\begin{figure}
    \centering
    \includegraphics[width=0.8\linewidth]{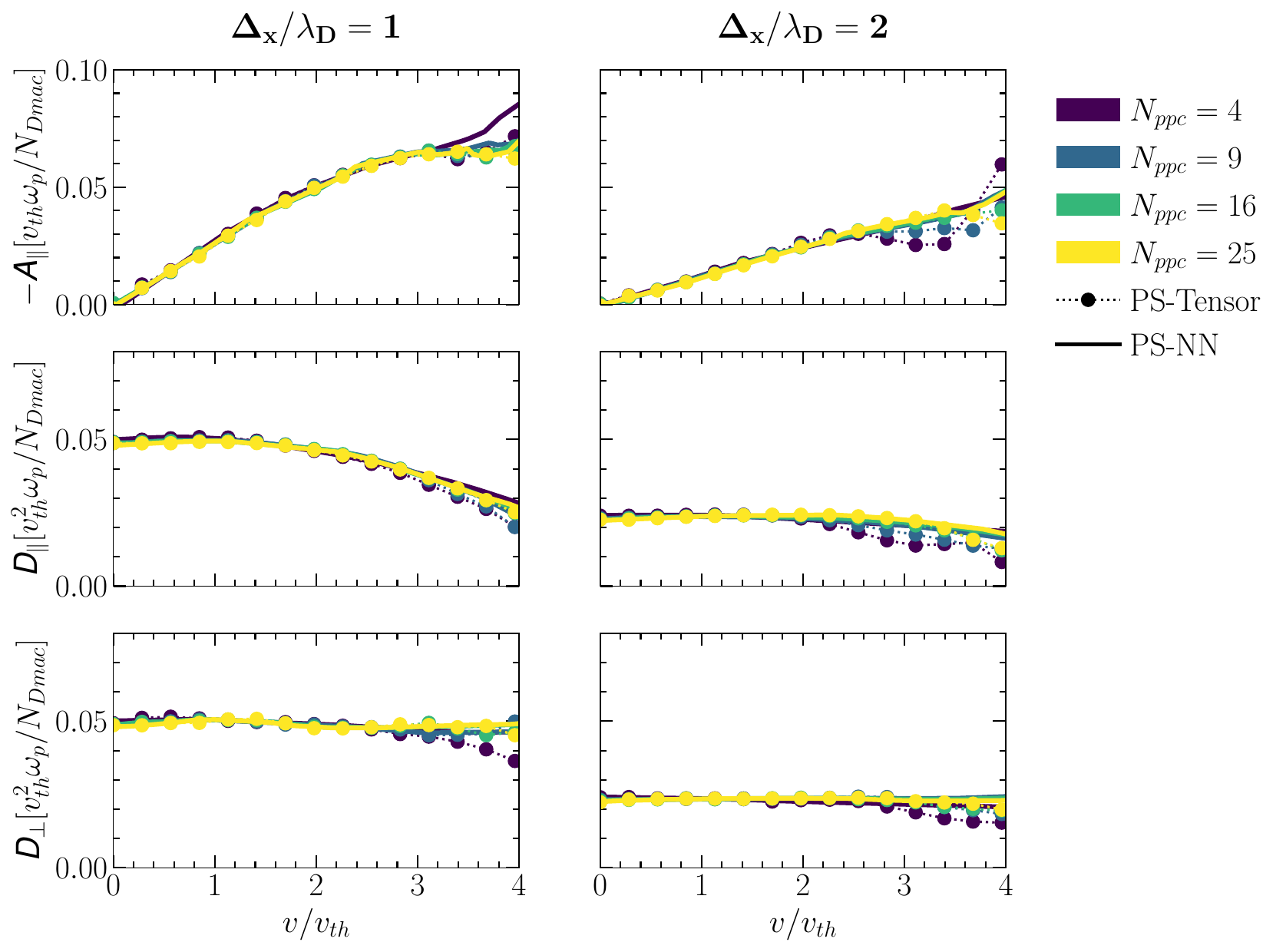}
    \caption{Comparison between $\mathrm{AD}_{\parallel\perp}$ operators for PS-Tensor and PS-NN models in function of the number of particles per cell. Curves are shown for $m=2$, $v_{th}=0.01c$, and varied grid resolutions. We observe that the advection and diffusion coefficients are inversely proportional to $N_{Dmac}$ as predicted by the theory~\citep{langdon1970nonphysical, langdon1970theory, touati2022kinetic}.}
    \label{fig:AD_parperp_tensor_vs_nn}
\end{figure}

The theoretical estimates in eqs. \eqref{eq:a_theory} and \eqref{eq:d_theory} state that the coefficients should be inversely proportional to $N_{Dmac} = N_{ppc} (\Delta_x/\lambda_D)^{-2}$. To verify this scaling, we plot in Figure~\ref{fig:AD_parperp_tensor_vs_nn} the advection and diffusion coefficients with respect to $N_{ppc}$ for varied grid resolutions, while fixing $m=2$ and $v_{th} = 0.01c$ (for readability purposes since the conclusions are equivalent for other tested values). 
We provide the results for both PS-Tensor and PS-NN models to showcase slight differences in behaviour at high $v$. We do not make use of Tracks since the coefficients do not reproduce the phase space dynamics of subpopulations with the same level of accuracy at smaller values of $N_{ppc}$ (c.f. Figure~\ref{fig:l1_model_comparison_per_simulation} in the Supplemental Material). It is clear that for smaller values of $v/v_{th} < 2.5$, both models recover the expected dependency, i.e. there exists no visible difference between the normalized curves. For $v/v_{th} > 2.5$ the conclusions are model-dependent. For PS-NN, the expected scaling is still observed, for PS-Tensor, we obtain slightly lower coefficients for lower $N_{ppc}$. We attribute the difference in behaviour to the lack of statistics at high $v$ and lower $N_{ppc}$, which makes the coefficient retrieval less accurate. While the PS-NN tends to extrapolate using a smoother estimate, the PS-Tensor model is more prone to noisy estimates.

\begin{figure}
    \centering
    \includegraphics[width=0.8\linewidth]{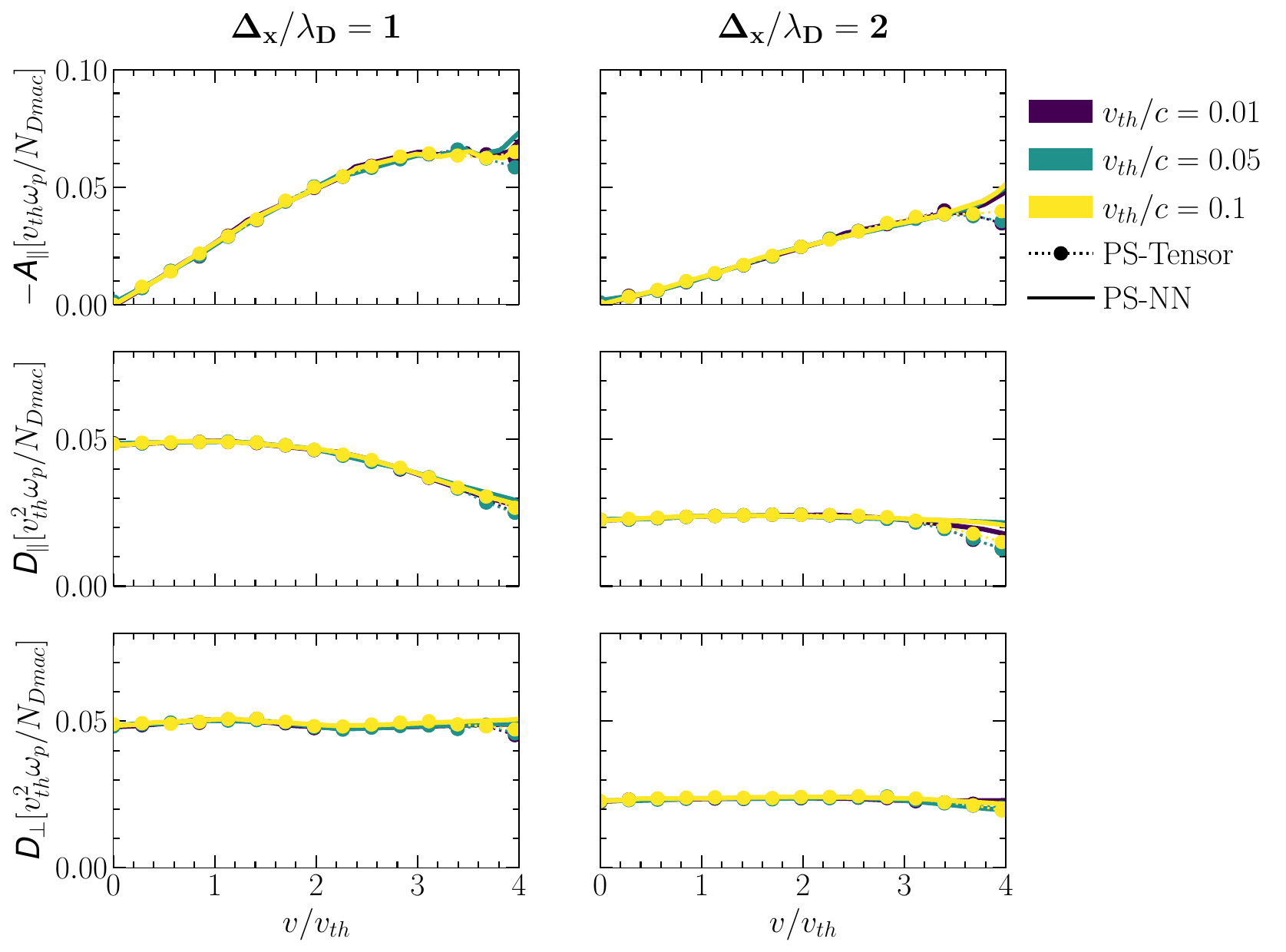}
    \caption{Comparison between $\mathrm{AD}_{\parallel\perp}$ operators for PS-Tensor and PS-NN  models in function of $v_{th}$. Curves are shown for $N_{ppc}=25$, $m=2$, and varied grid resolutions. We observe that advection is proportional to $v_{th}$ and diffusion is proportional to $v_{th}^2$ as predicted by the theory~\citep{langdon1970nonphysical, langdon1970theory, touati2022kinetic}.}
    \label{fig:AD_parperp_vth}
\end{figure}

Finally, in Figure~\ref{fig:AD_parperp_vth}, we plot the advection and diffusion coefficients with respect to $v_{th}$ for varied grid resolutions, while fixing $N_{ppc}=25$ and $m=2$. Once again, we recover the dependency predicted by theory, i.e. $\mathsfi{A}_\parallel \propto v_{th}$ and $\mathsfi{D}_{\parallel,\perp}\propto v_{th}^2$. This indicates that at these temperatures, electromagnetic interactions do not play a significant role in collisional dynamics since the collision operator is not modified.

These results allow us to demonstrate not only the accuracy of the existing theory, and its applicability to electromagnetic PIC simulations, but also the validity of the proposed approach to learn collision operators from phase space data. As far as we know, this is also the first time that such an in-depth verification of the theoretical PIC collision operator has been performed. This comparison also highlights that the methods proposed here can be used in physical regimes where analytical or closed-form expressions for $\mathsfbi{A}$ and $\mathsfbi{D}$ are not available or can not easily be determined. This will be explored in future publications.

\section{Conclusions}
\label{sec:conclusions}

In this paper we proposed a methodology which combines a differentiable simulator with gradient based optimization methods to learn Fokker-Planck collision operators from self-consistent Particle-in-Cell simulation data. We demonstrated that our method can efficiently learn an optimal operator based on phase space dynamics of subpopulations of particles across a wide range of simulations. The learned operators are on average more accurate than those estimated from particle tracks since they are optimized to predict the long-term phase space dynamics, and the method does not make assumptions about the relevant time scales of different plasma processes.
We have also shown that the recovered operators are in excellent agreement with the theoretical PIC collisional operator for a wide variety of numerical simulation parameters, further validating the theory and the proposed machine learning approach.

This work is a proof of principle, where we learn a collision operator which captures the self-consistent collisional dynamics of thermal plasmas composed of finite-size particles. The same approach can be used to learn collisional operators for self-consistent point-like particle dynamics, e.g., using MD simulations~\citep{zhao2025data, zhao2025fast} or PIC simulations which resolve the inter-particle distance~\citep{decker1994nonlinear, acciarri2024should}. These data-driven operators could then be integrated into Vlasov simulators~\citep{thomas2012review} or collisional modules in PIC codes~\citep{takizuka1977binary, manheimer1997langevin}, to statiscally reproduce the collisional effects without the need to self-consistently resolve the collisions. Furthermore, we note that in turbulent collisionless systems, the evolution of the distribution function can be described by a kinetic equation with an anomalous collision operator. Therefore our approach can also be used to learn reduced models of turbulent systems when provided with data of such setups~\citep{wong2020first, wong2025energy, camporeale2022data}. Finally, while a FP description is used in this work, the differentiable simulator framework here proposed can be easily extended to include other operator forms~\citep{zhao2025data, zhao2025fast}.

Since our method relies on phase space diagnostics rather than particle tracks, it is substantially more memory efficient than other post-processing approaches based on tracking individual particles. This advantage is particularly important for future studies employing large-scale, state-of-the-art 3-D simulations, where storing complete particle trajectories is prohibitively expensive and where the relevant time scales or operator forms are not known a priori and therefore cannot be identified during run time. Our approach enables diagnostics to be stored at significantly higher temporal cadence under the same memory constraints, allowing for a more comprehensive and better informed analysis during post-processing. Using phase space diagnostics also makes it possible to learn operators from observational~\citep{camporeale2022data} or experimental~\citep{bergeson2025experimental} data, although this might require extending the methodology to address the presence of (possibly significant) stochastic noise in the measurements.

In future works, we will focus both on the generalisation of the methodology and on exploring challenging scenarios for which the operator form is not known or the existing theory is expected to fail. We will extend the method to encompass time-varying background distributions, more general operator forms, the inclusion of external field contributions, and non-homogeneous plasmas. Our goal is to then apply this method to recover reduced models for non-thermal particle acceleration and collision operators for relativistic, electromagnetically dominated scenarios using self-consistent, first-principle simulations.

\vspace{+10pt}
\textbf{Acknowledgements}

The authors would like to thank S. Degen, G. Guttormsen, V. Decyk, and I. Kaganovich for valuable discussions, and the anonymous referees for their help improving the quality of the manuscript. The authors acknowledge the OSIRIS Consortium, consisting of UCLA, University of Michigan, and IST (Portugal) for the use of the
OSIRIS 4.0 framework. 

\vspace{+10pt}
\textbf{Funding}

Simulations and machine learning workloads were performed in Deucalion (Portugal) within the FCT I.P. project Masers in Astrophysical Plasmas (MAPs, P.B.) 2024.11062.CPCA.A3, FCT I.P. project Machine-learned closures for plasma simulations 2024.12682.CPCA.A1 (D.C.); and EuroHPC proposal No. EHPC-DEV-2025D02-069 (D.C.).
This work was supported by the FCT (Portugal) Grants No.~2022.13261.BD (D.C.), No.~2022.02230.PTDC (X-Maser, L.S.), and No.~UIDB/FIS/50010/2020-PESTB 2020-23 (L.S.); and by the National Science Foundation Grant No. PHY-2108089 (E.A.). D.C. research visit to the UCLA was sponsored by a Fulbright Grant for Research with the support of FCT and by the Mani L. Bhaumik Institute for Theoretical Physics at UCLA.

\vspace{+10pt}
\textbf{Declaration of interests}

The authors report no conflict of interest.

\vspace{+10 pt}
\textbf{Data availability}

The data that support the findings of this study are openly available in Zenodo at: \href{https://doi.org/10.5281/zenodo.18865875}{https://doi.org/10.5281/zenodo.18865875}, reference number 18865875. The software developed is openly available at: \href{https://github.com/diogodcarvalho/ml-pic-collision-operators}{https://github.com/diogodcarvalho/ml-pic-collision-operators}.

\putbib
\end{bibunit}

\begin{bibunit}
\include{supplementary}
\putbib
\end{bibunit}

\end{document}

%% file: supplementary.tex
%%%%%%%%%% Merge with supplemental materials %%%%%%%%%%
%%%%%%%%%% Prefix a "S" to all equations, figures, tables and reset the counter 
\setcounter{equation}{0}
\setcounter{figure}{0}
\setcounter{table}{0}
\setcounter{page}{1}
\setcounter{section}{0}
\makeatletter
\renewcommand{\theequation}{S\arabic{section}.\arabic{equation}}
\renewcommand{\thefigure}{S\arabic{figure}}
\renewcommand{\thetable}{S\arabic{table}}
\renewcommand{\thesection}{S\arabic{section}}
\renewcommand{\bibnumfmt}[1]{[S#1]}
\renewcommand{\citenumfont}[1]{S#1}

\shorttitle{}

\begin{center}
\textbf{\Large Learning collision operators from plasma phase space data using differentiable simulators \\ \vspace*{+5pt} Supplementary Materials} \\
\vspace{+10 pt}
D. D. Carvalho$^{1,2}$\footnote{\corresp{\email{diogo.d.carvalho@tecnico.ulisboa.pt}}}, P. J. Bilbao$^{1,3}$, W. B. Mori$^{2}$, L. O. Silva$^1$ and E. P. Alves$^{2,4}$ \\
\vspace{+10 pt}
\small
$^1$ GoLP/Instituto de Plasmas e Fus\~ao Nuclear, Instituto Superior T\'ecnico,
Universidade de Lisboa, 1049-001 Lisbon, Portugal \\
$^2$ Mani L. Bhaumik Institute for Theoretical Physics, University of California, Los Angeles, USA \\
$^3$ The Rudolf Peierls Centre for Theoretical Physics, University of Oxford, Oxford OX1 3NP, UK \\
$^4$ Department of Physics and Astronomy 
University of California, Los Angeles, USA
\end{center}
    
\vspace*{+20pt}
\addcontentsline{toc}{section}{Supplementary Materials}
\titlecontents{section}
  [3em]        % left margin
  {\bfseries}           % above code
  {\contentslabel{2em}} % number width
  {}           % unnumbered
  {\textbf{\titlerule*[0.5pc]{.}}\contentspage}
\titlecontents{subsection}
  [6em]        % left margin
  {}           % above code
  {\contentslabel{3em}} % number width
  {}           % unnumbered
  {\titlerule*[0.5pc]{.}\contentspage}
% Start collecting supplement contents
\startcontents[supp]
% Print supplement-only ToC
\printcontents[supp]{l}{1}{\setcounter{tocdepth}{2}}

\clearpage

\shorttitle{Supl. Material: Learning collision operators from plasma phase space data}

\section{Derivation of Advection-Diffusion Coefficients}
\label{app:derivation_pic_collision_operator}

Starting from eq. (86) in~\citet{touati2022kinetic} we obtain the general form of the PIC collision operator from the right-hand side of the equation defined as:
\begin{equation}
\begin{aligned}
\left(\frac{\partial f_a(\boldsymbol{v}_a)}{\partial t}\right)_{col}
= -\frac{1}{m_a}\,
\nabla_{\boldsymbol{v}_a} \boldsymbol{\cdot}
\Biggl[
 &\sum_{b}^{N_b} n_b
\int_{\mathbb{R}^N}\mathbf{d\boldsymbol{v}}_b\mathsfbi{Q}(\boldsymbol{v}_a,\boldsymbol{v}_b)
\\
& \boldsymbol{\cdot}
\Biggl(
\frac{\delta N_b}{m_a}\, f_b(\boldsymbol{v}_b)\,
\nabla_{\boldsymbol{v}_a} f_a(\boldsymbol{v}_a) 
\\
&-
\frac{\delta N_a}{m_b}\, f_a(\boldsymbol{v}_a)\,
\nabla_{\boldsymbol{v}_b} f_b(\boldsymbol{v}_b)
\Biggr)
\, 
\Biggr] .
\end{aligned}
\label{eq:touati_86}
\end{equation}
which considers the interaction of a beam of particles of specie $a$, and initial velocity $\boldsymbol{v}_a$, interacting with $N_b$ species. For each specie $i$, $f_i(\boldsymbol{v}_i)$ corresponds to the probability distribution function in velocity space such that $\int_{\mathbb{R}^2} \mathbf{d\boldsymbol{v}}_i  f_i(\boldsymbol{v}_i) = 1$,\footnote{\citet{touati2022kinetic} defines instead $\int_{\mathbb{R}^2} \mathbf{d\boldsymbol{v}}_i f_i(\boldsymbol{v}_i) = n_i$ and that is why an extra multiplying factor of $n_b$ appears in~\eqref{eq:touati_86}.} $m_i$ corresponds to the species mass, $\delta N_i = n_i \Delta_x^N/N_{i, ppc}$ is the specie macroparticle weight where $n_i$ is the specie density, $\Delta_x$ is the grid resolution (assumed to be equal across the $N$ dimensions), and $N_{i,ppc}$ is the number of particles per cell. The PIC \textit{collision tensor} $\mathsfbi{Q}(\boldsymbol{v}_a,\boldsymbol{v}_b)$ in 2-D is obtained from eq. (83) in~\citet{touati2022kinetic} as:
\begin{equation}
\mathsfbi{Q}(\boldsymbol{v}_a, \boldsymbol{v}_b) 
= 
- \frac{q_a^2 q_b^2}{4\pi \varepsilon_0^2} 
\int_{\mathbb{R}^2} \mathbf{d\boldsymbol{k}} 
    \frac{S_\rho^m(\boldsymbol{k})^4}{|\epsilon (\boldsymbol{k} \boldsymbol{\cdot} \boldsymbol{v}_a, \boldsymbol{k})|^2 k_s^4} 
    \boldsymbol{k}_s\boldsymbol{\otimes} \boldsymbol{k}_s 
    \delta(\boldsymbol{k} \boldsymbol{\cdot} \boldsymbol{v}_a - \boldsymbol{k} \boldsymbol{\cdot} \boldsymbol{v}_b) \ , 
\label{eq:touati_83}
\end{equation}
where we have discarded temporal and spatial aliasing effects.  The multiplying factors are different from eq. (83) in~\citet{touati2022kinetic} since we have changed the integration from $\mathbb{R}^3$ to  $\mathbb{R}^2$, which adds an extra multiplying factor of $2\pi$~\citep{touati2022kinetic}, and the original formula is in cgs units, therefore introducing an extra factor of $1/(4\pi\varepsilon_0)^2$ when writing the operator in SI units. In this formula $q_i$ corresponds to the species charge, $S_\rho^m(\boldsymbol{k})$ is the charge deposition shape function of order $m$, and for a FDTD solver $\boldsymbol{k}_s =  (k_x \mathrm{sinc}(k_x\Delta_x/2), k_y\mathrm{sinc}(k_y\Delta_y/2))$ with $\mathrm{sinc}(x) = \sin(x)/x$. Finally, $\delta(.)$ corresponds to the delta function, and $\epsilon(.)$ is the dielectric function defined as~\citep{touati2022kinetic}:
\begin{equation}
\epsilon(\boldsymbol{k} \boldsymbol{\cdot} \boldsymbol{v}_a, \boldsymbol{k}) = 
1 + \sum_{c}^{N_c} \frac{\omega_{pc}^2}{k_s^2}
S_\rho^m(\boldsymbol{k})^2
\int_{\mathbb{R}^2} \mathbf{d\boldsymbol{v}}_c \,
\nabla_{\boldsymbol{v}_c} f_c(\boldsymbol{v}_c) \boldsymbol{\cdot}
\frac{\boldsymbol{k}_s}{\boldsymbol{k} \boldsymbol{\cdot} \boldsymbol{v}_a - \boldsymbol{k} \boldsymbol{\cdot} \boldsymbol{v}_c} \ ,
\label{eq:touati_63}
\end{equation}
where the sum is defined over all the $N_c$ species in the plasma and $\omega_{pc} = \sqrt{n_c q_c^2 / \varepsilon_o m_c}$.

One can re-write eq.~\eqref{eq:touati_86} in the form of a Fokker-Planck (FP) equation:
\begin{equation}
\left( \frac{\partial f_a}{\partial t} \right)_{col}
=
-\,\nabla_{\boldsymbol{v}_a}
\boldsymbol{\cdot}
\left[
\mathsfbi{A}^{(1)}(\boldsymbol{v}_a)\,f_a(\boldsymbol{v}_a)
\right]
+
\frac{1}{2}\,
\nabla_{\boldsymbol{v}_a}
\boldsymbol{\cdot}
\left[
\mathsfbi{D}(\boldsymbol{v}_a)
\boldsymbol{\cdot}
\nabla_{\boldsymbol{v}_a} f_a(\boldsymbol{v}_a)
\right] \ ,
\label{eq:fp_v2}
\end{equation}
where
\begin{equation}
\mathsfbi{A}^{(1)}(\boldsymbol{v}_a)
=
-\sum_b^{N_b}
\frac{n_b \delta N_a}{m_am_b}
\int_{\mathbb{R}^2} \mathbf{d\boldsymbol{v}}_b \,
\mathsfbi{Q}(\boldsymbol{v}_a,\boldsymbol{v}_b)\,
\boldsymbol{\cdot}\nabla_{\boldsymbol{v}_b} f_b(\boldsymbol{v}_b)
\end{equation}
\begin{equation}
\mathsfbi{D}(\boldsymbol{v}_a)
=
-\sum_b^{N_b}
\frac{2 n_b \delta N_b}{m_a^2}
\int_{\mathbb{R}^2} \mathbf{d\boldsymbol{v}}_b \,
\mathsfbi{Q}(\boldsymbol{v}_a,\boldsymbol{v}_b)\,
f_b(\boldsymbol{v}_b) \ .
\end{equation}

Assuming that we have a one-component electron plasma, then the sum over species can be disregarded and the species under-scripts removed (i.e. $a\equiv b$ for all quantities except the velocity vectors where they are maintained). By defining the macroparticle density $n_{mac}=N_{ppc}/\Delta_x^N$ we obtain a simplified expression for the operators as:
\begin{equation}
\mathsfbi{A}^{(1)}(\boldsymbol{v}_a)
=
-\frac{n^2}{n_{mac}m^2}
\int_{\mathbb{R}^2} \mathbf{d\boldsymbol{v}}_b \,
\mathsfbi{Q}(\boldsymbol{v}_a,\boldsymbol{v}_b)\,
\boldsymbol{\cdot} \nabla_{\boldsymbol{v}_b} f(\boldsymbol{v}_b)
\label{eq:app_a1}
\end{equation}
\begin{equation}
\mathsfbi{D}(\boldsymbol{v}_a)
=
-\frac{2 n^2}{n_{mac}m^2}
\int_{\mathbb{R}^2} \mathbf{d\boldsymbol{v}}_b \,
\mathsfbi{Q}(\boldsymbol{v}_a,\boldsymbol{v}_b)\,
f(\boldsymbol{v}_b) \ .
\label{eq:app_d}
\end{equation}

To re-write eq.~\eqref{eq:fp_v2} as eq.~\eqref{eq:fp_equation} (the one used in the main body of the paper):
\begin{equation}
    \left(\frac{\partial f(\boldsymbol{v}_a)}{\partial t}\right)_{col} = - \nabla_{\boldsymbol{v}_a}\boldsymbol{\cdot}(\mathsfbi{A}(\boldsymbol{v}_a)f(\boldsymbol{v}_a)) + \frac{1}{2}\nabla_{\boldsymbol{v}_a}\boldsymbol{\cdot}\left[\nabla_{\boldsymbol{v}_a}\boldsymbol{\cdot}(\mathsfbi{D}(\boldsymbol{v}_a)f(\boldsymbol{v}_a))\right] \ ,
\label{eq:fp_v1}
\end{equation}
we define a new advection coefficient:
\begin{equation}
    \mathsfbi{A}(\boldsymbol{v}_a) = \mathsfbi{A}^{(1)}(\boldsymbol{v}_a) + \mathsfbi{A}^{(2)}(\boldsymbol{v}_a)
\end{equation}
\begin{equation}
\mathsfbi{A}^{(2)}(\boldsymbol{v}_a) = \frac{1}{2}\nabla_{\boldsymbol{v}} \boldsymbol{\cdot} \mathsfbi{D}(\boldsymbol{v}_a) \ .
\label{eq:app_a_2}
\end{equation}

The benefit of writing the FP equation in this form is that $\mathsfbi{A}(\boldsymbol{v}_a) \equiv d< \Delta \boldsymbol{v}>_{\boldsymbol{v}=\boldsymbol{v}_a} / dt$ and $\mathsfbi{D}(\boldsymbol{v}_a) \equiv d<\Delta \boldsymbol v\Delta \boldsymbol{v}^{T}>_{\boldsymbol{v}=\boldsymbol{v}_a} / dt$ which is the assumption we use in Section~\ref{sec:methods} to estimate the coefficients from particle tracks. Note however that the vast majority of previous works (e.g.,~\citet{langdon1970theory, langdon1970nonphysical, okuda1970collisions}) use the FP equation of the form~\eqref{eq:fp_v2} and therefore consider the advection coefficients to be given by what we define as $\mathsfbi{A}^{(1)}$.

To obtain the final expressions for the advection \eqref{eq:a_theory}, diffusion \eqref{eq:d_theory}, and dielectric function \eqref{eq:dieletric_function} one needs to: assume a thermal distribution $f(\boldsymbol{v}) = 1/(2\pi v_{th}^2) \exp{(-v^2/2v_{th}^2)}$; compute analytically the integrals in eqs. \eqref{eq:app_a1}, \eqref{eq:app_d}, and \eqref{eq:touati_63}; normalize all wavenumbers to the inverse of the Debye length $\lambda_D$, and velocities to the thermal velocity $v_{th}$. These steps are provided in detail in the following sub-sections.

\subsection{Advection Term}

For a normalized 2-D thermal distribution $f(\boldsymbol{v})$ given by:
\begin{equation}
    f(\boldsymbol{v}) = \frac{1}{2\pi v_{th}^2} e^{-\frac{v^2}{2v_{th}^2}} \ ,
\end{equation}
one obtains its gradient as:
\begin{equation}
\nabla_{\boldsymbol{v}} f(\boldsymbol{v}) = -\frac{\boldsymbol{v}}{2\pi v_{th}^4} e^{-\frac{v^2}{2v_{th}^2}} \ .
\label{eq:app_nabla_f}
\end{equation}

Eq.~\eqref{eq:app_a1} can then be written using eqs.~\eqref{eq:touati_83} and \eqref{eq:app_nabla_f}, and substituting $n^2q^4/m^2\varepsilon_0^2=\omega_p^4$ as:
\begin{equation}
\mathsfbi{A}^{(1)}(\boldsymbol{v}_a)
=
-\frac{\omega_p^4}{2^3\pi^2 n_{mac}v_{th}^4}
\int_{\mathbb{R}^2}
\mathbf{d\boldsymbol{k}} \,
\frac{
    S_\rho^m(\boldsymbol{k})^4
}{
    |\epsilon (\boldsymbol{k} \boldsymbol{\cdot} \boldsymbol{v}_a, \boldsymbol{k})|^2 k_s^4
}
(\boldsymbol{k}_s\boldsymbol{\otimes} \boldsymbol{k}_s) 
\boldsymbol{\cdot}
\int_{\mathbb{R}^2}
\mathbf{d\boldsymbol{v}}_b \,
\delta(\boldsymbol{k} \boldsymbol{\cdot} \boldsymbol{v}_a - \boldsymbol{k} 
\boldsymbol{\cdot}
\boldsymbol{v}_b) \,
\boldsymbol{v}_b \, e^{-\frac{v_b^2}{2v_{th}^2}}
\label{eq:app_a1_derivation_1}
\end{equation}

By defining $\boldsymbol{k} \boldsymbol{\cdot} \boldsymbol{v} = k v_\parallel$ and using the fact that $\delta(ax)  = \delta(x)/|a|$, the second integral can be rewritten as:
\begin{equation}
\int_{\mathbb{R}^2}
\mathbf{d\boldsymbol{v}}_b \,
\delta(\boldsymbol{k} \boldsymbol{\cdot} \boldsymbol{v}_a - \boldsymbol{k} 
\boldsymbol{\cdot}
\boldsymbol{v}_b) \,
\boldsymbol{v}_b \, e^{-\frac{v_b^2}{2v_{th}^2}} = 
    \frac{1}{k}\int_{\mathbb{R}^2} 
    dv_{b,\parallel}dv_{b,\perp} \,
    \delta(v_{a,\parallel} - v_{b,\parallel}) \,
(v_{b,\parallel}\boldsymbol{\hat{e}}_{\parallel} + v_{b,\perp} \boldsymbol{\hat{e}}_{\perp}) \, 
e^{-\frac{v_{b,\parallel}^2 +  v_{b,\perp}^2}{2v_{th}^2}} \ ,
\end{equation}
where $\boldsymbol{\hat{e}}_{\parallel}$ and $\boldsymbol{\hat{e}}_{\perp}$  represent the unit vectors parallel and perpendicular to $\boldsymbol{k}$. Integrating first over $dv_{b,\parallel}$ results in:
\begin{equation}
\frac{1}{k} e^{-\frac{v_{a,\parallel}^2}{2v_{th}^2}}\int_{\mathbb{R}} 
    dv_{b,\perp} \,\,
(v_{a,\parallel}\boldsymbol{\hat{e}}_{\parallel} + v_{b,\perp} \boldsymbol{\hat{e}}_{\perp}) \, 
e^{-\frac{v_{b,\perp}^2}{2v_{th}^2}} \ .
\end{equation}

The integral associated with $\boldsymbol{\hat{e}}_{\perp}$ is zero since the function is odd. By applying a change of variables such that $u = v_{b,\perp}/\sqrt{2}v_{th}$, noting that $\int_\mathbb{R} du \,\exp(-u^2) = \sqrt{\pi}$, and recovering the definition that $v_{a,\parallel} = \boldsymbol{k} \boldsymbol{\cdot} \boldsymbol{v}_a / k$ and $\boldsymbol{\hat{e}}_{\parallel}=\boldsymbol{k}/k$ one obtains the result for the parallel component as:
\begin{align}
\frac{1}{k} e^{-\frac{v_{a,\parallel}^2}{2v_{th}^2}}\int_{\mathbb{R}} 
    dv_{b,\perp} \,\,
v_{a,\parallel}\boldsymbol{\hat{e}}_{\parallel} \, 
e^{-\frac{v_{b,\perp}^2}{2v_{th}^2}} 
&= 
\frac{\sqrt{2\pi} v_{th}}{k} v_{a,\parallel} \boldsymbol{\hat{e}}_{\parallel} \, e^{-\frac{v_{a,\parallel}^2}{2v_{th}^2}} \\
&=\sqrt{2\pi} v_{th} \frac{\boldsymbol{k} (\boldsymbol{k} \boldsymbol{\cdot} \boldsymbol{v}_a)}{k^3} \, e^{-\frac{(\boldsymbol{k} \boldsymbol{\cdot} \boldsymbol{v}_a) ^2}{2k^2v_{th}^2}} 
\ .
\end{align}

We can then substitute this result in eq.~\eqref{eq:app_a1_derivation_1} to obtain:
\begin{equation}
    \mathsfbi{A}^{(1)}(\boldsymbol{v}_a)
=
-\frac{\omega_p^4}{2^{5/2}\pi^{3/2} n_{mac}v_{th}^3}
\int_{\mathbb{R}^2}
\mathbf{d\boldsymbol{k}} \,
\frac{
    S_\rho^m(\boldsymbol{k})^4
}{
    |\epsilon (\boldsymbol{k} \boldsymbol{\cdot} \boldsymbol{v}_a, \boldsymbol{k})|^2 k_s^4 
}
(\boldsymbol{k}_s\boldsymbol{\otimes} \boldsymbol{k}_s)
\boldsymbol{\cdot}
\frac{\boldsymbol{k} (\boldsymbol{k} \boldsymbol{\cdot} \boldsymbol{v}_a)}{k^3} \, e^{-\frac{(\boldsymbol{k} \boldsymbol{\cdot} \boldsymbol{v}_a) ^2}{2k^2v_{th}^2}} \ ,
\end{equation}
and use the property that for column vectors $(\boldsymbol{a}\boldsymbol{\otimes} \boldsymbol{b}) \boldsymbol{\cdot}  \boldsymbol{c} = \boldsymbol{a}(\boldsymbol{b}\boldsymbol{\cdot} \boldsymbol{c})$ to arrive at:
\begin{equation}
    \mathsfbi{A}^{(1)}(\boldsymbol{v}_a)
=
-\frac{\omega_p^4}{2^{5/2}\pi^{3/2} n_{mac}v_{th}^3}
\int_{\mathbb{R}^2}
\mathbf{d\boldsymbol{k}} \,
\frac{
    S_\rho^m(\boldsymbol{k})^4
}{
    |\epsilon (\boldsymbol{k} \boldsymbol{\cdot} \boldsymbol{v}_a, \boldsymbol{k})|^2 k_s^4 
}\frac{\boldsymbol{k}_s (\boldsymbol{k}_s \boldsymbol{\cdot} \boldsymbol{k}) (\boldsymbol{k} \boldsymbol{\cdot} \boldsymbol{v}_a)}{k^3} \, e^{-\frac{(\boldsymbol{k} \boldsymbol{\cdot} \boldsymbol{v}_a) ^2}{2k^2v_{th}^2}} \ .
\end{equation}

Finally, by normalizing all quantities with units of wavenumber to the inverse of the Debye length ($\boldsymbol{\tilde{k}} = \boldsymbol{k}\lambda_D$, $\boldsymbol{\tilde{k}}_s=\boldsymbol{k}_s\lambda_D$), the velocities by the thermal velocity ($\boldsymbol{\tilde{v}}_i = \boldsymbol{v}_i / v_{th}$), and by defining the number of macroparticles per Debye square as $N_{Dmac}=n_{mac}\lambda_D^2$ we obtain:
\begin{equation}
    \mathsfbi{A}^{(1)}(\boldsymbol{v}_a)
=
-\frac{\omega_p v_{th}}{2^{5/2}\pi^{3/2} N_{Dmac}}
\int_{\mathbb{R}^2}
\mathbf{d\boldsymbol{\tilde{k}}} \,
\frac{
    S_\rho^m(\boldsymbol{\tilde{k}})^4
}{
    |\epsilon (\boldsymbol{\tilde{k}} \boldsymbol{\cdot} \boldsymbol{\tilde{v}}_a, \boldsymbol{\tilde{k}})|^2 \tilde{k}_s^4 
}\frac{\boldsymbol{\tilde{k}}_s (\boldsymbol{\tilde{k}}_s \boldsymbol{\cdot} \boldsymbol{\tilde{k}}) (\boldsymbol{\tilde{k}} \boldsymbol{\cdot} \boldsymbol{\tilde{v}}_a)}{\tilde{k}^3} \, e^{-\frac{(\boldsymbol{\tilde{k}} \boldsymbol{\cdot} \boldsymbol{\tilde{v}}_a)^2}{2\tilde{k}^2}} \ ,
\end{equation}
which matches eq.~\eqref{eq:a1_theory} in the main body of the paper. Since eq.~\eqref{eq:a_2_theory} simply corresponds to a normalized version of eq.~\eqref{eq:app_a_2} (where $\boldsymbol{\tilde{v}} = \boldsymbol{v}/v_{th}$ leads to $\nabla_{\boldsymbol{v}} = \nabla_{\boldsymbol{\tilde{v}}}/v_{th}$), the equation provided for the full advection term in the main body of the paper \eqref{eq:a_theory} is therefore recovered.

\subsection{Diffusion Term}

Starting from eqs.~\eqref{eq:app_d} and~\eqref{eq:touati_83} we obtain for a thermal distribution:
\begin{equation}
\mathsfbi{D}(\boldsymbol{v}_a)
=
\frac{\omega_p^4}{2^2\pi^2 n_{mac}v_{th}^2}
\int_{\mathbb{R}^2}
\mathbf{d\boldsymbol{k}} \,
\frac{
    S_\rho^m(\boldsymbol{k})^4
}{
    |\epsilon (\boldsymbol{k} \boldsymbol{\cdot} \boldsymbol{v}_a, \boldsymbol{k})|^2 k_s^4
}
\boldsymbol{k}_s\boldsymbol{\otimes} \boldsymbol{k}_s 
\int_{\mathbb{R}^2}
\mathbf{d\boldsymbol{v}}_b \,
\delta(\boldsymbol{k} \boldsymbol{\cdot} \boldsymbol{v}_a - \boldsymbol{k} 
\boldsymbol{\cdot}
\boldsymbol{v}_b) \, e^{-\frac{v_b^2}{2v_{th}^2}} \ .
\label{eq:ap_d_derivation}
\end{equation}

Similarly to the advection case we can first handle the integral over $\boldsymbol{v}_b$ as follows:
\begin{align}
\int_{\mathbb{R}^2}
\mathbf{d\boldsymbol{v}}_b \,
\delta(\boldsymbol{k} \boldsymbol{\cdot} \boldsymbol{v}_a - \boldsymbol{k} 
\boldsymbol{\cdot}
\boldsymbol{v}_b) \,
e^{-\frac{v_b^2}{2v_{th}^2}} 
&= 
\frac{1}{k}\int_{\mathbb{R}^2} 
dv_{b,\parallel}dv_{b,\perp} \,
\delta(v_{a,\parallel} - v_{b,\parallel}) \, 
e^{-\frac{v_{b,\parallel}^2 +  v_{b,\perp}^2}{2v_{th}^2}} \\
&=
\frac{1}{k}e^{-\frac{v_{a,\parallel}^2}{2v_{th}^2}} 
\int_{\mathbb{R}} 
    dv_{b,\perp} \,
    e^{-\frac{v_{b,\perp}^2}{2v_{th}^2}} \\
&= \frac{\sqrt{2\pi} v_{th}}{k}e^{-\frac{v_{a,\parallel}^2}{2v_{th}^2}}  \\
&= \frac{\sqrt{2\pi} v_{th}}{k}e^{-\frac{(\boldsymbol{k}\boldsymbol{\cdot}\boldsymbol{v}_a)^2}{2k^2v_{th}^2}} \ .
\end{align}

Substituting this result in eq.~\eqref{eq:ap_d_derivation} leads to:
\begin{equation}
\mathsfbi{D}(\boldsymbol{v}_a)
=
\frac{\omega_p^4}{2^{3/2}\pi^{3/2} n_{mac}v_{th}}
\int_{\mathbb{R}^2}
\mathbf{d\boldsymbol{k}} \,
\frac{
    S_\rho^m(\boldsymbol{k})^4
}{
    |\epsilon (\boldsymbol{k} \boldsymbol{\cdot} \boldsymbol{v}_a, \boldsymbol{k})|^2 k_s^4
}
\frac{\boldsymbol{k}_s\boldsymbol{\otimes} \boldsymbol{k}_s}{k}
e^{-\frac{(\boldsymbol{k}\boldsymbol{\cdot}\boldsymbol{v}_a)^2}{2 k^2 v_{th}^2}} \ ,
\end{equation}
which after normalizing all wavenumbers and velocities results in:
\begin{equation}
\mathsfbi{D}(\boldsymbol{v}_a)
=
\frac{\omega_p v_{th}^2}{2^{3/2}\pi^{3/2} N_{Dmac}}
\int_{\mathbb{R}^2}
\mathbf{d\boldsymbol{\tilde{k}}} \,
\frac{
    S_\rho^m(\boldsymbol{\tilde{k}})^4
}{
    |\epsilon (\boldsymbol{\tilde{k}} \boldsymbol{\cdot} \boldsymbol{\tilde{v}}_a, \boldsymbol{\tilde{k}})|^2 \tilde{k}_s^4
}
\frac{\boldsymbol{\tilde{k}}_s\boldsymbol{\otimes} \boldsymbol{\tilde{k}}_s}{\tilde{k}}
e^{-\frac{(\boldsymbol{\tilde{k}}\boldsymbol{\cdot}\boldsymbol{\tilde{v}}_a)^2}{2 \tilde{k}^2}} \ ,
\end{equation}
the same result presented in eq.~\eqref{eq:d_theory} in the main body of the paper.

\subsection{Dielectric Function}

For a one component plasma in thermal equilibrium the dielectric function in eq.~\eqref{eq:touati_63} simplifies to:
\begin{align}
\epsilon(\boldsymbol{k} \boldsymbol{\cdot} \boldsymbol{v}_a, \boldsymbol{k}) 
&= 
1 + \frac{\omega_{p}}{k_s^2}
S_\rho^m(\boldsymbol{k})^2
\int_{\mathbb{R}^2} \mathbf{d\boldsymbol{v}}_c \,
\nabla_{\boldsymbol{v}_c} f(\boldsymbol{v}_c) \boldsymbol{\cdot}
\frac{\boldsymbol{k}_s}{\boldsymbol{k} \boldsymbol{\cdot} \boldsymbol{v}_a - \boldsymbol{k} \boldsymbol{\cdot} \boldsymbol{v}_c} \\
&=
1 - \frac{\omega_{p}}{k_s^2}
S_\rho^m(\boldsymbol{k})^2
\int_{\mathbb{R}^2} \mathbf{d\boldsymbol{v}}_c \,
\frac{1}{2\pi v_{th}^4}
e^{-\frac{v_c^2}{2v_{th}^2}}
\frac{\boldsymbol{k}_s \boldsymbol{\cdot} \boldsymbol{v}_c}{\boldsymbol{k} \boldsymbol{\cdot} \boldsymbol{v}_a - \boldsymbol{k} \boldsymbol{\cdot} \boldsymbol{v}_c} \ .
\end{align}

Using $v_{i,\parallel} = \boldsymbol{k} \boldsymbol{\cdot} \boldsymbol{v}_i /k$ and, noting that $\boldsymbol{k}_s \boldsymbol{\cdot} \boldsymbol{v}_c = (\boldsymbol{k}_s\boldsymbol{\cdot}\boldsymbol{\hat{e}}_{\parallel})v_{c,\parallel} + (\boldsymbol{k}_s\boldsymbol{\cdot}\boldsymbol{\hat{e}}_{\perp})v_{c,\perp} $, we obtain:
\begin{align}
\epsilon(\boldsymbol{k} \boldsymbol{\cdot} \boldsymbol{v}_a, \boldsymbol{k}) 
&= 
1 - \frac{\omega_{p}}{k_s^2}
S_\rho^m(\boldsymbol{k})^2
\frac{1}{2\pi v_{th}^4}
\int_{\mathbb{R}^2} \mathrm{d}v_{c,\parallel}\mathrm{d}v_{c,\perp} \,
e^{-\frac{v_{c,\parallel}^2 + v_{c,\perp}^2}{2v_{th}^2}}
\frac{(\boldsymbol{k}_s\boldsymbol{\cdot}\boldsymbol{\hat{e}}_{\parallel})v_{c,\parallel} + (\boldsymbol{k}_s\boldsymbol{\cdot}\boldsymbol{\hat{e}}_{\perp})v_{c,\perp}}{kv_{a,\parallel} - kv_{c,\parallel}} \\
&=
1 - \frac{\omega_{p}}{k_s^2}
S_\rho^m(\boldsymbol{k})^2
\frac{1}{2\pi v_{th}^4}
\int_{\mathbb{R}^2} \mathrm{d}v_{c,\parallel}\mathrm{d}v_{c,\perp} \,
e^{-\frac{v_{c,\parallel}^2 + v_{c,\perp}^2}{2v_{th}^2}}
\frac{(\boldsymbol{k}_s\boldsymbol{\cdot}\boldsymbol{\hat{e}}_{\parallel})v_{c,\parallel}}{kv_{a,\parallel} - kv_{c,\parallel}} \\
&=
1 - \frac{\omega_{p}}{k_s^2}
S_\rho^m(\boldsymbol{k})^2
\frac{1}{\sqrt{2\pi} v_{th}^3}
\frac{(\boldsymbol{k}_s\boldsymbol{\cdot}\boldsymbol{\hat{e}}_{\parallel})}{k}
\int_{\mathbb{R}} \mathrm{d}v_{c,\parallel} \,
e^{-\frac{v_{c,\parallel}^2}{2v_{th}^2}}
\frac{v_{c,\parallel}}{v_{a,\parallel} - v_{c,\parallel}} \,
\label{eq:app_dieletric_function_derivation_2}
\end{align}
where in the second step we made use of the fact that the term $(\boldsymbol{k}_s\boldsymbol{\cdot}\boldsymbol{\hat{e}}_{\perp})v_{c,\perp}$ gives rise to an odd function in $v_{c,\perp}$, whose integral is therefore zero. We now define the variables $t=v_{c,\parallel} / \sqrt{2}v_{th}$ and $\xi =v_{a,\parallel}/\sqrt{2}v_{th}$ so that the integral can be re-written as:
\begin{align}
\int_{\mathbb{R}} \mathrm{d}v_{c,\parallel} \,
e^{-\frac{v_{c,\parallel}^2}{2v_{th}^2}}
\frac{v_{c,\parallel}}{v_{a,\parallel} - v_{c,\parallel}} \ 
& =
\sqrt{2} v_{th} 
\int_{\mathbb{R}} \mathrm{d}t \, \frac{t e^{-t^2}}{\xi - t} \\
&=
\frac{v_{th}}{\sqrt{2}}
\int_{\mathbb{R}} \mathrm{d}t \, \frac{e^{-t^2}}{(\xi - t)^2} \\
& = 
\frac{\sqrt{\pi} v_{th}}{2} Z'(\xi) \ ,
\end{align}
where we recover the derivative of the plasma dispersion function $Z'(\xi)$~\citep{fried2015plasma}. We can then re-write eq.~\eqref{eq:app_dieletric_function_derivation_2} as:
\begin{align}
\epsilon(\boldsymbol{k} \boldsymbol{\cdot} \boldsymbol{v}_a, \boldsymbol{k})  
&=
1 - \frac{\omega_{p}^2}{k_s^2}
S_\rho^m(\boldsymbol{k})^2
\frac{1}{2 v_{th}^2}
\frac{(\boldsymbol{k}_s\boldsymbol{\cdot}\boldsymbol{\hat{e}}_{\parallel})}{k}
Z'(\xi) \\
&= 1 - \frac{\omega_p^2}{v_{th}^2}\frac{S_\rho^m(\boldsymbol{k})^2 (\boldsymbol{k}_s \boldsymbol{\cdot} \boldsymbol{k})}{2k_s^2k^2}
Z'\left(\frac{\boldsymbol{k} \boldsymbol{\cdot} \boldsymbol{v}_a}{\sqrt{2}kv_{th}}\right) \ ,
\end{align}
and finally normalize all wavenumbers and velocities to obtain:
\begin{equation}
\epsilon(\boldsymbol{\tilde{k}} \boldsymbol{\cdot} \boldsymbol{\tilde{v}}_a, \boldsymbol{\tilde{k}})  
=
1 - \frac{S_\rho^m(\boldsymbol{\tilde{k}})^2 (\boldsymbol{\tilde{k}}_s \boldsymbol{\cdot} \boldsymbol{\tilde{k}})}{2\tilde{k}_s^2\tilde{k}^2}
Z'\left(\frac{\boldsymbol{\tilde{k}} \boldsymbol{\cdot} \boldsymbol{\tilde{v}}_a}{\sqrt{2}\tilde{k}}\right) \ , 
\end{equation}
which matches eq.~\eqref{eq:dieletric_function} in the main body of the paper.

\clearpage
\section{PIC Simulation Parameters}
\label{app:simulation_parameters}

To generate our dataset, we vary four simulation parameters: the number of particles per cell~($N_{ppc}\in [4, 9, 16, 25]$), the current deposition shape function interpolation order~($m\in[1,2,3,4]$), the ratio of the grid-resolution over the electron Debye length~($\Delta_x/\lambda_D\in[1,2]$), and the electron thermal velocity~($v_{th}/c\in[0.01, 0.05, 0.1]$). In total, we perform 96 simulations that correspond to all possible combinations. The number of macroparticles per simulation varies between $10^6-10^7$ depending on the numerical parameters.

Each simulation was run until a maximum time $t_{max}$ given by:
\begin{equation}
t_{max}\omega_p = 50 \frac{N_{ppc}}{4} \left(1 + \left(\frac{\lambda_D}{\Delta_x}\right)^2 \right) \left(\frac{5}{6} + \frac{m}{6}\right),   
\label{eq:tmax}
\end{equation}
which corresponds to $t_{max}\omega_p=100$ for $N_{ppc}=4$, $\Delta_x/\lambda_D=1$, and $m=1$. The time-step was chosen as $c\Delta t = 0.98\Delta_x/\sqrt{2}$ to ensure that the CFL stability condition was fulfilled~\citep{courant1928partiellen}. The varying maximum time ensures that we will observe similar phase space dynamics across the full duration of the simulation for all numerical parameters, which facilitates test comparisons across the full dataset. The final formula was motivated by the results of~\citet{hockney1971measurements}, which empirically demonstrated that the thermalization time for 2-D simulations was proportional to $\propto N_{ppc} (1 + (\lambda_D/\Delta_x^2))$ while the additional shape function factor term was an estimated ratio from our initial tests based on the measured advection/diffusion coefficients for a reduced set of simulations.

For each simulation, we output the raw data of all macroparticles at equally spaced time intervals to collect a minimum of 100 snapshots over the full simulation\footnote{For some simulations, a slightly larger value might be saved due to the ratio of the total simulation time over the time step.} while also keeping track of the total energy of the system. For downstream tasks, we will disregard any data collected up to $t\omega_p = 5$ to avoid introducing artifacts due to the non-physical initialization of the fields (which are initialized to zero).

Finally, since in this work we are interested in studying advection and diffusion coefficients that do not change over time, we decided to disregard simulations for which numerical heating was an issue~(c.f. \citet{birdsall2018plasma} for more details on the possible causes of numerical heating in PIC). For this purpose, we filtered out all simulations for which we measured a maximum energy variation larger than $1\%$ of the initial energy. This reduced the final number of simulations used to 76 (mostly the simulations that correspond to a linear shape function, $m=1$, were removed). The detailed list of simulations kept and disregarded, as well as their corresponding numerical parameters and measured energy variations, are provided in table~\ref{tab:simulation_parameters}.

\begin{longtable}[c]{ccccccc}
\hline
\textbf{Index} & \textbf{$\Delta_x / \lambda_D$} & \textbf{$N_{ppc}$} & \textbf{$m$} & \textbf{$v_{th}/c$} & \textbf{$t_{max}$} & \textbf{$\Delta E (\%)$} \\ \hline
\endfirsthead
\multicolumn{7}{c}%
{} \\
%{{\bfseries table \thetable\ continued from previous page}} \\
\endhead
\rowcolor[HTML]{D6F5D5} 
0 & 1 & 4 & 1 & 0.01 & 100 & 5E-01 \\
\rowcolor[HTML]{FFCCC9} 
1 & 1 & 4 & 1 & 0.05 & 100 & 1E+00 \\
\rowcolor[HTML]{FFCCC9} 
2 & 1 & 4 & 1 & 0.10 & 100 & 3E+00 \\
\rowcolor[HTML]{D6F5D5} 
3 & 1 & 4 & 2 & 0.01 & 117 & 1E-01 \\
\rowcolor[HTML]{D6F5D5} 
4 & 1 & 4 & 2 & 0.05 & 117 & 1E-01 \\
\rowcolor[HTML]{D6F5D5} 
5 & 1 & 4 & 2 & 0.10 & 117 & 2E-01 \\
\rowcolor[HTML]{D6F5D5} 
6 & 1 & 4 & 3 & 0.01 & 133 & 6E-02 \\
\rowcolor[HTML]{D6F5D5} 
7 & 1 & 4 & 3 & 0.05 & 133 & 7E-02 \\
\rowcolor[HTML]{D6F5D5} 
8 & 1 & 4 & 3 & 0.10 & 133 & 7E-02 \\
\rowcolor[HTML]{D6F5D5} 
9 & 1 & 4 & 4 & 0.01 & 150 & 5E-02 \\
\rowcolor[HTML]{D6F5D5} 
10 & 1 & 4 & 4 & 0.05 & 150 & 5E-02 \\
\rowcolor[HTML]{D6F5D5} 
11 & 1 & 4 & 4 & 0.10 & 150 & 5E-02 \\
\rowcolor[HTML]{D6F5D5} 
12 & 1 & 9 & 1 & 0.01 & 225 & 4E-01 \\
\rowcolor[HTML]{FFCCC9} 
13 & 1 & 9 & 1 & 0.05 & 225 & 1E+00 \\
\rowcolor[HTML]{FFCCC9} 
14 & 1 & 9 & 1 & 0.10 & 225 & 3E+00 \\
\rowcolor[HTML]{D6F5D5} 
15 & 1 & 9 & 2 & 0.01 & 262 & 7E-02 \\
\rowcolor[HTML]{D6F5D5} 
16 & 1 & 9 & 2 & 0.05 & 262 & 7E-02 \\
\rowcolor[HTML]{D6F5D5} 
17 & 1 & 9 & 2 & 0.10 & 262 & 1E-01 \\
\rowcolor[HTML]{D6F5D5} 
18 & 1 & 9 & 3 & 0.01 & 300 & 4E-02 \\
\rowcolor[HTML]{D6F5D5} 
19 & 1 & 9 & 3 & 0.05 & 300 & 4E-02 \\
\rowcolor[HTML]{D6F5D5} 
20 & 1 & 9 & 3 & 0.10 & 300 & 4E-02 \\
\rowcolor[HTML]{D6F5D5} 
21 & 1 & 9 & 4 & 0.01 & 338 & 3E-02 \\
\rowcolor[HTML]{D6F5D5} 
22 & 1 & 9 & 4 & 0.05 & 338 & 3E-02 \\
\rowcolor[HTML]{D6F5D5} 
23 & 1 & 9 & 4 & 0.10 & 338 & 3E-02 \\
\rowcolor[HTML]{D6F5D5} 
24 & 1 & 16 & 1 & 0.01 & 400 & 4E-01 \\
\rowcolor[HTML]{FFCCC9} 
25 & 1 & 16 & 1 & 0.05 & 400 & 1E+00 \\
\rowcolor[HTML]{FFCCC9} 
26 & 1 & 16 & 1 & 0.10 & 400 & 3E+00 \\
\rowcolor[HTML]{D6F5D5} 
27 & 1 & 16 & 2 & 0.01 & 467 & 5E-02 \\
\rowcolor[HTML]{D6F5D5} 
28 & 1 & 16 & 2 & 0.05 & 467 & 6E-02 \\
\rowcolor[HTML]{D6F5D5} 
29 & 1 & 16 & 2 & 0.10 & 467 & 1E-01 \\
\rowcolor[HTML]{D6F5D5} 
30 & 1 & 16 & 3 & 0.01 & 533 & 3E-02 \\
\rowcolor[HTML]{D6F5D5} 
31 & 1 & 16 & 3 & 0.05 & 533 & 3E-02 \\
\rowcolor[HTML]{D6F5D5} 
32 & 1 & 16 & 3 & 0.10 & 533 & 3E-02 \\
\rowcolor[HTML]{D6F5D5} 
33 & 1 & 16 & 4 & 0.01 & 600 & 2E-02 \\
\rowcolor[HTML]{D6F5D5} 
34 & 1 & 16 & 4 & 0.05 & 600 & 2E-02 \\
\rowcolor[HTML]{D6F5D5} 
35 & 1 & 16 & 4 & 0.10 & 600 & 2E-02 \\
\rowcolor[HTML]{D6F5D5} 
36 & 1 & 25 & 1 & 0.01 & 625 & 3E-01 \\
\rowcolor[HTML]{FFCCC9} 
37 & 1 & 25 & 1 & 0.05 & 625 & 1E+00 \\
\rowcolor[HTML]{FFCCC9} 
38 & 1 & 25 & 1 & 0.10 & 625 & 3E+00 \\
\rowcolor[HTML]{D6F5D5} 
39 & 1 & 25 & 2 & 0.01 & 729 & 5E-02 \\
\rowcolor[HTML]{D6F5D5} 
40 & 1 & 25 & 2 & 0.05 & 729 & 5E-02 \\
\rowcolor[HTML]{D6F5D5} 
41 & 1 & 25 & 2 & 0.10 & 729 & 9E-02 \\
\rowcolor[HTML]{D6F5D5} 
42 & 1 & 25 & 3 & 0.01 & 833 & 2E-02 \\
\rowcolor[HTML]{D6F5D5} 
43 & 1 & 25 & 3 & 0.05 & 833 & 2E-02 \\
\rowcolor[HTML]{D6F5D5} 
44 & 1 & 25 & 3 & 0.10 & 833 & 2E-02 \\
\rowcolor[HTML]{D6F5D5} 
45 & 1 & 25 & 4 & 0.01 & 938 & 1E-02 \\
\rowcolor[HTML]{D6F5D5} 
46 & 1 & 25 & 4 & 0.05 & 938 & 1E-02 \\
\rowcolor[HTML]{D6F5D5} 
47 & 1 & 25 & 4 & 0.10 & 938 & 1E-02 \\
\rowcolor[HTML]{FFCCC9} 
48 & 2 & 4 & 1 & 0.01 & 62 & 2E+00 \\
\rowcolor[HTML]{FFCCC9} 
49 & 2 & 4 & 1 & 0.05 & 62 & 3E+00 \\
\rowcolor[HTML]{FFCCC9} 
50 & 2 & 4 & 1 & 0.10 & 62 & 5E+00 \\
\rowcolor[HTML]{D6F5D5} 
51 & 2 & 4 & 2 & 0.01 & 73 & 4E-01 \\
\rowcolor[HTML]{D6F5D5} 
52 & 2 & 4 & 2 & 0.05 & 73 & 4E-01 \\
\rowcolor[HTML]{D6F5D5} 
53 & 2 & 4 & 2 & 0.10 & 73 & 5E-01 \\
\rowcolor[HTML]{D6F5D5} 
54 & 2 & 4 & 3 & 0.01 & 83 & 2E-01 \\
\rowcolor[HTML]{D6F5D5} 
55 & 2 & 4 & 3 & 0.05 & 83 & 2E-01 \\
\rowcolor[HTML]{D6F5D5} 
56 & 2 & 4 & 3 & 0.10 & 83 & 2E-01 \\
\rowcolor[HTML]{D6F5D5} 
57 & 2 & 4 & 4 & 0.01 & 94 & 1E-01 \\
\rowcolor[HTML]{D6F5D5} 
58 & 2 & 4 & 4 & 0.05 & 94 & 1E-01 \\
\rowcolor[HTML]{D6F5D5} 
59 & 2 & 4 & 4 & 0.10 & 94 & 1E-01 \\
\rowcolor[HTML]{FFCCC9} 
60 & 2 & 9 & 1 & 0.01 & 141 & 2E+00 \\
\rowcolor[HTML]{FFCCC9} 
61 & 2 & 9 & 1 & 0.05 & 141 & 2E+00 \\
\rowcolor[HTML]{FFCCC9} 
62 & 2 & 9 & 1 & 0.10 & 141 & 5E+00 \\
\rowcolor[HTML]{D6F5D5} 
63 & 2 & 9 & 2 & 0.01 & 164 & 2E-01 \\
\rowcolor[HTML]{D6F5D5} 
64 & 2 & 9 & 2 & 0.05 & 164 & 3E-01 \\
\rowcolor[HTML]{D6F5D5} 
65 & 2 & 9 & 2 & 0.10 & 164 & 3E-01 \\
\rowcolor[HTML]{D6F5D5} 
66 & 2 & 9 & 3 & 0.01 & 188 & 1E-01 \\
\rowcolor[HTML]{D6F5D5} 
67 & 2 & 9 & 3 & 0.05 & 188 & 1E-01 \\
\rowcolor[HTML]{D6F5D5} 
68 & 2 & 9 & 3 & 0.10 & 188 & 1E-01 \\
\rowcolor[HTML]{D6F5D5} 
69 & 2 & 9 & 4 & 0.01 & 211 & 7E-02 \\
\rowcolor[HTML]{D6F5D5} 
70 & 2 & 9 & 4 & 0.05 & 211 & 7E-02 \\
\rowcolor[HTML]{D6F5D5} 
71 & 2 & 9 & 4 & 0.10 & 211 & 8E-02 \\
\rowcolor[HTML]{FFCCC9} 
72 & 2 & 16 & 1 & 0.01 & 250 & 1E+00 \\
\rowcolor[HTML]{FFCCC9} 
73 & 2 & 16 & 1 & 0.05 & 250 & 2E+00 \\
\rowcolor[HTML]{FFCCC9} 
74 & 2 & 16 & 1 & 0.10 & 250 & 5E+00 \\
\rowcolor[HTML]{D6F5D5} 
75 & 2 & 16 & 2 & 0.01 & 292 & 2E-01 \\
\rowcolor[HTML]{D6F5D5} 
76 & 2 & 16 & 2 & 0.05 & 292 & 2E-01 \\
\rowcolor[HTML]{D6F5D5} 
77 & 2 & 16 & 2 & 0.10 & 292 & 3E-01 \\
\rowcolor[HTML]{D6F5D5} 
78 & 2 & 16 & 3 & 0.01 & 333 & 8E-02 \\
\rowcolor[HTML]{D6F5D5} 
79 & 2 & 16 & 3 & 0.05 & 333 & 8E-02 \\
\rowcolor[HTML]{D6F5D5} 
80 & 2 & 16 & 3 & 0.10 & 333 & 9E-02 \\
\rowcolor[HTML]{D6F5D5} 
81 & 2 & 16 & 4 & 0.01 & 375 & 5E-02 \\
\rowcolor[HTML]{D6F5D5} 
82 & 2 & 16 & 4 & 0.05 & 375 & 5E-02 \\
\rowcolor[HTML]{D6F5D5} 
83 & 2 & 16 & 4 & 0.10 & 375 & 5E-02 \\
\rowcolor[HTML]{FFCCC9} 
84 & 2 & 25 & 1 & 0.01 & 391 & 1E+00 \\
\rowcolor[HTML]{FFCCC9} 
85 & 2 & 25 & 1 & 0.05 & 391 & 2E+00 \\
\rowcolor[HTML]{FFCCC9} 
86 & 2 & 25 & 1 & 0.10 & 391 & 5E+00 \\
\rowcolor[HTML]{D6F5D5} 
87 & 2 & 25 & 2 & 0.01 & 456 & 2E-01 \\
\rowcolor[HTML]{D6F5D5} 
88 & 2 & 25 & 2 & 0.05 & 456 & 2E-01 \\
\rowcolor[HTML]{D6F5D5} 
89 & 2 & 25 & 2 & 0.10 & 456 & 2E-01 \\
\rowcolor[HTML]{D6F5D5} 
90 & 2 & 25 & 3 & 0.01 & 521 & 7E-02 \\
\rowcolor[HTML]{D6F5D5} 
91 & 2 & 25 & 3 & 0.05 & 521 & 7E-02 \\
\rowcolor[HTML]{D6F5D5} 
92 & 2 & 25 & 3 & 0.10 & 521 & 7E-02 \\
\rowcolor[HTML]{D6F5D5} 
93 & 2 & 25 & 4 & 0.01 & 586 & 4E-02 \\
\rowcolor[HTML]{D6F5D5} 
94 & 2 & 25 & 4 & 0.05 & 586 & 4E-02 \\
\rowcolor[HTML]{D6F5D5} 
95 & 2 & 25 & 4 & 0.10 & 586 & 4E-02 \\ \hline \\
\caption{PIC simulation parameters for the generated dataset. It includes: $\Delta_x/\lambda_D$ - ratio of the grid resolution over the electron Debye length; $N_{ppc}$ - number of particles per cell; $m$ - current deposition shape function order; $v_{th}/c$ - electron thermal velocity normalized to the speed of light; $t_{max}$ - maximum simulation time computed according to \eqref{eq:tmax}; $\Delta E$ - percentage of energy variation over the full simulation. Green rows represent simulation with energy variation below the chosen threshold value of $1\%$.  Red rows represent the ones discarded due to a larger energy variation.}
\label{tab:simulation_parameters}\\
\end{longtable}

\clearpage
\section{Subpopulation Sampling}
\label{app:subpopulation_sampling}

For each simulation, we sample 28 subpopulations of macroparticles (a macroparticle might be used in multiple subpopulations). These subpopulations are sampled at $t=t_0$ (where $t_0$ is the first time step that is not discarded) based on a requested probability density function $p_\theta(\boldsymbol{\tilde{v}})$:
\begin{equation}
    P(\boldsymbol{v}_i | p_{\theta})  = \frac{P(\boldsymbol{v}_i | p_{\theta}, \{\boldsymbol{v}_j\})}{P(\boldsymbol{v}_i | \{\boldsymbol{v}_j\})}
\end{equation}
where $\theta$ are the distribution parameters associated with $p_\theta$, $\{\boldsymbol{v}_j\}$ the set of velocities of all particle at $t_0$. The denominator $P(\boldsymbol{v}_i | \{\boldsymbol{v}_j\})$ is computed by discretizing the phase space $\mathcal{V}_{max}:[-v_x^{max},v_x^{max}] \times [-v_y^{max}, v_y^{max}]$ in a $100\times100$ grid ($v_i^{max}$ is the maximum velocity over all particles along the $i$-axis) and computing the ratio of the bin count over the total number of particles.

The chosen distributions $p_{\theta}$ are (considering velocities $\boldsymbol{\tilde{v}} = \boldsymbol{v}/v_{th}$):
\begin{itemize}
    \item \textbf{Normals-0 ($\times 1$)}: Normal distribution centered at $\boldsymbol{\mu} = (0,0)$ with covariance $\Sigma_{ii} = 0.05$ and $\Sigma_{ij} = 0$
    \item \textbf{Normals-1 ($\times 8$)}: Normal distributions centered at $\boldsymbol{\mu} = (\{0,\pm 1\},\{0,\pm 1\})$ except $(0,0)$ with covariance $\Sigma_{ii} = 0.05$ and $\Sigma_{ij} = 0$
    \item \textbf{Normals-2 ($\times 8$)}: Normal distributions centered at $\boldsymbol{\mu} = (\{0,\pm 2\},\{0,\pm 2\})$ except $(0,0)$ with covariance $\Sigma_{ii} = 0.05$ and $\Sigma_{ij} = 0$
    \item \textbf{Normals-Rot ($\times 4$)}: Normal distributions centered at $\boldsymbol{\mu} = (0,0)$ with covariance matrices with entries $\Sigma_{xx}= 1$, $\Sigma_{yy}=0.05$, and $\Sigma_{xy} = 0$ rotated by $\theta \in [0^{\circ}, 22.5^{\circ}, 45^{\circ}, 67.5^{\circ}]$
    \item \textbf{Rings ($\times 3$)}: Ring centered at radius $\mathrm{v} \in [1, 2, 3]$ with standard deviation $\sigma_\mathrm{v} = 0.2$
    \item \textbf{Quadrants ($\times 4$)}: Waterbag distribution covering each of the 4 quadrants of the phase space $\mathcal{V}: [-5,5] \times [-5,5]$
\end{itemize}

\begin{figure}
    \centering
    \includegraphics[width=0.7\linewidth]{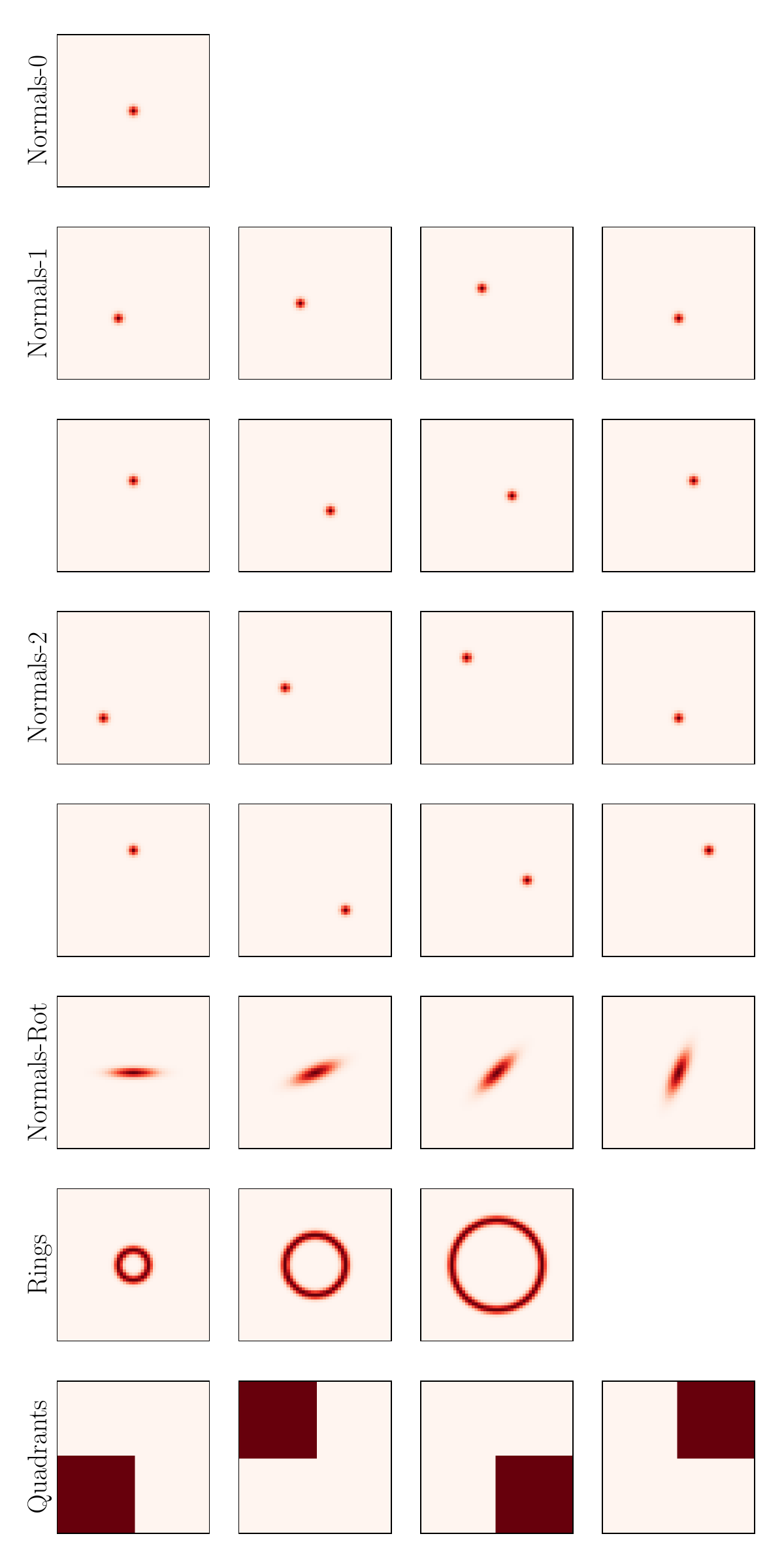}
    \caption{Probability distribution functions $p_\theta$ used to sample subpopulations of macroparticles. All plots show the density over the phase space $\mathcal{V}: [-5,5]v_{th} \times [-5, 5]v_{th}$ discretized over $51\times51$ bins.}
    \label{fig:pdf_examples}
\end{figure}

\begin{figure}
    \centering
    \includegraphics[width=0.7\linewidth]{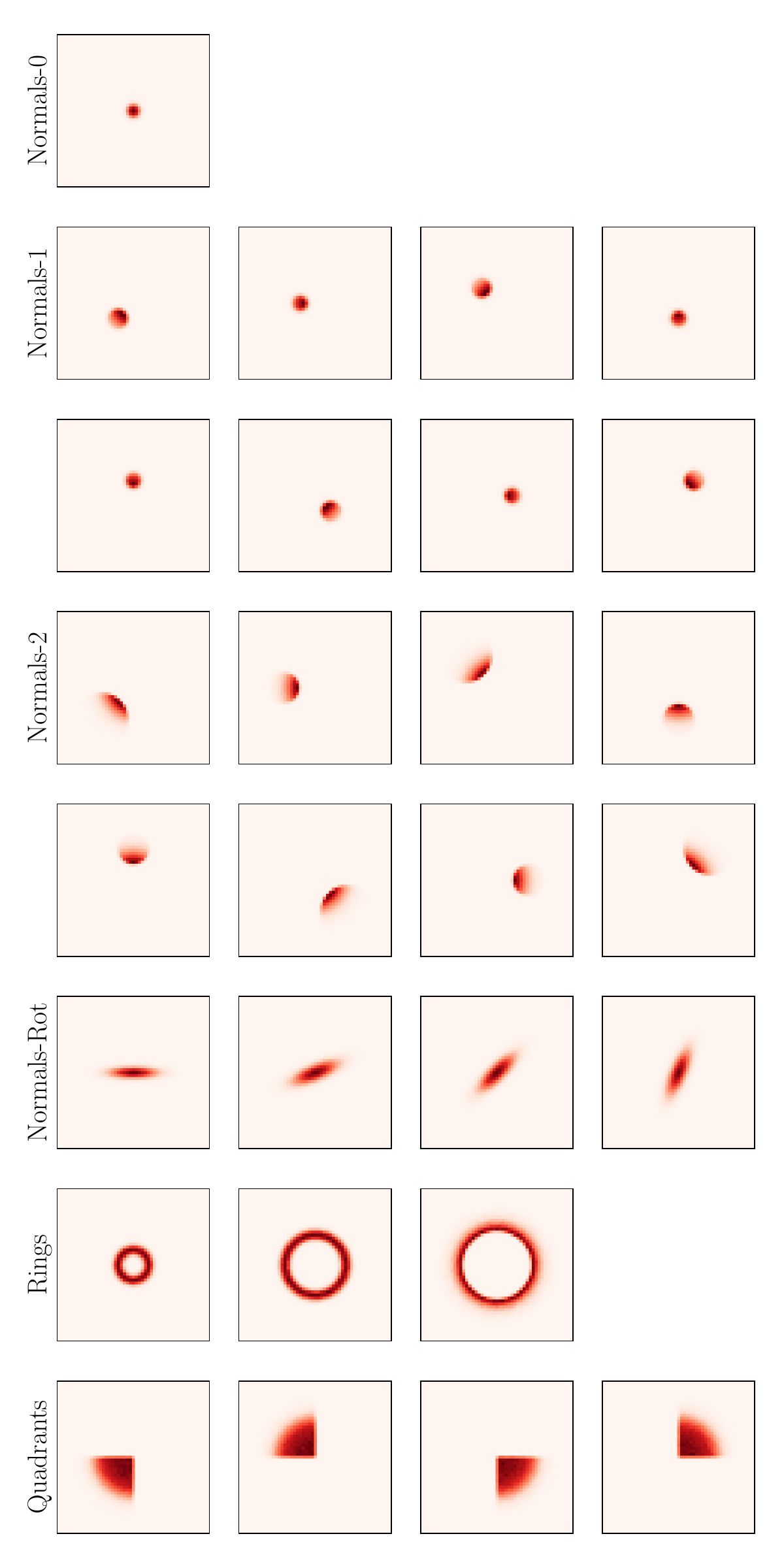}
    \caption{Sampled subpopulations at $t=t_0$ for simulation $index=0$ (simulation parameters in table~\ref{tab:simulation_parameters}). All plots show the density over the phase space $\mathcal{V}: [-5,5]v_{th} \times [-5, 5]v_{th}$ discretized over $51\times51$ bins. The total number of macroparticles in this simulation is $N=4 \times 10^6$. This is approximately the average number of macroparticles in the full dataset.}
    \label{fig:pdf_examples_index_0}
\end{figure}

\begin{figure}
    \centering
    \includegraphics[width=0.7\linewidth]{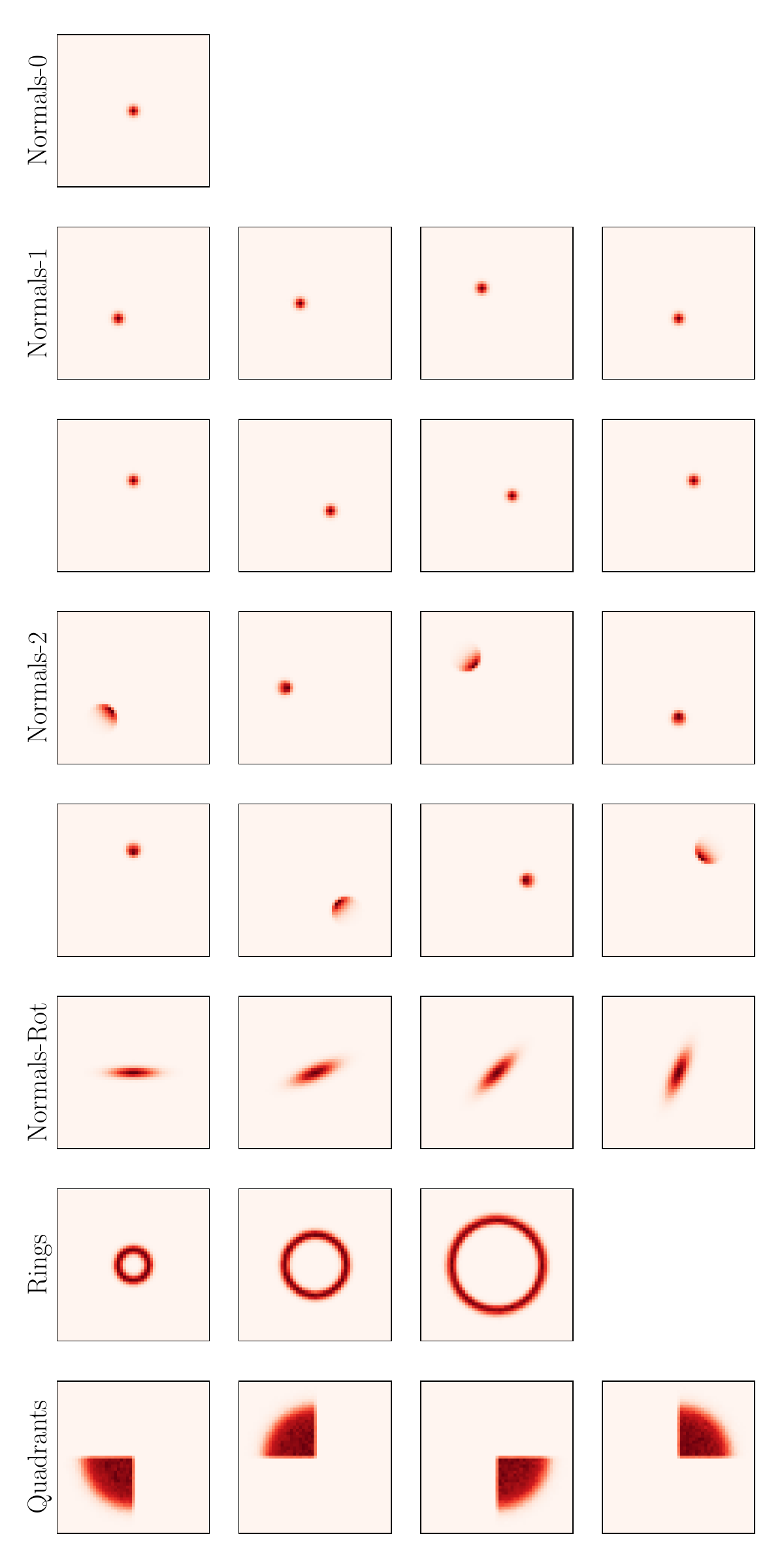}
    \caption{Sampled subpopulations at $t=t_0$ for simulation $index=47$ (simulation parameters in table~\ref{tab:simulation_parameters}). All plots show the density over the phase space $\mathcal{V}: [-5,5]v_{th} \times [-5, 5]v_{th}$ discretized over $51\times51$ bins. The total number of macroparticles in this simulation is $N=2.5 \times 10^7$. This is the maximum number of macroparticles in the full dataset.}
    \label{fig:pdf_examples_index_47}
\end{figure}

\begin{figure}
    \centering
    \includegraphics[width=0.7\linewidth]{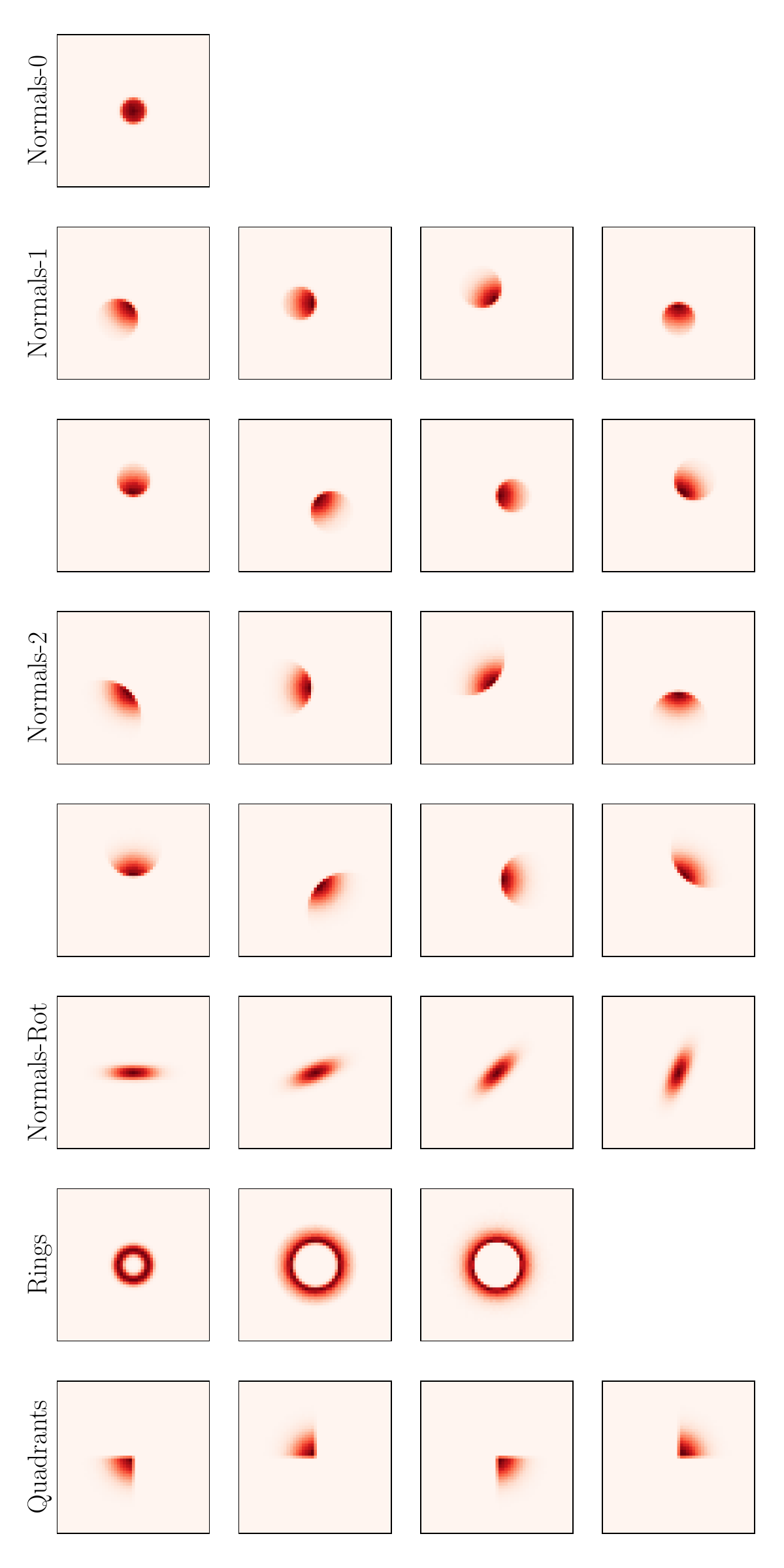}
    \caption{Sampled subpopulations at $t=t_0$ for simulation $index=51$ (simulation parameters in table~\ref{tab:simulation_parameters}). All plots show the density over the phase space $\mathcal{V}: [-5,5]v_{th} \times [-5, 5]v_{th}$ discretized over $51\times51$ bins. The total number of macroparticles in this simulation is $N=10^6$. This is the minimum number of macroparticles in the full dataset.}
    \label{fig:pdf_examples_index_51}
\end{figure}

We sample $3\times10^5$ macroparticles per distribution, except for the \textit{Quadrants} case where only $2.5\times10^5$ macroparticles are sampled to account for the simulations with the least amount of macroparticles ($N_{ppc}=2$ and $\Delta_x/\lambda_D=2$).
For completion, we plot the chosen distributions $p_\theta$ in Figure~\ref{fig:pdf_examples} and examples of the corresponding sampled subpopulations for simulations with different numbers of macroparticles in Figures~\ref{fig:pdf_examples_index_0} to \ref{fig:pdf_examples_index_51}. The latter examples are used to highlight the distinct initial subpopulations depending on the average ($index=0$, Figure~\ref{fig:pdf_examples_index_0}), maximum ($index=46$, Figure~\ref{fig:pdf_examples_index_47}), and minimum ($index=51$, Figure~\ref{fig:pdf_examples_index_51}) number of macroparticles across all simulations in our dataset.

After sampling the subpopulations at $t=t_0$ (first stored time-step with $t\omega_p > 5$ as explained in Supplemental Material~\ref{app:simulation_parameters}), we compute their phase space evolution based on the particle tags present in the stored particle tracks. Once this process is finished, we delete all tracks due to their significant memory costs.

Throughout the paper, we divide the 28 subpopulations into two sets:

\begin{itemize}
    \item \textbf{Train} ($\times 9$): Normals-0 + Normals-2
    \item \textbf{Test} ($\times 19$): All others
\end{itemize}

The training subpopulations were chosen such that they cover a good portion of the phase space during their full dynamics. We do not include subpopulations with larger density at $v \gtrapprox 4 v_{th}$ (e.g., Rings and Quadrants) into the train set to assess generalization issues at higher $v$. Future works not interested in assessing generalization should not do this split, since it is detrimental to the extraction of the operator at high $v$. Note that the train/test split is only relevant for operators learned using the differentiable simulator approach. Advection and diffusion models learned from particle tracks have access to statistics over all regions of the phase space where macroparticles were present.

\clearpage
\section{Advection-Diffusion Coefficients from Particle Tracks}
\label{app:ad_from_tracks}

In this Supplemental Material, we elaborate further on the technicalities and difficulties of estimating advection and diffusion coefficients from particle tracks, in particular, the impact of the time interval over which velocity changes are computed.

Equations \eqref{eq:a_tracks} and \eqref{eq:d_tracks} only provide accurate estimates of the advection and diffusion coefficients if the evolution of the quantities $<\Delta v_i>$ and $<\Delta v_i \Delta v_j>$ is linear in time within that interval. Therefore, computing the average velocity changes over longer periods will result in either an underestimation or an overestimation of the value (depending on whether the behaviour is sub-linear or super-linear). Furthermore, the time scales over which the linear approximation holds depend on the macroparticle velocity. 

Intuitively, one would then choose the shortest possible time interval to measure the coefficients. However, if the time scale is too small, other effects (e.g., plasma oscillations) can lead to a wrong estimation of the coefficients since they might not average to zero even if statistics over multiple (independent) time-intervals are gathered (e.g., $<\Delta v^2>$ would never average out if there is an underlying time-resolved oscillatory motion). Similarly, the coefficients might be wrongly estimated if the time interval is too short with respect to the collisional time scales (which we might not know \textit{a priori}), since the statistics are not independent across consecutive time intervals. Therefore, manually tuning and verifying that we used an acceptable time interval across all simulations and phase space bins was deemed counter-productive given our dataset size (and we expect future similar studies to face equivalent issues). 

To tackle this problem, we then devised the automatised process briefly mentioned in Section~\ref{sec:ad_tracks}, which we make more explicit here. First, we split the full simulation into contiguous equally sized time-intervals of size $\Delta t = N_{d}\Delta t_{dump}$ where $\Delta t_{dump}$ is the raw data dumping period (values per simulation are provided in table~\ref{tab:simulation_parameters}) and $N_d \in [1, 5, 10, 20]$ is the number of dumped time-steps used for each time-interval. 

For each value of $N_d$, we then compute $\mathsfi{A}_i(\boldsymbol{v}_{bin})$ and $\mathsfi{D}_{ij}(\boldsymbol{v}_{bin})$ as per~\eqref{eq:a_tracks} and \eqref{eq:d_tracks}. That is, for each value of $N_d$, we perform a weighted average across all time intervals based on the macroparticles inside the phase space region at the beginning of each time interval. With this strategy, we end up computing four different estimates of the advection and diffusion coefficients per simulation from particle tracks (one for each value of $N_d$). We can then evaluate their accuracy in reproducing phase space dynamics of subpopulations and pick, for each individual simulation, the best performing coefficients to compare against the differentiable simulator based approaches. 

\begin{figure}
    \centering
    \includegraphics[width=0.4\linewidth]{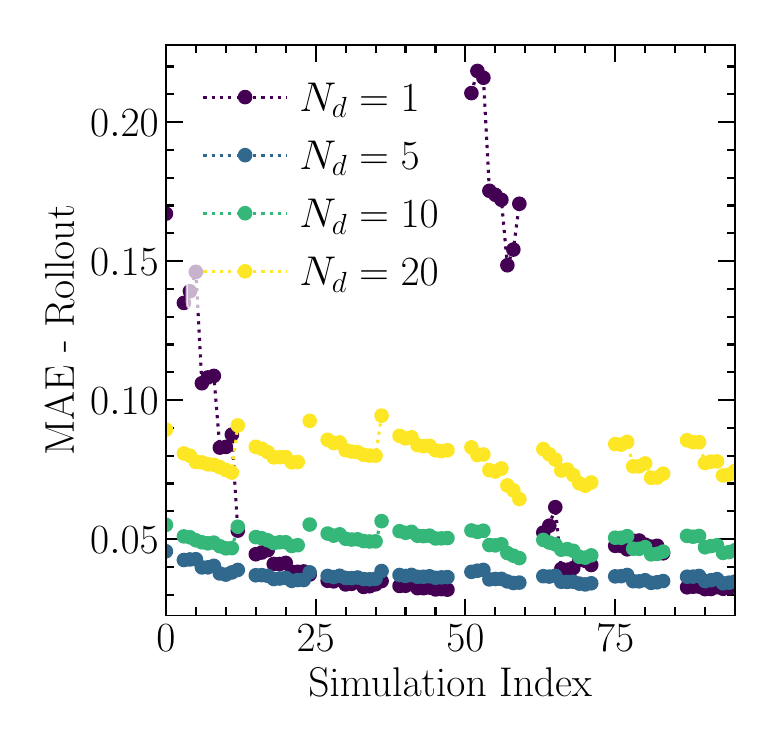}
    \caption{Rollout error averaged over training and test subpopulations per simulation. In general, $N_d=5$ provides better estimates for simulations with lower macroparticle count (index $\in [0,25] \cup [50,75]$), while $N_t=1$ performs slightly better for simulations with a larger number of macroparticles (index $\in [25,50] \cup [75,95]$). The best performing value of $N_d$ per simulation is used for the comparisons against the differentiable simulator results presented in the main body of the paper.}
    \label{fig:stats_l1_nt_simulation}
\end{figure}

On average, we find that the value $N_d=5$ provides the best estimates for simulations with $\Delta t_{dump}\omega_p\lesssim  3-4$ while $N_t=1$ performs slightly better for larger values (see Figure \ref{fig:stats_l1_nt_simulation}) . We believe this difference in behaviour is mainly caused by the simulations with $\Delta t_{dump}\omega_p\lesssim  3-4$ resolving the plasma oscillations.

\begin{figure}
    \centering
    \includegraphics[width=\textwidth]{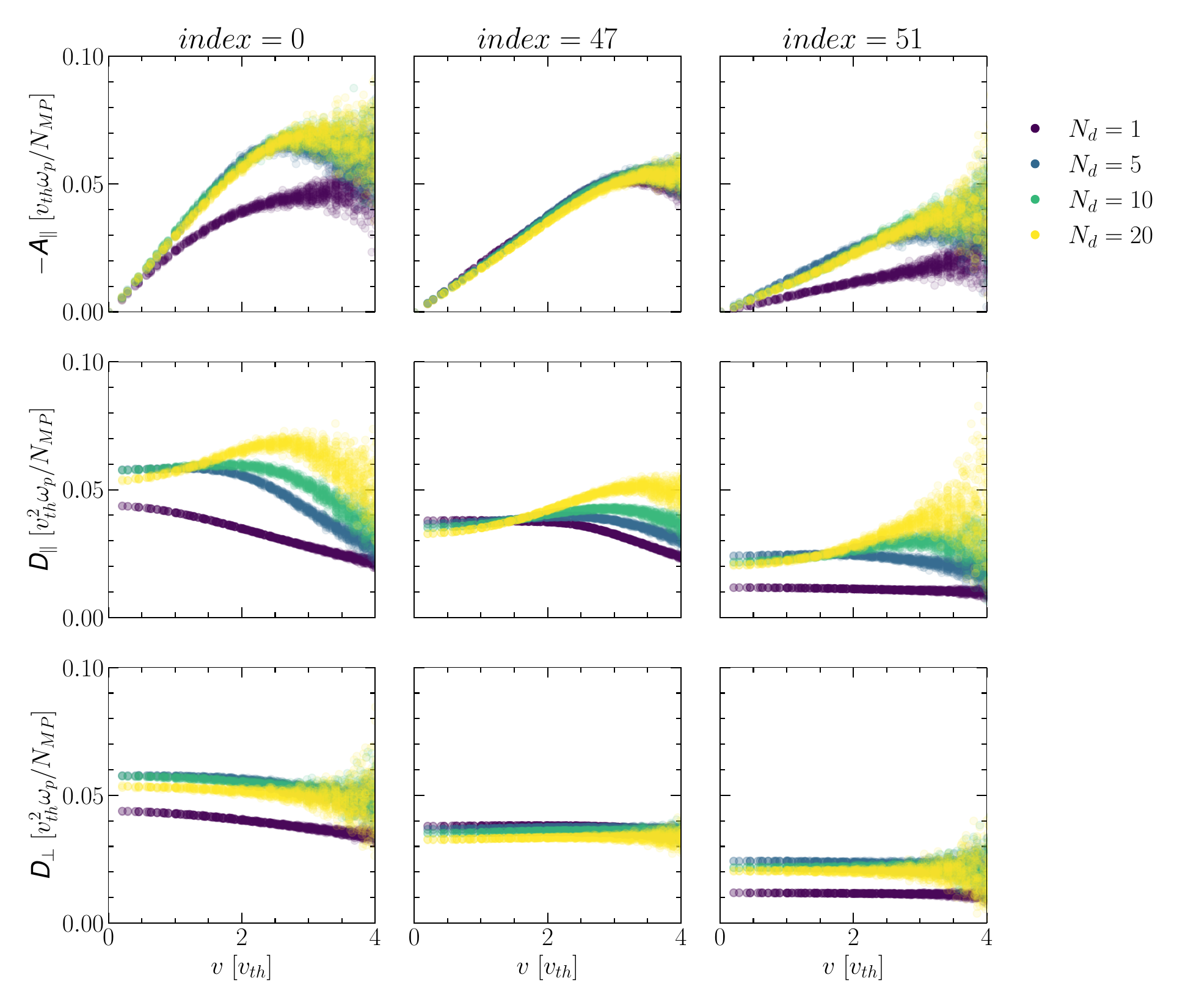}
    \caption{Parallel advection and parallel/perpendicular diffusion measured from particle tracks for different simulations in function of $N_d$. Parallel/perpendicular components were obtained by projecting the measured advection and diffusion along ($v_x$, $v_y$) onto the corresponding macroparticle velocity frame.  The best performing coefficients correspond to $N_d=5$ for $index=0$, $N_d=5$ for $index=47$, and $N_d=1$ for $index=51$.}
    \label{fig:stats_AD_nt_index}
\end{figure}

Examples of the predicted advection and diffusion values as a function of $N_d$ are presented in Figure~\ref{fig:stats_AD_nt_index}. It is clear that retrieved coefficients vary significantly depending on the value of $N_d$ used. These examples are used to highlight that if $N_t$ is smaller than the optimal value, we seem to be underestimating both advection and diffusion. On the other hand, if $N_d$ is too large, advection estimates do not change significantly while diffusion clearly varies.

\begin{figure}
    \centering
    \includegraphics[width=\textwidth]{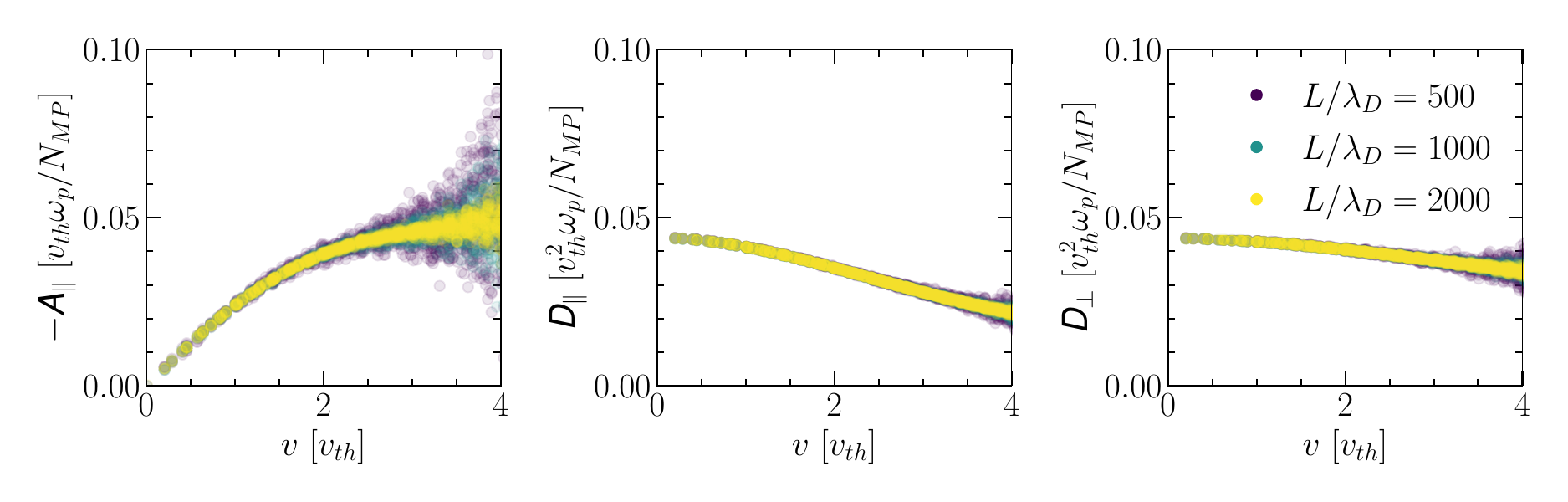}
    \caption{Advection and diffusion coefficients measured from particle tracks for the same simulation parameters as $index=0$ with varying box lengths $L/\lambda_D$ and $N_d=1$. Changing the amount of collected statistics does not change the average values of the coefficients, it only reduces the spread at higher $v$.}
    \label{fig:stats_AD_nt_L_index}
\end{figure}

We want to highlight that performing simulations with more macroparticles to increase the statistics (e.g., using a larger box size while keeping all other parameters fixed or simulating for a longer time) does not change the overall trends observed. A larger number of statistics only reduces the noise levels at higher $v$ and not the estimated average values of the coefficients (example in Figure~\ref{fig:stats_AD_nt_L_index}).

\clearpage
\section{Non-Uniqueness of A/D solution}
\label{ap:non_uniqueness_ad}

For simplicity, let us consider the 1-D version of the Fokker-Planck Equation in~\eqref{eq:fp_equation}:
\begin{equation}
\frac{\partial f(v, t)}{\partial t} = -\frac{\partial}{\partial v} \left( A(v, t) f(v, t) \right) + \frac{1}{2}\frac{\partial^2}{\partial v^2} \left( D(v, t) f(v, t) \right).
\end{equation}

We now define two sets of advection/diffusion operators, respectively $(A_1, D_1)$ and $({A}_2, D_2)$ such that:
\begin{align}
    A_2(v,t) &= A_1(v, t) + \alpha(v, t) \label{eq:a_2}\\ 
    D_2(v,t) &= D_1(v, t) + \beta(v,t) \label{eq:d_2} \ ,
\end{align}
where $\alpha(v,t), \beta(v,t)$ are unknown perturbations.

For the two sets of operators to produce the same phase space dynamics, the following equality must hold true:
\begin{equation}
    -\frac{\partial}{\partial v} \left( A_1 f \right) + \frac{1}{2}\frac{\partial^2}{\partial v^2} \left( D_1 f \right) = -\frac{\partial}{\partial v} \left( A_2 f \right) + \frac{1}{2}\frac{\partial^2}{\partial v^2} \left( D_2 f \right), \quad \forall f(v,t).
\end{equation}

If we substitute equations~\eqref{eq:a_2} and \eqref{eq:d_2} we obtain the condition:
\begin{equation}
    -\frac{\partial}{\partial v}\left(\alpha f\right) + \frac{1}{2}\frac{\partial^2}{\partial^2v}\left(\beta f\right) = 0, \quad \forall f(v,t) \ ,
\label{eq:ad_family_solutions}
\end{equation}
which defines the family of infinite solutions for the advection/diffusion operators that capture the same dynamics.

The practical consequence is that, depending on the initial guess for the coefficient values, an optimization algorithm might find very different solutions. This is exactly the cause of the results reported in~\citet{camporeale2022data}, where PINNs initialized with different random seeds, obtained visibly different advection/diffusion coefficients that equally described the phase space dynamics.

To mitigate the ill-posedness of the problem, we must introduce further constraints. This can be done by decreasing the number of degrees of freedom via the enforcement of known symmetries (e.g. assume no dependence on $t$ or introduce spatial symmetries). Similarly, if we assume that the advection/diffusion coefficients do not depend on the particular initial distribution function $f$ (which is a valid assumption if $f$ corresponds to a subpopulation of the global distribution), we can provide the evolution of multiple distribution functions $f_i(v,t)$ since, for each of new distribution introduced,  an extra eq.~\eqref{eq:ad_family_solutions} must be verified. In the particular scenario where the coefficients do not change over time (which should be the case  for a thermal plasma), each individual time-step of a subpopulation $f_i(v,t)$ is equivalent to considering a new subpopulation $f_{i,t}(v)$.

Further difficulties can occur if the distributions functions $f_i$ are not smooth (in our case due to the limited number of particles tracked) since this causes noisy estimates of derivatives and leads to non-disappearing residuals of the FP equation. A possible way to mitigate this issue is to optimize for the long term prediction of the dynamics, which further acts as a soft-constraint on the operator form. Hard constraints such as imposing symmetries also reduce the possibility of overfitting to the noisy dynamics. We note that state-of-the-art PIC simulations are commonly run with a large number of particles (commonly in the range $10^9-19^{11}$) which should enable the collection of subpopulations with significantly large statistics. In this work, the statistics were limited solely by the memory requirements necessary to store the particle tracks used for a fair comparison against the phase space approach. Future works should not be constrained by this decision and, empirically, we observe that a lack of statistics is the main cause for variability in the coefficients estimated from phase space dynamics.

Finally, we highlight that, for the problem under study in this work, we consistently recover equivalent advection-diffusion coefficients when using the combined approach of tracking multiple subpopulations and imposing strict symmetries. This is clearly illustrated, e.g., in the comparisons between the PS-Tensor and PS-NN models in Section~\ref{sec:comparisons_against_theory}. These results empirically demonstrate the viability of the proposed approach to mitigate the ill-posedness of the problem.

\clearpage
\section{Fokker-Planck Solver Implementation}
\label{app:fp_solver_implementation}

The Fokker-Planck solver is implemented in PyTorch~\citep{ansel2024pytorch} using a standard Euler integrator:
\begin{equation}
    f^{(t+1)} = f^{(t)} + \Delta t\left(-\nabla_{\boldsymbol{v}}\boldsymbol{\cdot}\left(\mathsfbi{A} f^{(t)}\right) + \frac{1}{2}\nabla_{\boldsymbol{v}} \boldsymbol{\cdot} \left[ \nabla_{\boldsymbol{v}}\boldsymbol{\cdot}\left(\mathsfbi{D}f^{(t)}\right) \right] \right)
\label{eq:fp_step}
\end{equation}
The gradients are expanded in 2-D as:
\begin{align}
    \nabla_{\boldsymbol{v}}\boldsymbol{\cdot}\left(\mathsfbi{A} f^{(t)}\right) &= \partial_{v_x}\left(\mathsfbi{A}_x \odot \mathsfbi{F}^{(t)}\right) + \partial_{v_y}\left(\mathsfbi{A}_y \odot \mathsfbi{F}^{(t)}\right) \label{eq:fp_gradients_A} \\ 
    \nabla_{\boldsymbol{v}}\boldsymbol{\cdot}\left[\nabla_{\boldsymbol{v}}\boldsymbol{\cdot}\left(\mathsfbi{D} f^{(t)}\right)\right] &=
    \partial^2_{v_x}\left(\mathsfbi{D}_{xx} \odot \mathsfbi{F}^{(t)}\right) + \partial^2_{v_y}\left(\mathsfbi{D}_{yy} \odot \mathsfbi{F}^{(t)}\right) \label{eq:fp_gradients_D} \\
    &+ \partial_{v_x}\partial_{v_y}\left(\mathsfbi{D}_{xy} \odot \mathsfbi{F}^{(t)}\right) + \partial_{v_y}\partial_{v_x}\left(\mathsfbi{D}_{xy} \odot \mathsfbi{F}^{(t)}\right)
    \nonumber
\end{align}
where $\mathsfbi{F}^{(t)}$, $\mathsfbi{A}_{i}$, and $\mathsfbi{D}_{ij}$ correspond to the distribution function and advection/diffusion coefficients represented over the uniform grid of size $N_v\times N_v$, and $\odot$ the elementwise product between the matrices. All the derivatives are computed using centred second-order finite difference stencils without dividing by the grid resolution ($\Delta_{v_i}, \Delta_{v_i}^2$), which means that $\mathsfbi{A}_i$ and $\mathsfbi{D}_{ij}$ are learned in units of the grid resolution (respectively $\Delta_{v_i}\omega_p$ and $\Delta_{v_i}\Delta_{v_j}\omega_p$). To mimic the effect of an open boundary, we pad the phase space with an extra row/column of zeros whose values are not updated throughout the simulation.

When the advection and diffusion coefficients are parameterized by a NN, we obtain the matrices as follows. We define (at initialization) an auxiliary matrix $\mathsfbi{V}_{grid} \in \mathbb{R}^{N_v} \times \mathbb{R}^{N_v}$ which contains the center values of the bins normalized to the range $[-1,1]$ (normalization facilitates the learning process). The matrices $\mathsfbi{A}_{i}$, and $\mathsfbi{D}_{ij}$ are then obtained by simply calling the NN models on the matrix of grid values: $\mathsfbi{A}_{i} = \mathrm{NN}_{\mathsfi{A}_i}(\mathsfbi{V}_{grid})$ and $\mathsfbi{D}_{ij} = \mathrm{NN}_{\mathsfi{D}_{ij}}(\mathsfbi{V}_{grid})$. When accounting for symmetries, this process is slightly changed (details in Supplemental Material~\ref{app:ad_t} to \ref{app:ad_parperp}). In the case where we train a single model on multiple simulations (with different numerical parameters), we also provide as input to the NNs the numerical parameters, i.e. $\mathsfbi{A}_{i} = \mathrm{NN}_{\mathsfi{A}_i}(\mathsfbi{V}_{grid}, \boldsymbol{p})$ and $\mathsfbi{D}_{ij} = \mathrm{NN}_{\mathsfi{D}_{ij}}(\mathsfbi{V}_{grid}, \boldsymbol{p})$, where $\boldsymbol{p} = [N_{ppc}, \Delta_x/\lambda_D, m,v_{th}/c]$. All numerical parameters $\boldsymbol{p}$ are (independently) normalized to the region $[-1, 1]$ with respect to the minimum and maximum values over the full train set. 

To ensure the numerical stability of the integrator and the calculation of physical solutions, we enforce two constraints. First, we enforce that $f^{(t+1)} \geq 0$ by applying the equivalent of a ReLU function after the update \eqref{eq:fp_step}. Secondly, when producing test rollouts, we enforce that $D_{ii} \geq 0$ everywhere, even if the estimated values (obtained via the particle tracks or phase space evolution) are negative. 

While simulation rollouts performed with the operators learned via the differentiable simulator were stable at the PIC tracks/phase space diagnostic frequency ($\Delta t_{dump}$), the same was not observed for operators extracted from particle tracks. To produce stable rollouts using the operators retrieved from the particle statistics, we were forced to use a smaller time step ($\Delta t_{dump}/10$) due to the very noisy coefficient estimates at larger $v$. We found this to be the most stable and general solution when compared to alternatives, e.g., artificially smoothing or zeroing the coefficient values above a certain velocity magnitude. Implementing a higher-order integration scheme could also have avoided the problem. However, since the differentiable simulator approach did not suffer from any issues, we did not see the need to implement a higher-order scheme. In future works, if the need arises, a more accurate option, e.g., the Crank-Nicolson algorithm~\citep{crank1947practical}, could be considered. 

To enforce symmetries into the recovered operators, we proceed as summarized in the following subsections. For additional details regarding the simulator implementation, we refer the reader to the code repository~\href{https://github.com/diogodcarvalho/ml-pic-collision-operators.git}{https://github.com/diogodcarvalho/ml-pic-collision-operators.git}

\subsection{$\mathrm{AD}_{T}$}
\label{app:ad_t}

\textit{Tensor}: The only free parameters are the matrices $\mathsfbi{A}_x$, $\mathsfbi{D}_{xx}$, and $\mathsfbi{D}_{xy}$ (each with $N_v\times N_v$ parameters). The remaining matrices are obtained by simply applying transpositions $\mathsfbi{A}_y = \mathsfbi{A}_x^T$ and $\mathsfbi{D}_{yy} = \mathsfbi{D}_{xx}^T$.

\textit{NN}: Only three NN models are used $\mathsfbi{A}_x(\mathsfbi{V}_{grid}) = \mathrm{NN}_{\mathsfi{A}_x}(\mathsfbi{V}_{grid})$, $\mathsfbi{D}_{xx}(\mathsfbi{V}_{grid}) = \mathrm{NN}_{\mathsfi{D}_{xx}}(\mathsfbi{V}_{grid})$, $\mathsfbi{D}_{xy}(\mathsfbi{V}_{grid}) = \mathrm{NN}_{\mathsfi{D}_{xy}}(\mathsfbi{V}_{grid})$. The remaining matrices are obtained by transposition $\mathsfbi{A}_y = \mathsfbi{A}_x^T$ and $\mathsfbi{D}_{yy} = \mathsfbi{D}_{xx}^T$.

\subsection{$\mathrm{AD}_{sym}$}
\label{app:ad_sym}

In addition to the transposed symmetry imposed by $\mathrm{AD}_T$, this symmetry additionally enforces (anti-)axi-symmetry in the following way.

\textit{Tensor}: We half the number of free parameters by allowing coefficients to be defined only on half the grid along the $v_x$-axis, i.e. $\mathsfbi{V}_{grid}^{free} = [v^{min}_x, 0] \times [v^{min}_y, v^{max}_y]$. If the grid has an uneven number of grid points, the bins centered at zero are also free parameters. The remaining portion of the tensors is obtained by forcing that $\mathsfi{A}_x(v_x, v_y) = -\mathsfi{A}_x(-v_x, v_y)$, $\mathsfi{D}_{xx}(v_x, v_y) = \mathsfi{D}_{xx}(-v_x, v_y)$, and $\mathsfi{D}_{xy}(v_x, v_y) = -\mathsfi{D}_{xy}(-v_x, v_y)$.

\textit{NN}: Similar to the Tensor case, the NNs only receive as input half the grid ($\mathsfbi{V}_{grid}^{free}$), and the remaining values are enforced programmatically.

\subsection{$\mathrm{AD}_{\parallel,\perp}$}
\label{app:ad_parperp}

Given a unit velocity vector $\boldsymbol{\hat{v}} = \boldsymbol{v}/v$ that makes an angle $\theta$ with the $v_x$-axis, we define the advection and diffusion coefficients: $\mathsfi{A}_{\parallel}$, $\mathsfi{D}_{\parallel}$ - advection and diffusion along $\boldsymbol{\hat{v}}$; $\mathsfi{A}_{\perp}$, $\mathsfi{D}_{\perp}$ - advection and diffusion perpendicular to $\boldsymbol{\hat{v}}$; and the rotation matrix:
\begin{equation}
\mathsfbi{R}(\theta) = 
\begin{bmatrix}
\cos\theta & -\sin\theta \\
\sin\theta & \cos\theta
\end{bmatrix}.
\end{equation}

If we express the advection vector in the rotated frame as:
\begin{equation}
\mathsfbi{A}' =
\begin{bmatrix}
\mathsfi{A}_{\parallel} \\
\mathsfi{A}_{\perp}
\end{bmatrix},
\end{equation}
we can transform to/from the global $(v_x, v_y)$ coordinate system via:
\begin{equation}
\mathsfbi{A} = \mathsfbi{R}(\theta) \mathsfbi{A}'.
\end{equation}

Assuming that $\mathsfi{A}_{\perp} = 0$, since there should no net perpendicular drag force along this direction~\citep{okuda1970collisions, reynolds1997velocity}, the advection components in the global frame are:
\begin{equation}
\begin{aligned}
\mathsfi{A}_x &= \mathsfi{A}_{\parallel} \cos\theta \\
A_y &= \mathsfi{A}_{\parallel} \sin\theta \quad .
\end{aligned}
\label{eq:a_projection}
\end{equation}

The diffusion tensor in the rotated frame is:
\begin{equation}
\mathsfbi{D'} = 
\begin{bmatrix}
\mathsfi{D}_{\parallel} & 0 \\
0 & \mathsfi{D}_{\perp}
\end{bmatrix} \ .
\end{equation}

To express this in the original coordinate system, we apply a rotation:
\begin{equation}
\mathsfbi{D} = \mathsfbi{R}(\theta) \mathsfbi{D}' \mathsfbi{R}^\top(\theta) \ .  
\end{equation}

Multiplying out gives the diffusion tensor in the $(v_x, v_y)$ frame:
\begin{equation}
\mathsfbi{D} =
\begin{bmatrix}
\mathsfi{D}_{xx} & \mathsfi{D}_{xy} \\
\mathsfi{D}_{yx} & \mathsfi{D}_{yy}
\end{bmatrix} \ ,
\end{equation}
with components:
\begin{equation}
\begin{aligned}
\mathsfi{D}_{xx} &= \mathsfi{D}_{\parallel} \cos^2\theta + \mathsfi{D}_{\perp} \sin^2\theta \\
\mathsfi{D}_{yy} &= \mathsfi{D}_{\parallel} \sin^2\theta + \mathsfi{D}_{\perp} \cos^2\theta \\
\mathsfi{D}_{xy} = \mathsfi{D}_{yx} &= (\mathsfi{D}_{\parallel} - \mathsfi{D}_{\perp}) \sin\theta \cos\theta \ . 
\end{aligned}
\label{eq:d_projection}
\end{equation}

To implement this symmetry in our models, we proceed as follows. 

\textit{Tensor}: We define a new auxiliary 1-D grid, representing the range of velocity magnitudes $\boldsymbol{v}^{aux}_{grid} \in [0, v_{max}]$ divided into $N_v^{aux}=51$ bins. The free parameters are then parallel/perpendicular components of the advection/diffusion terms in these auxiliary grid points (a total of $3\times N_v^{aux}$ free parameters). The values of $\mathsfbi{A}_\parallel(\mathsfbi{V}_{grid})$, $\mathsfbi{D}_\parallel(\mathsfbi{V}_{grid})$, and $\mathsfbi{D}_\perp(\mathsfbi{V}_{grid})$ are obtained by linearly interpolating from the two nearest neighbors in the auxiliary 1-D grid (i.e. for each $\boldsymbol{v} \in \mathsfbi{V}_{grid}$ we find $v_i, v_{i+1} \in \boldsymbol{v}_{grid}^{aux}$ such that $v_i \leq v < v_{i+1}$). Finally, we use eqs.~\eqref{eq:a_projection} and \eqref{eq:d_projection} to obtain the coefficients in the original axis. We ensure that all these operations are fully differentiable.

\textit{NN}: The parallel/perpendicular coefficients are obtained using three distinct NN via $\mathsfi{A}_{\parallel}(\boldsymbol{v}) = \mathrm{NN}_{\mathsfi{A}_{\parallel}}(v)$, $\mathsfi{D}_{\parallel}(\boldsymbol{v}) = \mathrm{NN}_{\mathsfi{D}_{\parallel}}(v)$, and $\mathsfi{D}_{\perp}(\boldsymbol{v}) = \mathrm{NN}_{\mathsfi{D}_{\perp}}(v)$ for $\boldsymbol{v} \in \mathsfbi{V}_{grid}$. They are then projected into the original axis via \eqref{eq:a_projection} and \eqref{eq:d_projection}.

By default we assign $(v_x, v_y) = (0,0)$ to $\theta = 45^{\circ}$ (or equivalently $\cos{\theta} = \sin{\theta} =  \sqrt{2}/{2}$) to force the models to learn that $\mathsfi{A}_{\parallel}(0,0) = 0$ and $\mathsfi{D}_{\parallel}(0,0) = \mathsfi{D}_{\perp}(0,0)$.

\clearpage
\section{Training Procedure from the Phase Space Evolution}
\label{app:training_procedure}

All models trained using the differentiable simulator approach have access to the evolution of 9 subpopulations of macroparticles over the full simulation (approximately 100 time frames each). These subpopulations are sampled according to a normal distribution centered at $\mu_{\boldsymbol{v}} \in [\{0, \pm 2\}, \{0, \pm 2\}]v_{th}$ (see Supplemental Material~\ref{app:subpopulation_sampling}). 

We use different data splitting strategies depending on the type of operator learned (discrete or continuous). For the discrete operators (referred to as PS-Tensor in Section~\ref{sec:accuracy_different_methods}), we use all the timesteps of all subpopulations for training. For the continuous operators (PS-NN and PS-NN-Multi in Section~\ref{sec:accuracy_different_methods}), we randomly split the data into 90\% training and 10\% validation to account for overfitting and perform hyperparameter tuning.

To train the models, we use a curriculum learning approach, where the temporal unrolling length $N_u$ is consistently increased. This allows for a more stable training procedure than immediately starting with a longer temporal unrolling length. The different curriculum stages, their duration (number of epochs), and associated learning rates ($\alpha$) are summarized in table~\ref{tab:training_curriculum}.

The loss function at each stage is the mean absolute error over the temporal unrolled phase space dynamics:
\begin{equation}
    \mathcal{L}_{N_u} = \frac{1}{N_uN_v^2}\sum_{u=1}^{N_u}\sum_{v_x,v_y \in \mathcal{V}}\left|\hat{f}^{(t+u)}(v_x, v_y) - f^{(t+u)}(v_x,v_y)\right|
\end{equation}
averaged over the batch size, where $N_v^2$ corresponds to the total number of bins of phase space $\mathcal{V}$).

For all cases, we use the full training set per gradient update (i.e., the batch size equals the training set size) and the Adam optimizer~\citep{kingma2014adam}. For models trained on a single simulation (PS-Tensor, PS-NN), the training process lasts approximately 5-10 minutes on a single Nvidia Titan X GPU. For larger NN models trained across all simulations at once (PS-NN-Multi), the training process lasts approximately 3h on a node containing 4 Nvidia A100 GPUs. The final stored model weights correspond to the ones that achieved the lowest training loss for the PS-Tensor models, and the lowest validation loss for the PS-NN and PS-NN-Multi models.

\begin{table}
\begin{center}
\begin{tabular}{c|ccc|ccc|ccc}
 & \multicolumn{3}{c}{Tensor} & \multicolumn{3}{c}{NN} & \multicolumn{3}{c}{NN-Multi} \\ \hline
Stage & $N_u$ & $\alpha$ & Epochs & $N_u$ & $\alpha$ & Epochs & $N_u$ & $\alpha$ & Epochs  \\
1 & 1  & $10^{-3}$ & 1000 & 1  & $10^{-3}$ &  500 & 1  & $10^{-3}$ & 500 \\
2 & 2  & $10^{-3}$ & 200  & 1  & $10^{-4}$ &  500 & 1  & $10^{-4}$ & 500 \\
3 & 5  & $10^{-3}$ & 200  & 2  & $10^{-4}$ &  200 & 2  & $10^{-4}$ & 200 \\
4 & 10 & $10^{-3}$ & 500  & 5  & $10^{-4}$ &  200 & 5  & $10^{-4}$ & 200 \\
5 &    &           &      & 10 & $10^{-4}$ &  500 & 10 & $10^{-4}$ & 1000 \\
6 &    &           &      &    &           &      & 10 & $10^{-5}$ & 1000
\end{tabular}
\caption{Curriculum training stages for the different operator models used throughout this work. The same values are used independently of the symmetries embedded. The temporal unrolling length $N_u$ is increased progressively, and the learning rate $\alpha$ and the number of epochs are adapted accordingly to ensure convergence and stability across all tested simulations.}
\label{tab:training_curriculum}
\end{center}
\end{table}

\clearpage
\section{Example for a Single Simulation}
\label{app:example_single_simulation}

\subsection{Advection-Diffusion Operators Retrieved}
\label{app:ad_extra_examples}

\begin{figure}
    \centering
    \includegraphics[width=\linewidth]{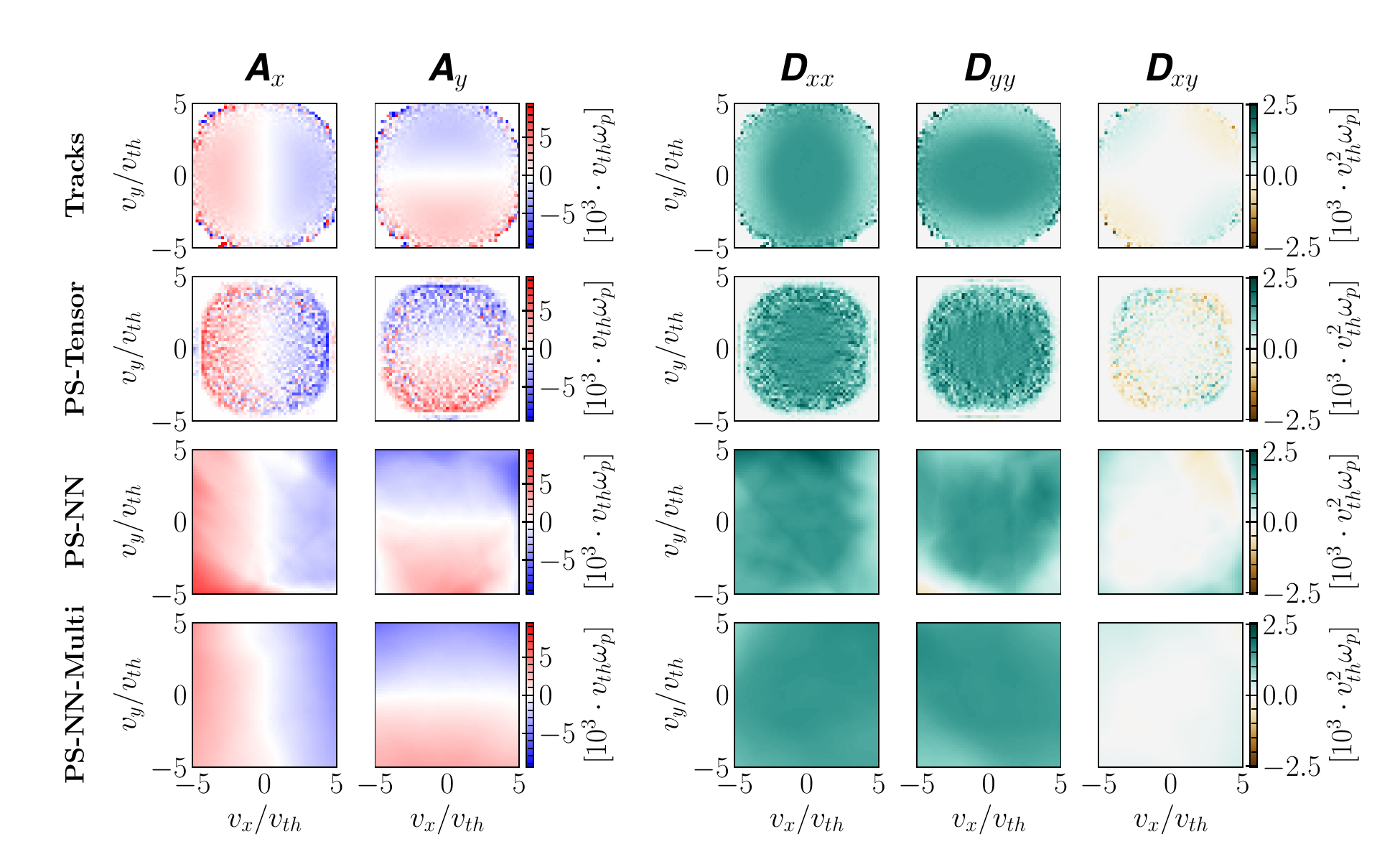}
    \includegraphics[width=\linewidth]{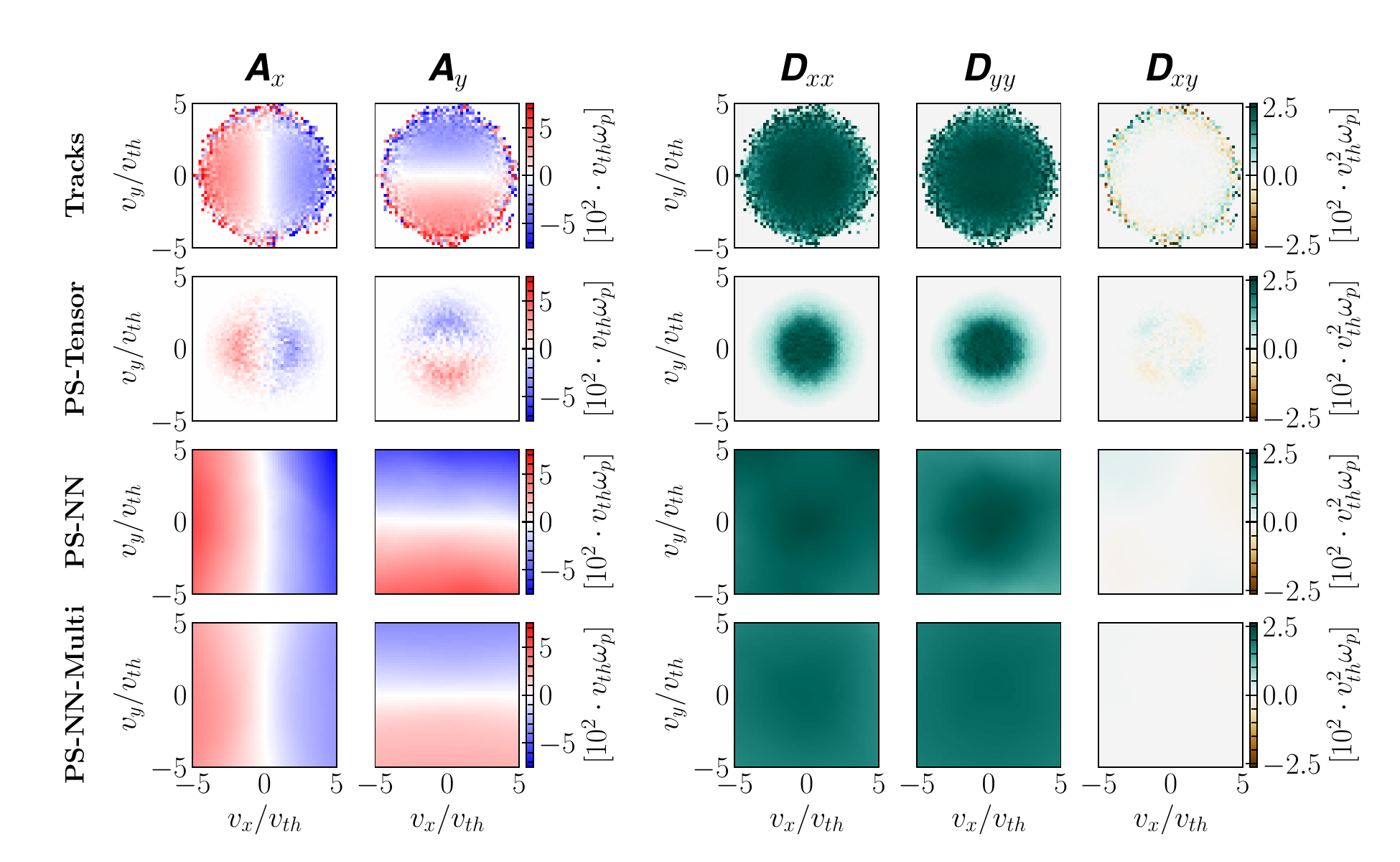}
    \caption{Advection and diffusion coefficients retrieved with different approaches for simulation $index=47$ ($N_{ppc}=25$, $m=4$, $\Delta_x/\lambda_D=1, v_{th}=0.1c)$ and $index=51$ ($N_{ppc}=4$, $m=2$, $\Delta_x/\lambda_D=2, v_{th}=0.01c)$. These correspond to the extreme cases in terms of simulated macroparticles ($N=2.5\times10^7$ and $N=10^6$). Operators retrieved from particle tracks benefit directly from a larger number of macroparticles due to a larger number of statistics. On the other hand, operators retrieved from phase space diagnostics suffer indirectly from a reduced number of macro particles since there are not enough macroparticles to generate dense subpopulations at high $v$ regions.
    }
    \label{fig:AD_comparison_index_extra} 
\end{figure}

In Figure~\ref{fig:AD_comparison_index_extra}, we provide two extra examples of the operators retrieved for different simulation parameters. They represent the extremes of the dataset in terms of the number of macroparticles available in the simulation (in contrast with Figure~\ref{fig:AD_comparison_index_0} in the main body of the paper that represents the average). 

As expected, a larger number of macroparticles (up to $2.5\times10^7$) leads to a smoother operator calculated from particle tracks. On the other hand, since we use a fixed number of macroparticles to construct subpopulations (approximately $3\times10^5$), the other methods do not benefit as clearly from this increase in the number of simulated macroparticles.  

For the PS-Tensor operators, when we have a very limited number of particles, the region of phase space recovered is considerably smaller. This happens because the particles sampled to construct the training subpopulations are significantly more concentrated in regions of small $v$ (because the total populations does not include as many macroparticles at high $v$). Across all scenarios, the retrieved operator is nonetheless noisy (including the examples shown Figure~\ref{fig:AD_comparison_index_0}).

For the PS-NN operators, the NN model guarantees some smoothness constraints across the different cases, but there are still clear artifacts at high $v$ (the regions where not enough training data exists). Nonetheless, it is important to point out that these models capture the phase space dynamics for the tested subpopulations as accurately as the operators learned from the particle tracks (c.f. corresponding index in Figure~\ref{fig:l1_model_comparison_boxplot_sym}).

Finally, the PS-NN-Multi models retrieve significantly simpler and smoother models, which at first glance might seem to be a good approximation. However, for $index=51$, both advection and diffusion coefficients are severely underestimated. This is one of three outliers across all simulations which we discuss in more detail in Supplemental Material~\ref {app:rollout_errors_per_simulation}.

\subsection{Phase Space Evolution Comparison}

\begin{figure}
    \centering
    \includegraphics[width=0.7\linewidth]{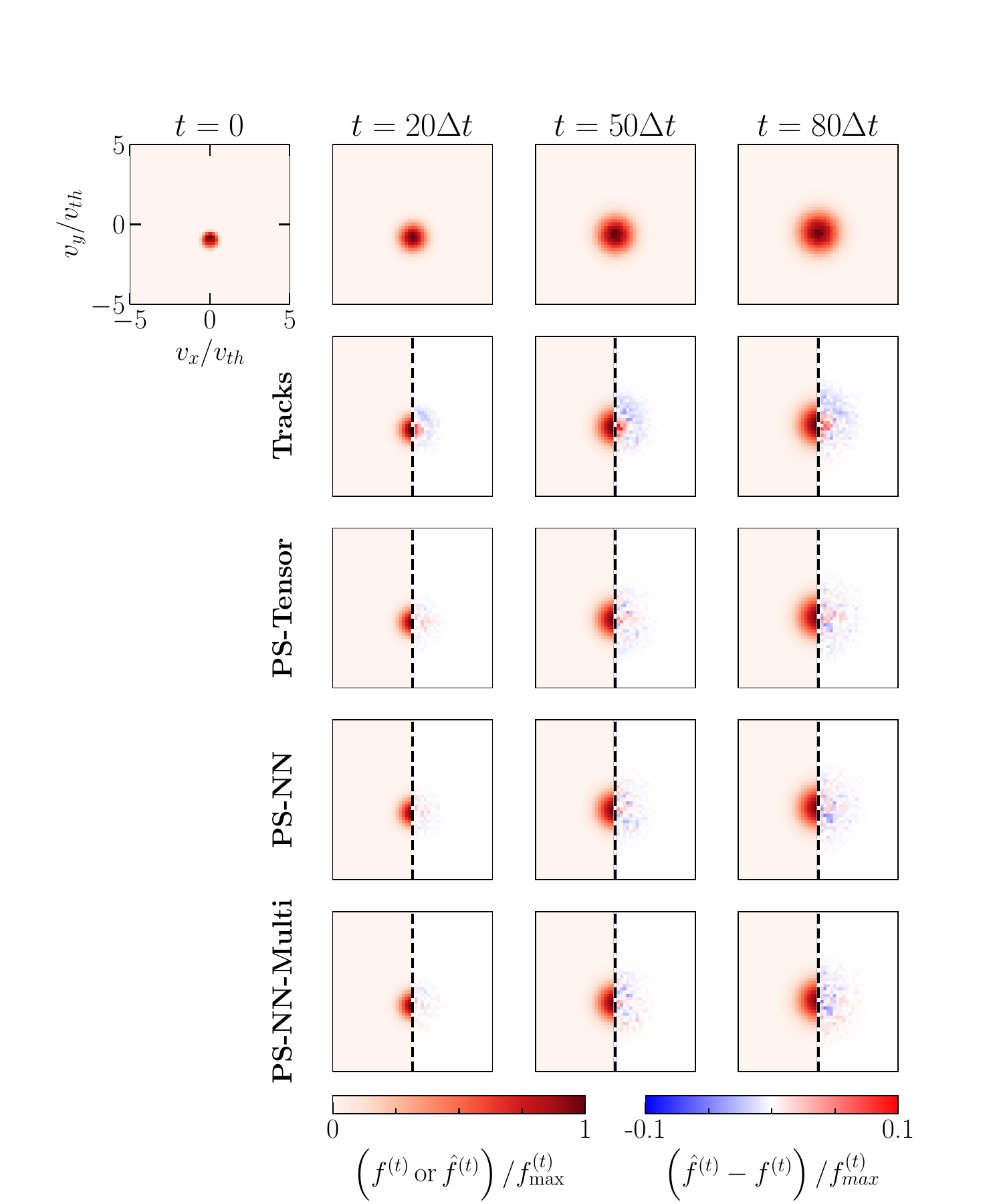}
    \caption{Comparison of phase space dynamics for a Normals-1 test subpopulation centred at $\boldsymbol{\mu}_v = (0, -1)v_{th}$ and simulation $index=0$. This test subpopulation is similar to the ones used for training when learning from phase space data. This is the reason why all phase space-based models perform equivalently well.}
    \label{fig:ex_rolllout_dif_0_normal_-1_0}
\end{figure}

\begin{figure}
    \centering
    \includegraphics[width=0.7\linewidth]{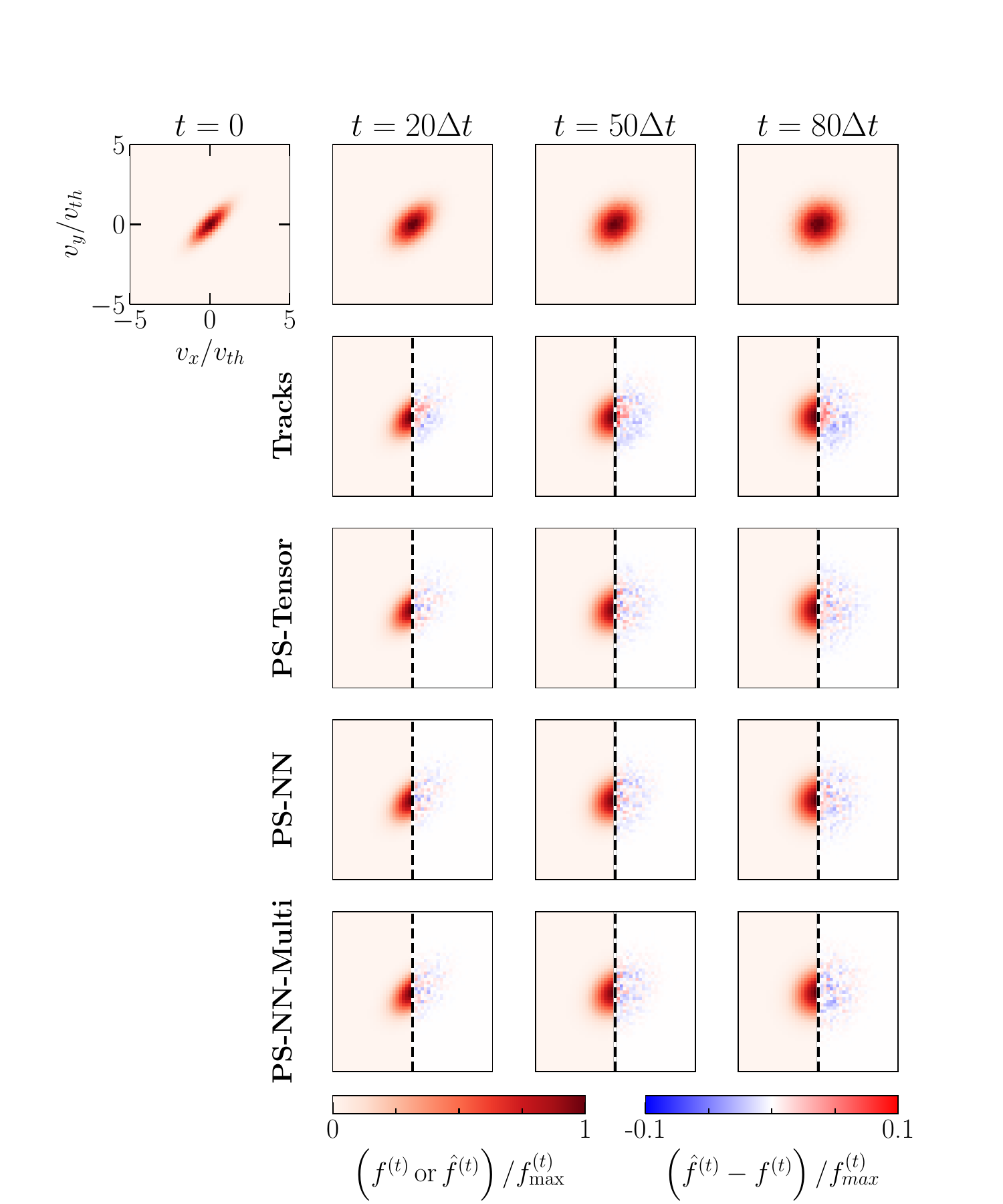}
    \caption{Comparison of phase space dynamics for a Normals-Rot test subpopulation (rotation of $\theta =45^{\circ}$) and simulation $index=0$. All models seem to equally capture the dynamics, with slightly higher errors for Tracks. The initial subpopulation is still mainly in regions where training phase space data was provided.}
    \label{fig:ex_rolllout_dif_0_normal_rot_45}
\end{figure}

\begin{figure}
    \centering
    \includegraphics[width=0.7\linewidth]{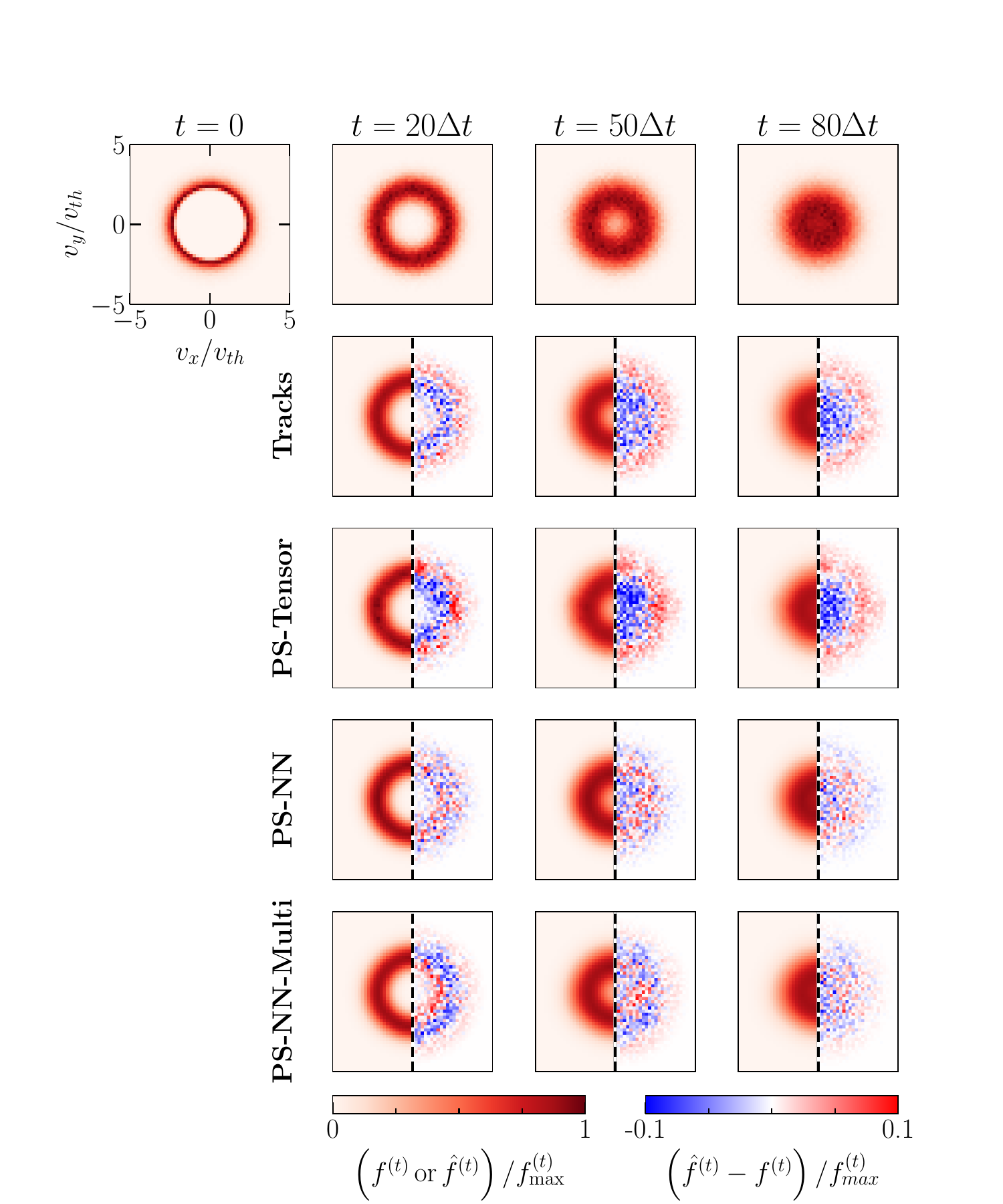}
    \caption{Comparison of phase space dynamics for a Rings test subpopulation centred at $v=3v_{th}$ and simulation $index=0$. Relative errors are higher when compared to other examples due to the normalization with respect to $f_{max}^{(t)}$ (which is lower in this case). Additionally, this subpopulation has a larger density of particles at higher $v$, and the phase space models were not provided with significant statistics in this region during training. Nonetheless, NN models, particularly PS-NN, are capable of accurately capturing the dynamics. Both Tracks and PS-Tensor significantly underestimate advection at high $v$.}
    \label{fig:ex_rolllout_0_ring_normal_3_0.2}
\end{figure}

\begin{figure}
    \centering
    \includegraphics[width=0.7\linewidth]{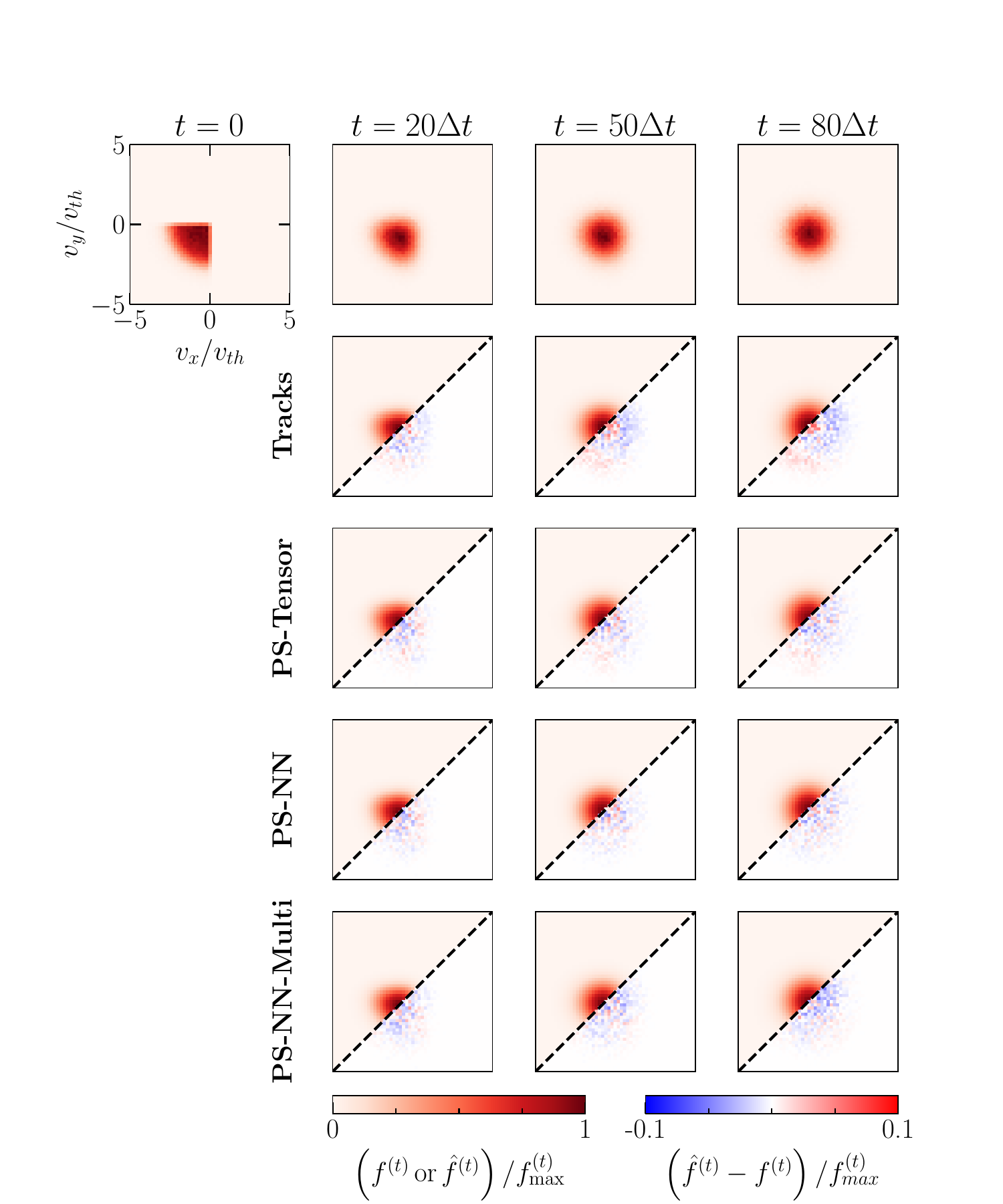}
    \caption{Comparison of phase space dynamics for a Quadrants distribution (bottom left quadrant) and simulation $index=0$. Overall, the operators learned from phase space dynamics seem to perform better than those learned from tracks, and they do not reveal systematic errors.}
    \label{fig:ex_rolllout_dif_0_uniform_-5.0_0_-5.0_0}
\end{figure}

In Figures~\ref{fig:ex_rolllout_dif_0_normal_-1_0} to \ref{fig:ex_rolllout_dif_0_uniform_-5.0_0_-5.0_0}, we provided additional comparisons between the phase space dynamics of different test subpopulations when simulated using the operators retrieved from the distinct methods proposed in this work. The same simulation ($index=0$) is used for all cases, which coincides with the one used in Figure~\ref{fig:ex_rolllout_dif_0_ring_normal_2_0.2}  in the main body of the paper.

Overall, the PS-NN operator leads to better reproductions of the dynamics (as made clear by the randomized error distribution) while the Tracks, PS-Tensor, and PS-NN-Multi operators might reveal a systematic underestimation of advection and diffusion (especially at higher values of $v$). These conclusions coincide with those made in Section~\ref{sec:example_single_simulation} where a different initial subpopulation was shown (c.f. Figure~\ref{fig:ex_rolllout_dif_0_ring_normal_2_0.2}).

\clearpage
\section{Long-term Accuracy of Different Methods}
\label{app:accuracy_different_methods_extra}

In this supplemental material, we provide complementary materials to those shown in Section~\ref{sec:accuracy_different_methods} of the main body of the paper. The models used henceforth correspond to those presented in the main body of the text.

\subsection{Rollout Errors per Subpopulation}
\label{app:rollout_errors_per_subpopulation}
\begin{figure}
    \centering
    \includegraphics[width=0.6\linewidth]{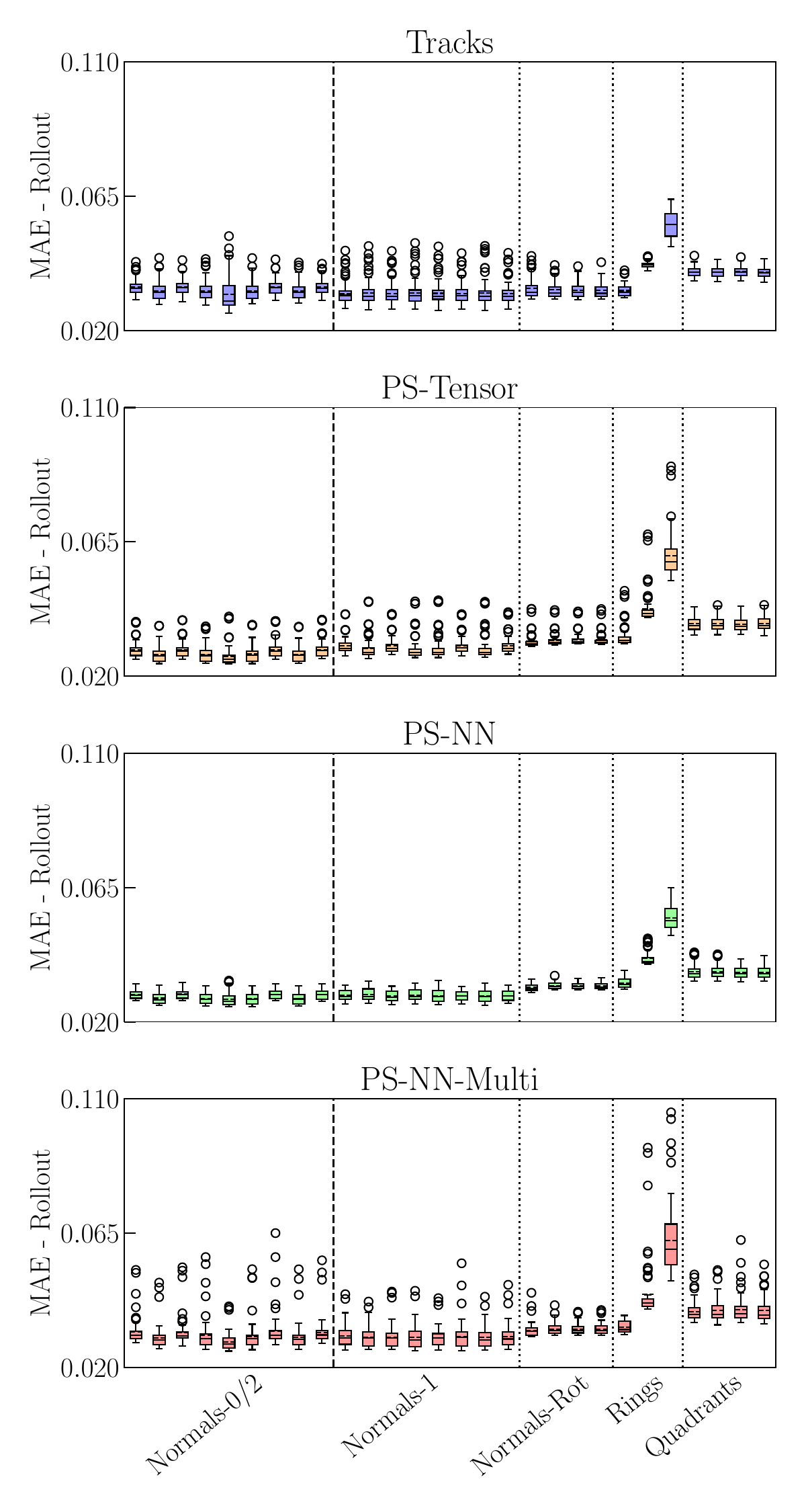}
    \caption{Rollout errors for different initial subpopulations (see Supplemental Material~\ref{app:subpopulation_sampling} for a full description). Statistics are computed over the full simulation dataset. Higher errors are observed for ring distributions with high radial velocity. For Tracks, this is caused by the lack of statistics to accurately estimate the coefficients in this region. For other methods, the training subpopulations chosen (Normals-0/2) do not cover this region of the phase space as accurately. This exercise highlights the requirement of gathering enough statistical information (via tracks or phase space data) to correctly assess the operators at different regions in phase space.}
    \label{fig:l1_model_comparison_per_subpopulation}
\end{figure}

To better understand the difference in average errors between train and test subpopulations, we plot in Figure~\ref{fig:l1_model_comparison_per_subpopulation} the error statistics per subpopulation (statistics computed over the full dataset of PIC simulations). 

Starting with the Normals-0/2 group (used for training), we see that each model performs similarly (on average) across the different subpopulations. When analyzing the Normals-1 group (first test group), we see that there is no significant change in the error values when compared with the train group. This happens because the evolution of the train subpopulations extensively covers similar areas in phase space.

The slightly higher error for Normals-Rot is attributed to a higher initial density of particles at higher $v$. For the Rings subpopulations, the same situation causes a significantly larger increase in error, with the ring centered at $ v=3v_{th}$ standing out as a clear outlier across all models. The Quadrants subpopulations denote similar issues, but these are also attributed to using a smaller number of macroparticles, which leads to a higher granularity of the distribution function. For the Tracks operators, this increased error in test subpopulations is caused by the lack of statistics to accurately estimate smooth coefficients at high $v$ (see e.g., Figure~\ref{fig:AD_comparison_index_0}). For all the PS operators, the training subpopulations chosen (Normals-0/2) do not cover this region of the phase space as accurately.

Furthermore, we would like to highlight that the PS-NN models are clearly more robust to outliers and perform equivalently to Tracks for subpopulations with larger densities at high $v$, even though they have access to significantly less data. This happens due to the smoothness bias introduced by the NN models.

The main goal of this exercise (using different subpopulations) was to understand the generalization capabilities of the different models for regions outside their training data. Based on the findings, we can conclude that extrapolation is not guaranteed, and one should be careful when applying an operator to regions of phase space where not enough statistics were provided. One possible way to mitigate this issue is to enforce known symmetries to artificially increase the statistics obtained for a phase space region. This is exactly what we do in Section~\ref{sec:enforcing_known_symmetries}.  

\subsection{Rollout Errors per Simulation}
\label{app:rollout_errors_per_simulation}

\begin{figure}
    \centering
    \includegraphics[width=0.8\linewidth]{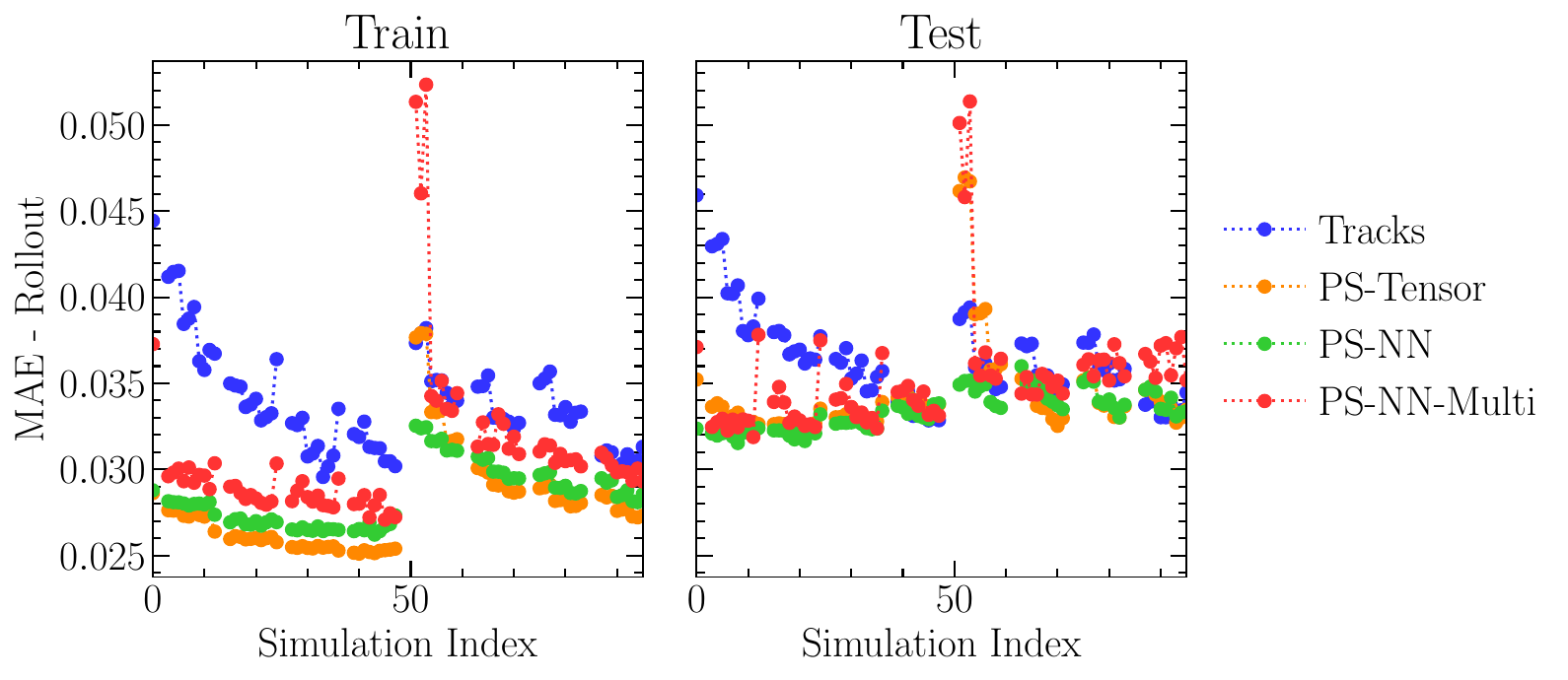}
    \caption{Rollout errors for advection and diffusion models obtained from tracks versus phase space dynamics per simulation of the dataset (table~\ref{tab:simulation_parameters}). Results are shown for the train and test subpopulations. Methods based on learning operators from phase space dynamics using a differentiable simulator consistently outperform operators learned from tracks. Outliers for PS-NN-Multi correspond to simulations with a smaller number of macroparticles.}
    \label{fig:l1_model_comparison_per_simulation}
\end{figure}

In Figure~\ref{fig:l1_model_comparison_per_simulation}, we show the average rollout error over subpopulations per simulation of the dataset. These are the values used to produce the boxplot in Figure~\ref{fig:l1_model_comparison_boxplot}. We observe that, across almost all simulations, the operators retrieved using phase space information perform better than the operators retrieved from tracks. This happens because they are optimized to predict long-term dynamics, while values from tracks can slightly vary depending on the time interval over which statistics are computed. These results highlight the benefit of using an optimization-based approach when we do not know a priori what are the relevant collisional time scales, and need to automatically process large amounts of simulations and phase space regions. 

The outliers observed for the PS-NN-Multi models in Figure~\ref{fig:l1_model_comparison_per_simulation} correspond to simulations with a smaller number of macroparticles ($N_{ppc}=2$, $\Delta_x/\lambda_D = 2$) and quadratic shape function ($index\in[51,52,53]$ in table~\ref{tab:simulation_parameters}). We did not find any specific reason for the poorer performance at these specific numerical parameters, since it performs well at equivalent numbers of macroparticles for other shape functions (i.e., it is not a problem of lack of phase space dynamics statistics), and other phase space-based methods (PS-Tensor and PS-NN) perform well for the same data. This leads us to believe that the only justification is a lack of representative power of the network and a limited size of the training simulation set (in terms of the variety of the simulation parameters). We conjecture that producing a larger set of simulations and training a larger model should result in better generalization capabilities.

\clearpage
\section{Impact of Temporal Unrolling during Training}
\label{app:impact_temporal_unrolling}

\begin{figure}
    \centering
    \includegraphics[width=0.7\linewidth]{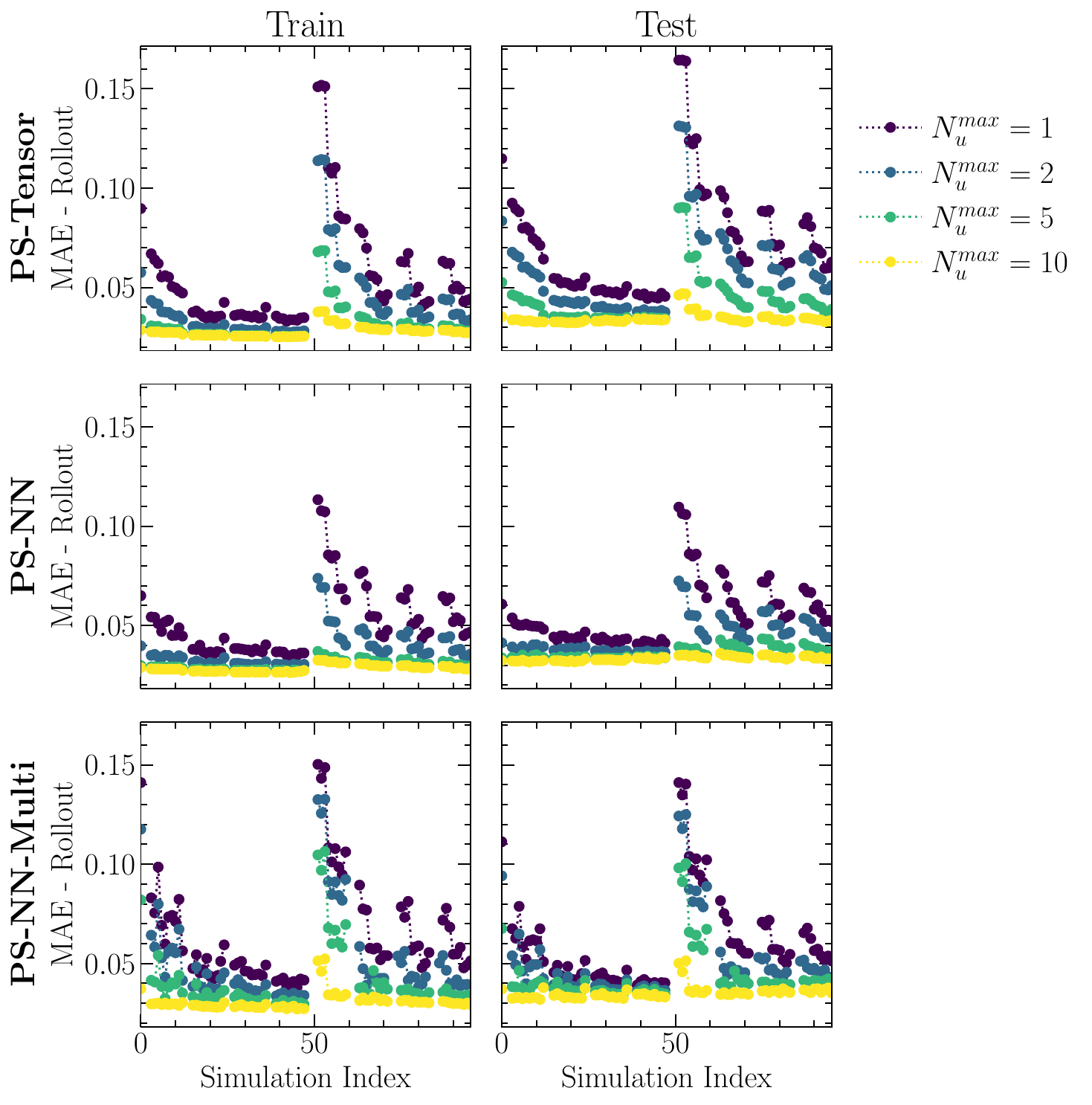}
    \caption{Impact of temporal unrolling length ($N_u^{max}$) on the rollout error (averaged over subpopulations) across the full dataset of simulations. Increasing the value of $N_u^{max}$ leads to a consistent improvement in long-term rollout performance (test rollout length is $\approx 100$ time-steps).}
    \label{fig:l1_model_comparison_per_simulation_unroll}
\end{figure}

In Figure~\ref{fig:l1_model_comparison_per_simulation_unroll}, we show the average rollout error over subpopulations per simulation of the dataset as a function of the maximum temporal unrolling length during training ($N_u^{max}$). These are the values used to produce the boxplot in Figure~\ref{fig:l1_model_comparison_boxplot_unroll}. It is clear that increasing $N_u^{max}$ decreases the rollout error across all simulations and operator types, with a higher impact on simulations with a reduced number of macroparticles. 

\begin{figure}
    \centering
    \includegraphics[width=\linewidth]{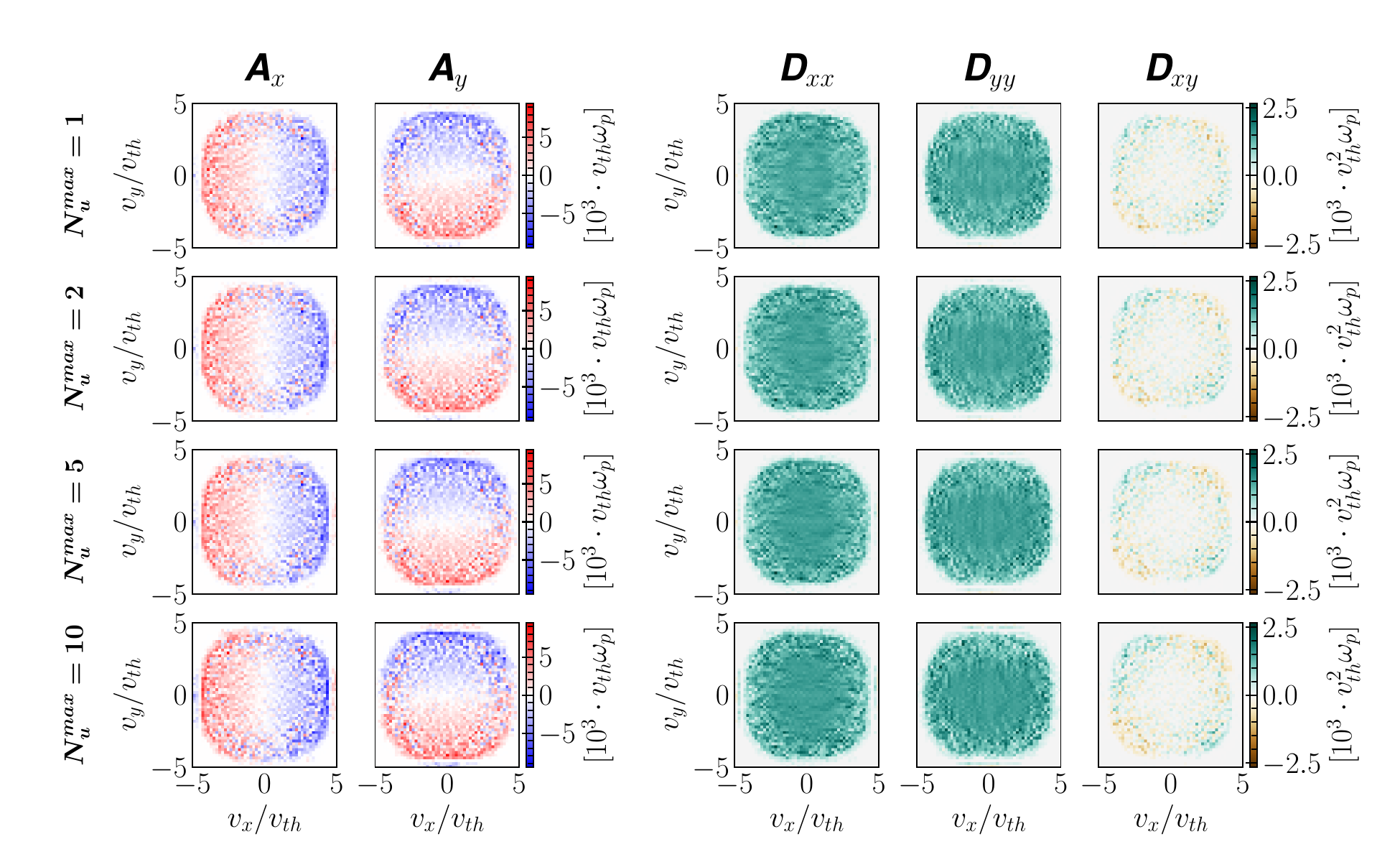}
    \includegraphics[width=\linewidth]{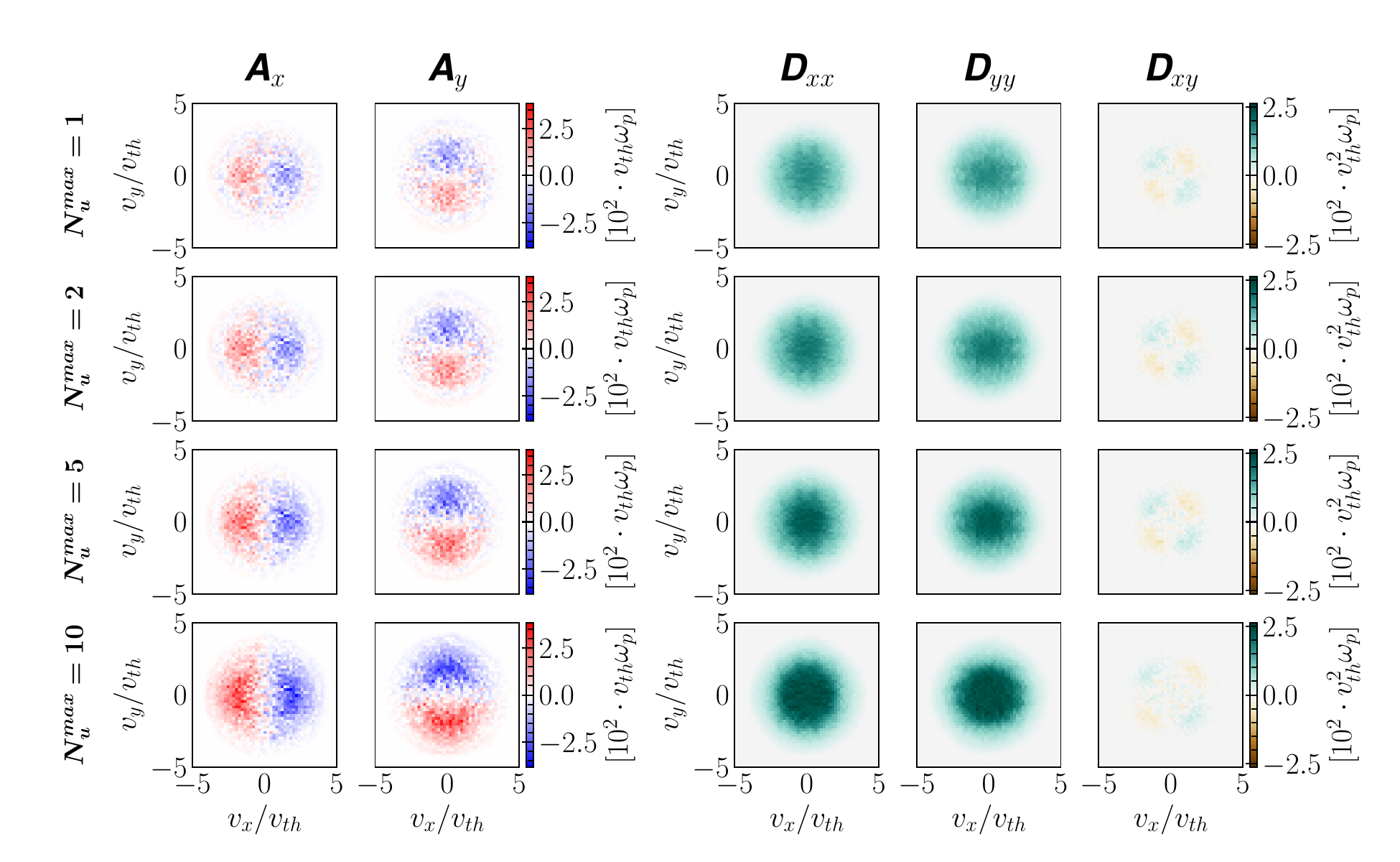}
    \caption{Illustration of the impact of the maximum training temporal unroll length on PS-Tensor models for $index=47$ (top) and $index=51$ (bottom). For $index=47$ no significant change is observed, while for $index=51$ we see significant changes. In both cases, a larger value of $N_u^{(max)}$ leads to improved rollout performance (see Figure~\ref{fig:l1_model_comparison_per_simulation_unroll}).}
    \label{fig:AD_comparison_unroll_tensor_index_extra}
\end{figure}

In Figure~\ref{fig:AD_comparison_unroll_tensor_index_extra}, we provide additional examples of the impact of the training temporal unrolling length $N_u^{max}$ on the form of PS-Tensor models for different simulation indices. For $index=47$, no significant major changes are observed (although performance slightly improves according to results in Figure~\ref{fig:l1_model_comparison_per_simulation_unroll}, meaning some fine-tuning is performed), while for $index=51$ it is clear the benefit of using a higher unrolling length (associated also with a significant improvement in performance in Figure~\ref{fig:l1_model_comparison_per_simulation_unroll}).

\begin{figure}
    \centering
    \includegraphics[width=\linewidth]{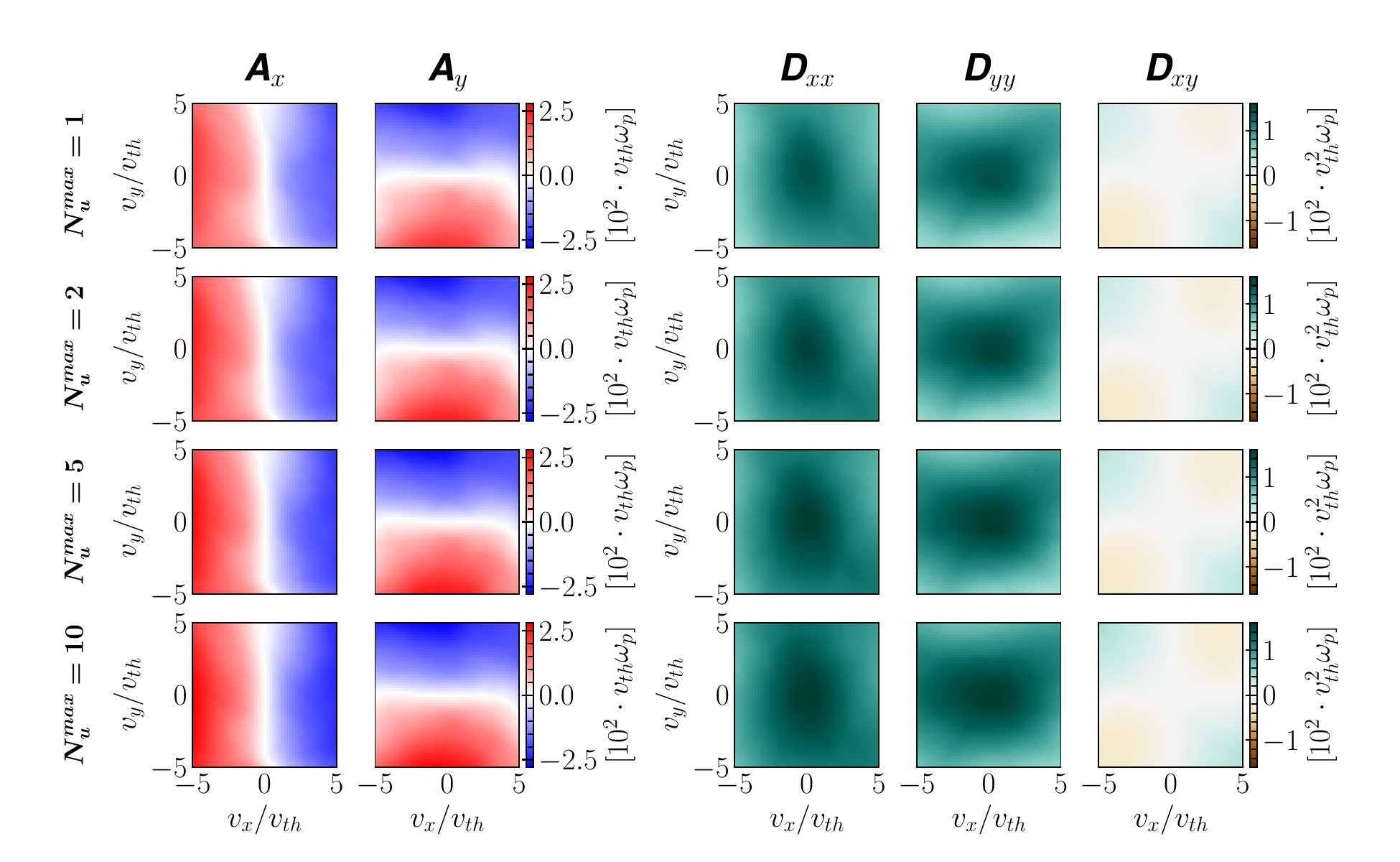}
    \includegraphics[width=\linewidth]{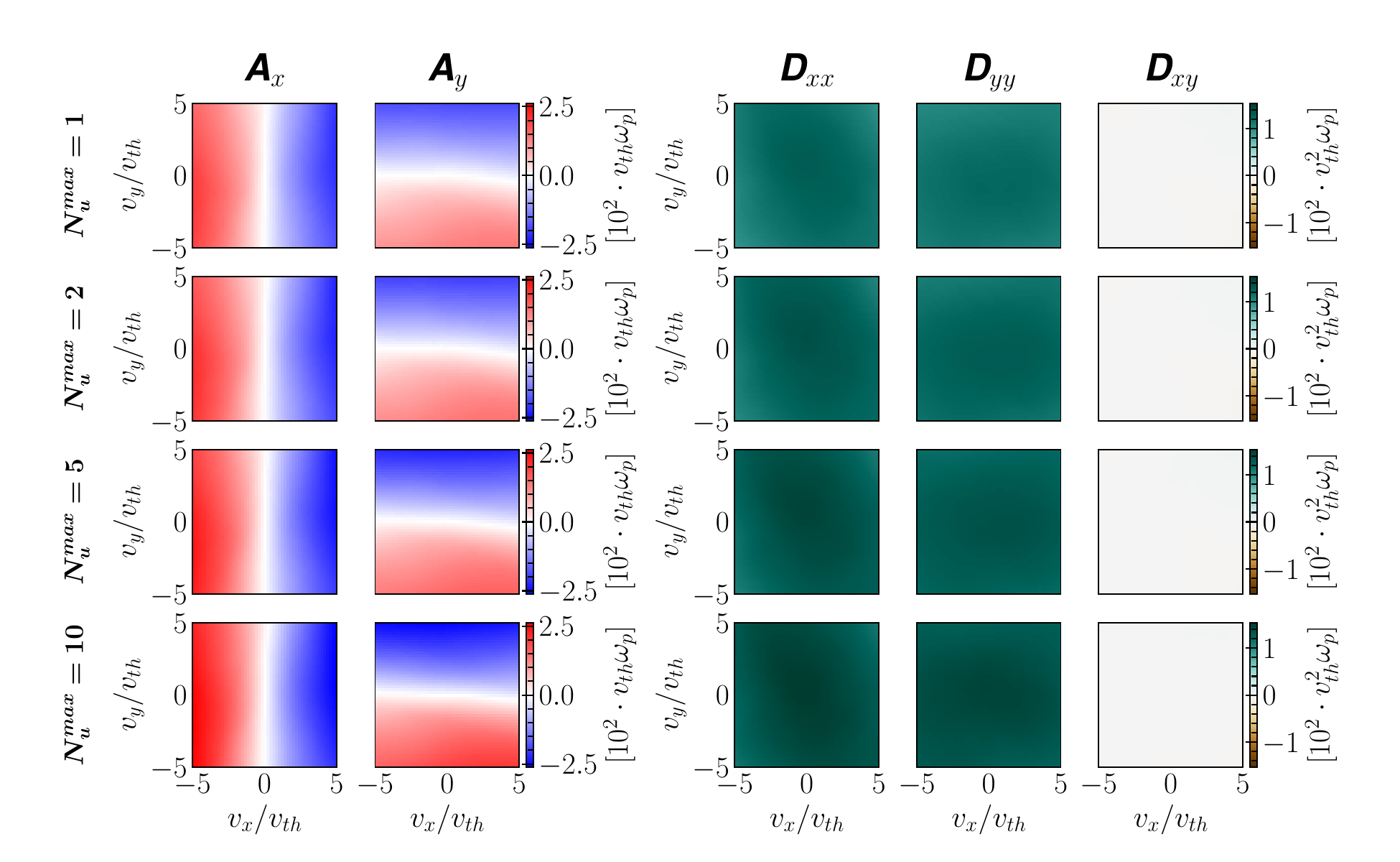}
    \caption{Illustration of the impact of the maximum training temporal unroll length on PS-NN (top) and PS-NN-Multi (bottom) models for $index=0$. Changes are not as clear as for PS-Tensor models, but overall, there is an increase in the predicted advection and diffusion values with higher $N_u^{max}$.}
    \label{fig:AD_comparison_unroll_nn_index_extra}
\end{figure}

Finally, in Figure~\ref{fig:AD_comparison_unroll_nn_index_extra} we provide the impact of temporal unrolling on PS-NN and PS-NN-Multi models for $index=0$ (same as used for Figure~\ref{fig:AD_comparison_unroll_tensor_index_0}). Overall, increasing $N_u^{max}$ leads to an increase in the predicted advection and diffusion coefficients but, unlike for PS-Tensor, it does not change the overall form of the operator. The difference in behavior is somehow expected given that NN models impose a smoothness bias.

\clearpage
\section{Enforcing Known Symmetries}
\label{app:enforcing_known_symmetries}

\begin{figure}
    \centering
    \includegraphics[width=0.7\linewidth]{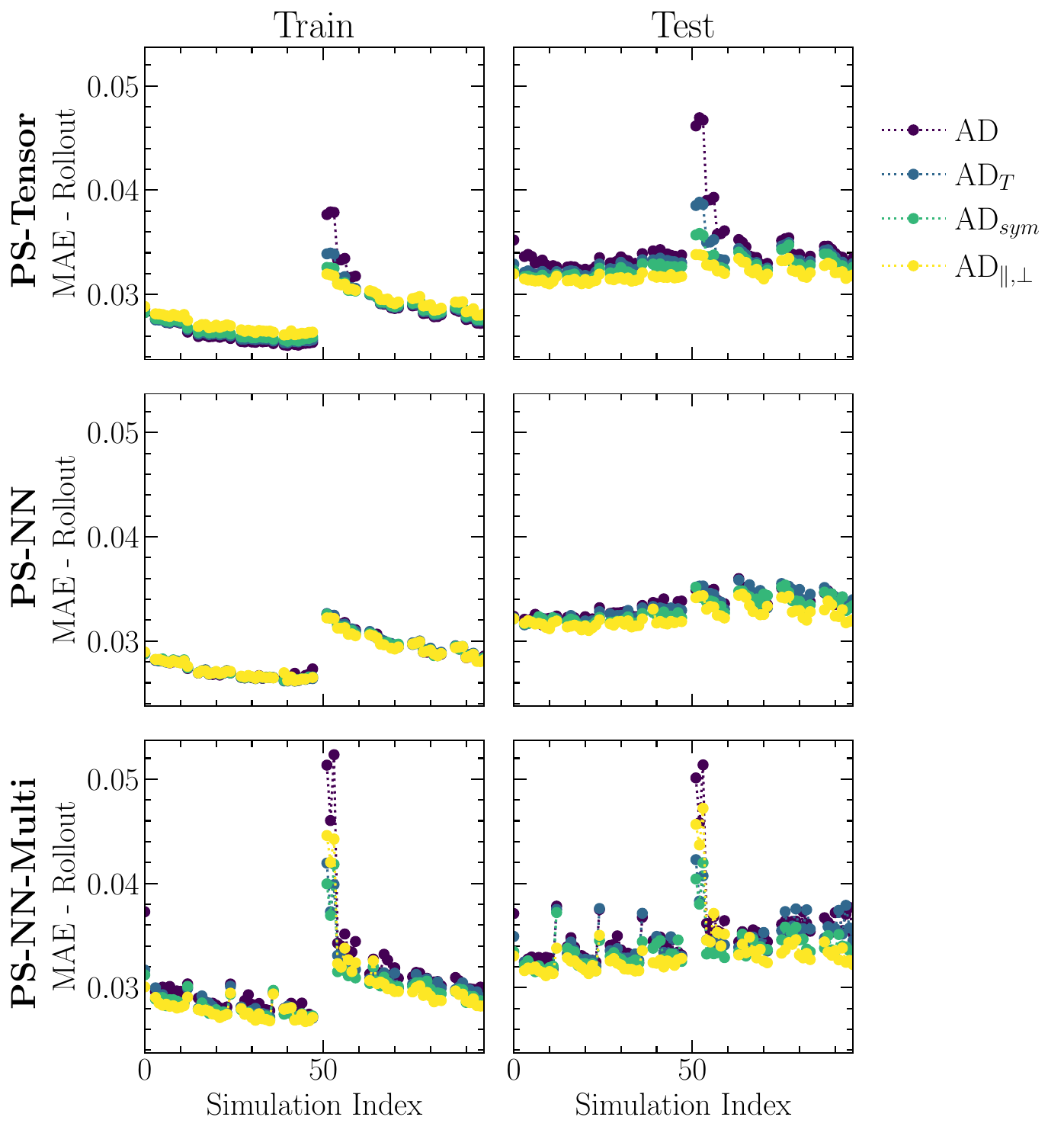}
    \caption{Rollout errors per simulation of the dataset (table~\ref{tab:simulation_parameters}) for phase space models imposing different symmetries. Results are shown for the train and test subpopulations. It is clear that imposing stricter symmetris leads to improved performance at test time.}
\label{fig:l1_model_comparison_per_simulation_sym}
\end{figure}

In Figure~\ref{fig:l1_model_comparison_per_simulation_sym}, we show the average rollout error over subpopulations per simulation of the dataset as a function of the enforced symmetries. These are the values used to produce the boxplot in Figure~\ref{fig:l1_model_comparison_boxplot_sym}. It is clear that enforcing stricter (and correct) symmetries improves generalization in the test set, even if it is associated with a slight error increase in the train set.

\begin{figure}
    \centering
    \includegraphics[width=0.6\linewidth]{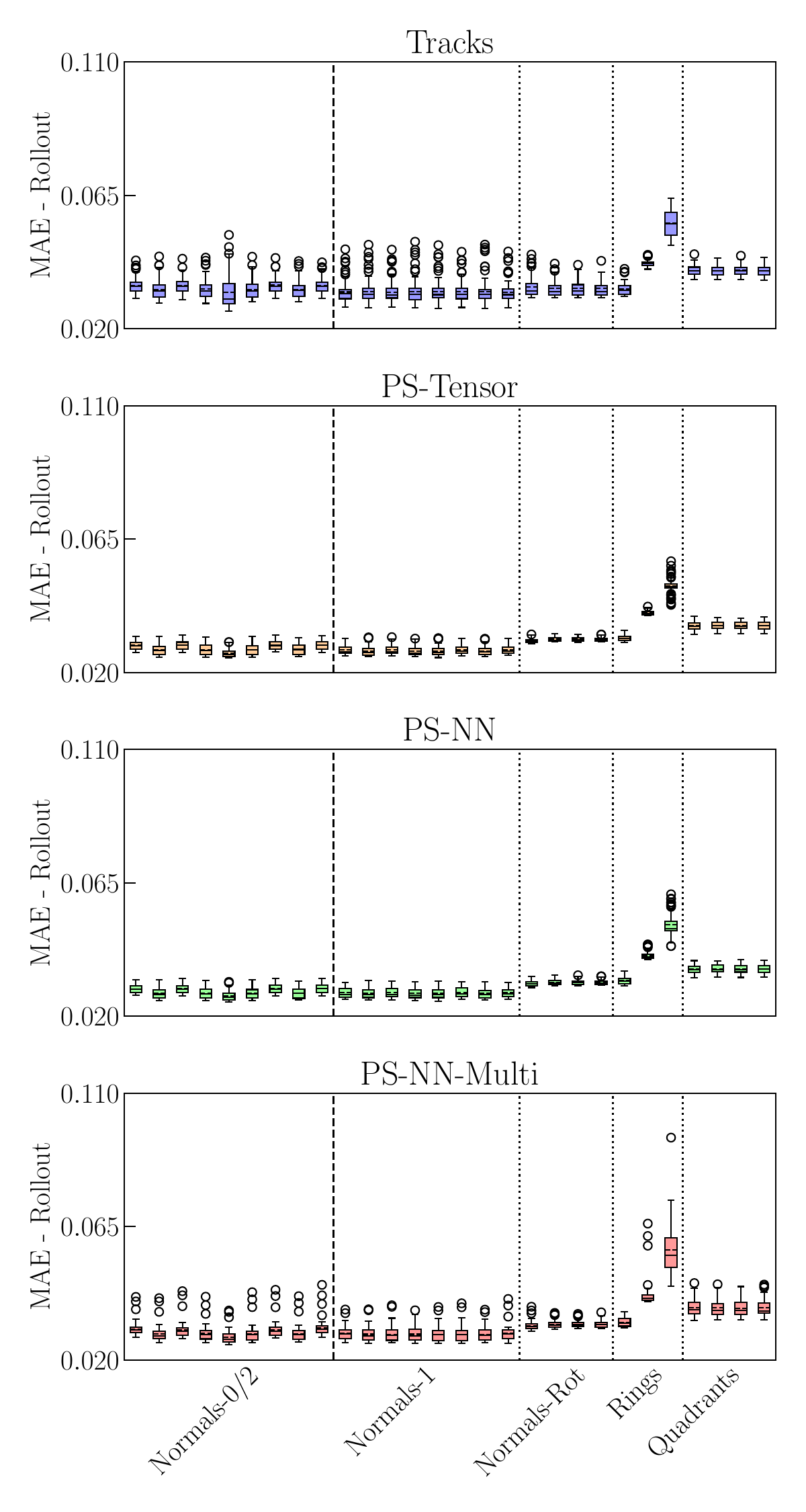}
    \caption{Rollout errors for different initial subpopulations for $\mathrm{AD}_{\parallel,\perp}$ models. The Tracks values are provided for quick reference. When compared with models that do not enforce any symmetries (Figure~\ref{fig:l1_model_comparison_per_subpopulation}), we observe a significant reduction in the number of outliers (especially for PS-Tensor and PS-NN-Multi). This is a consequence of restricting the degrees of freedom of the models, which enforces smoother solutions and better estimates in regions with originally less statistics (high $v$).}
    \label{fig:l1_model_comparison_per_subpopulation_sym}
\end{figure}

In Figure~\ref{fig:l1_model_comparison_per_subpopulation_sym}, we provide for $\mathrm{AD}_{\parallel,\perp}$ models, similar statistics but now with respect to the initial subpopulations used. When compared with models that do not enforce any symmetries (Figure~\ref{fig:l1_model_comparison_per_subpopulation}), we observe a significant reduction in the number of outliers across all subpopulations. This is mainly due to smoother estimates of operators at higher values of $v$, caused by a reduction in the degrees of freedom of the models.

\begin{figure}
    \centering
    \includegraphics[width=\linewidth]{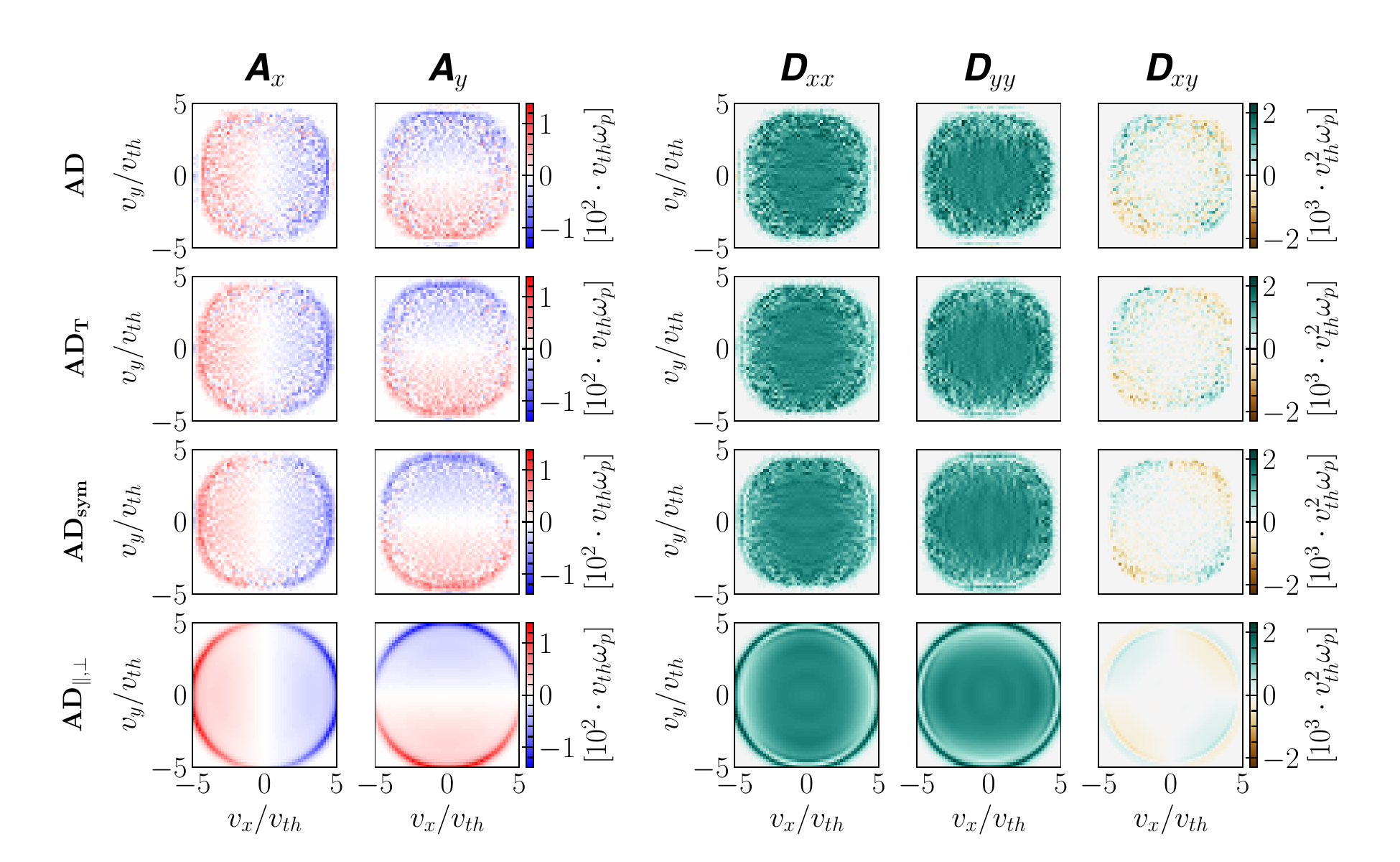}
    \includegraphics[width=\linewidth]{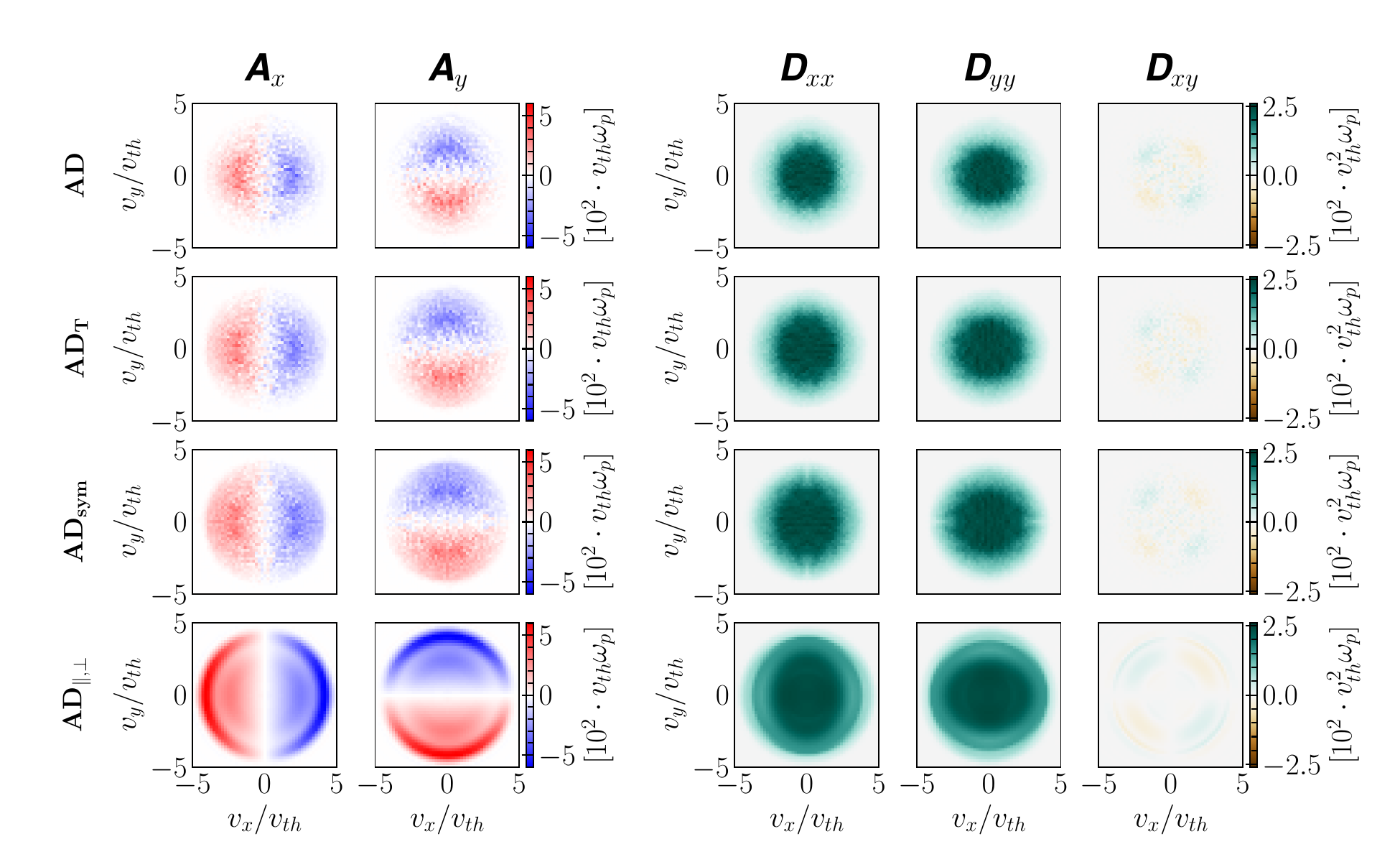}
    \caption{Advection and diffusion PS-Tensor models retrieved with different symmetries for simulation $index=47$ ($N_{ppc}=25$, $m=4$, $\Delta_x/\lambda_D=1, v_{th}=0.1c)$ and $index=51$ ($N_{ppc}=4$, $m=2$, $\Delta_x/\lambda_D=2, v_{th}=0.01c)$. Similar to Figure~\ref{fig:AD_comparison_sym_tensor_index_0}, enforcing symmetries leads to smoother operators defined over a larger range of the phase space. Artifacts at high $v$ are similarly observed, and they are caused by the reduced statistics provided during training for these phase space regions.
    }
    \label{fig:AD_comparison_sym_tensor_index_extra} 
\end{figure}

\begin{figure}
    \centering
    \includegraphics[width=\linewidth]{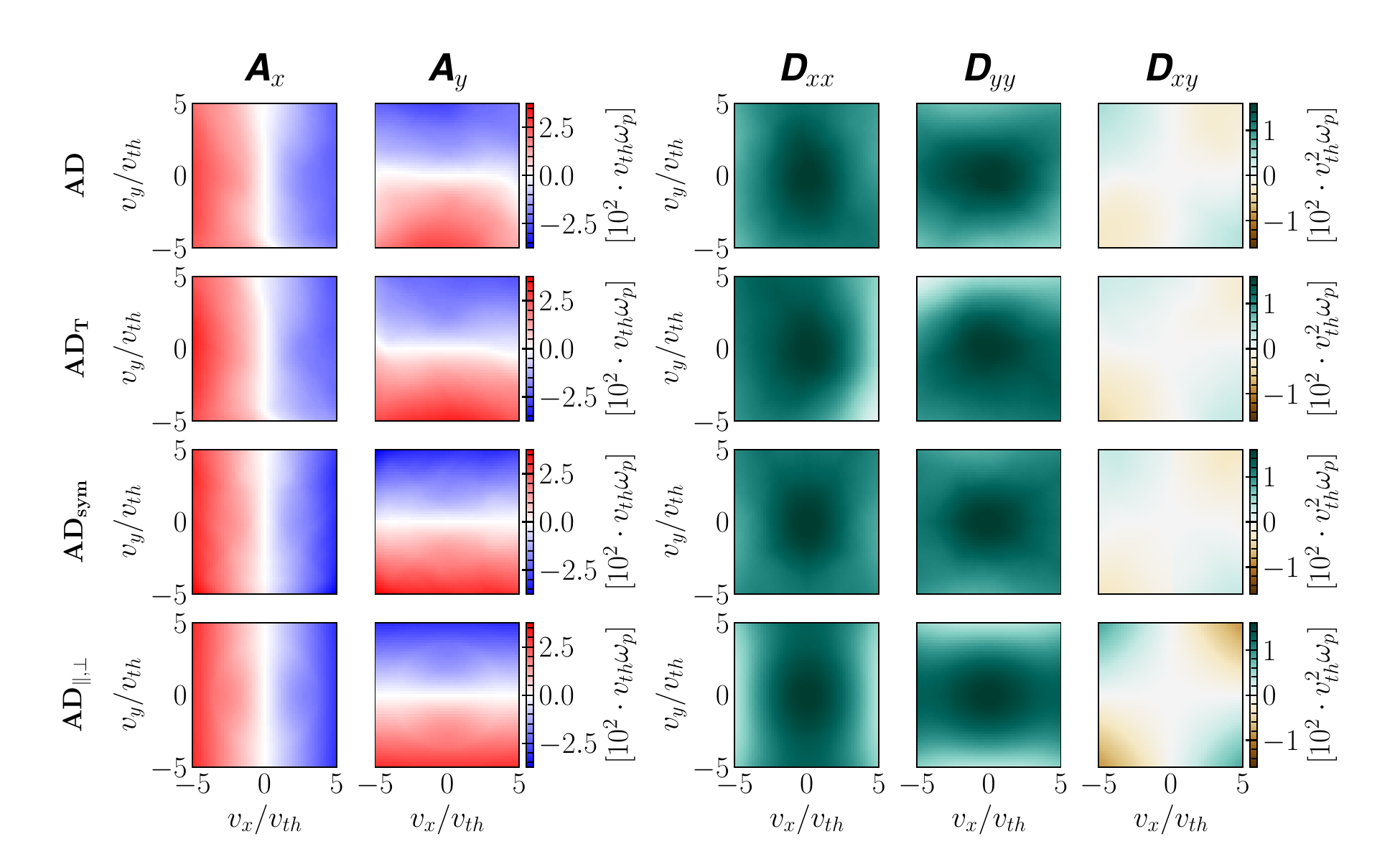}
    \includegraphics[width=\linewidth]{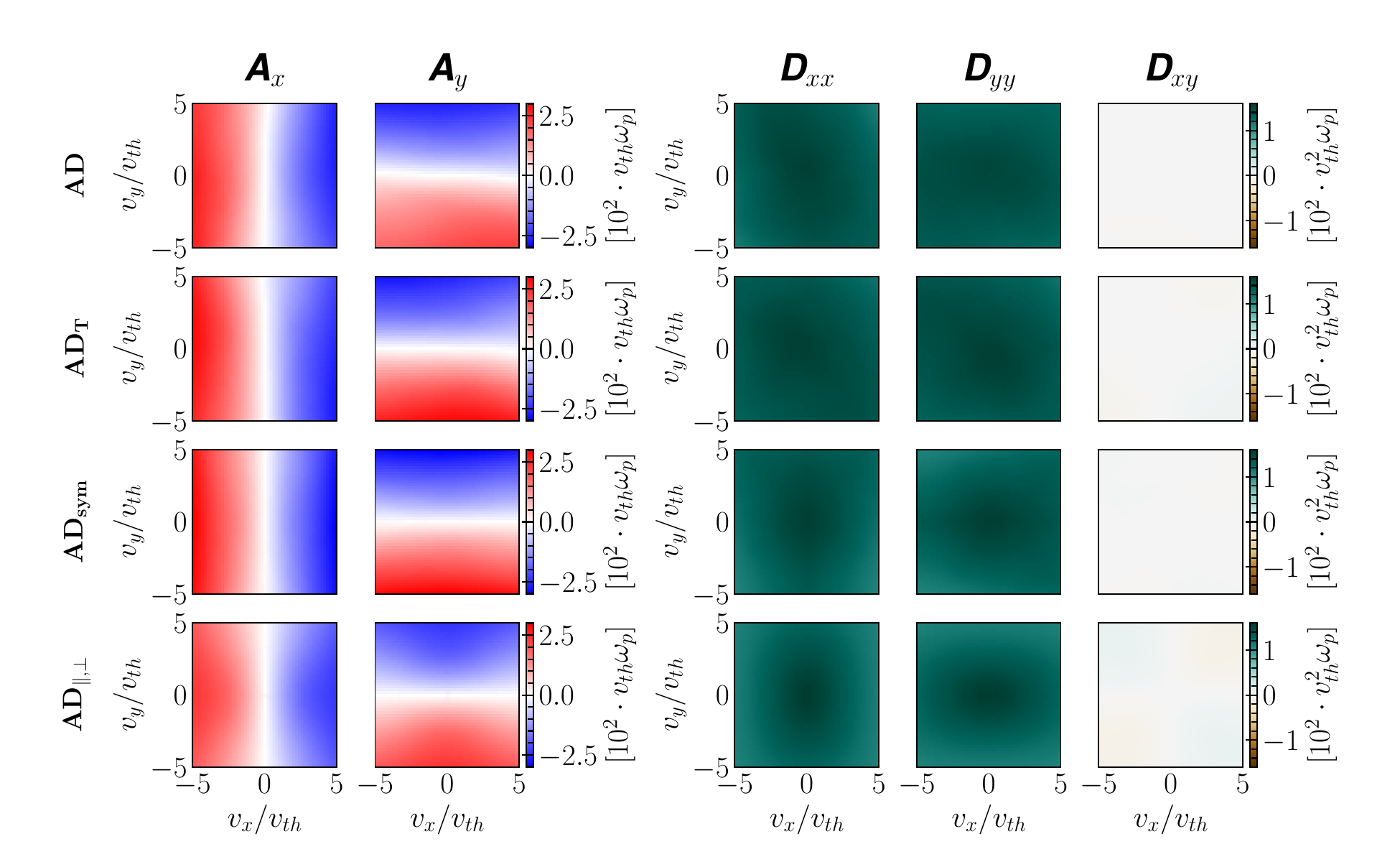}
    \caption{Retrieved PS-NN (top) and PS-NN-Multi (bottom) models for $index=0$ with different symmetries enforced. Overall, imposing symmetries leads to more accurate and smoother predictions at high $v$. For PS-NN-Multi, it seems to also facilitate model expressivity, leading us to believe that larger model sizes might in fact achieve better performance.}
    \label{fig:AD_comparison_nn_sym_index_0}
\end{figure}

In Figure~\ref{fig:AD_comparison_sym_tensor_index_extra}, we provide additional examples of the impact of symmetries on the PS-Tensor models retrieved for the different simulations. These examples reinforce the conclusions from Figure~\ref{fig:AD_comparison_sym_tensor_index_0}, where it was observed that imposing symmetries leads to a smoother operator, which covers a larger region of phase space. Finally, in Figure~\ref{fig:AD_comparison_nn_sym_index_0}, we provide examples of the impact of enforcing symmetries on NN models. For these models, imposing symmetries improves estimates at high $v$ and in the case of PS-NN-Multi also enhances the expressivity of the model.

Overall, these results reinforce the argument that imposing known symmetries is not only beneficial for performance and ensuring physical soundness but also helps to reduce the impact of a lack of statistics in the collected data.